\documentclass[12pt]{article}

\usepackage{PRIMEarxiv}

\usepackage{setspace}

\usepackage[utf8]{inputenc} 
\usepackage[T1]{fontenc}    

\usepackage{hyperref}       
\usepackage{url}            
\usepackage{booktabs}       
\usepackage{amsfonts}       
\usepackage{nicefrac}       
\usepackage{microtype}      
\usepackage{lipsum}
\usepackage{fancyhdr}       
\usepackage{graphicx}       
\graphicspath{{graphs/}}     
\usepackage{caption}
\usepackage{subcaption}
\usepackage[dvipsnames]{xcolor}
\usepackage{epsfig}
\usepackage{textcomp}
\usepackage{bm}

\usepackage{algorithm}
\usepackage{algorithmic}
\usepackage{dsfont}
\usepackage{natbib}
\usepackage{mathrsfs}
\usepackage{multirow, makecell}
\usepackage{tabularray}
\usepackage{soul}

 \usepackage{amsmath}

\title{Optimal Adaptive SMART Designs with Binary Outcomes
}

\author{
  Rik Ghosh \\
  Department of Mathematics \\
  Indian Institute of Technology Guwahati, Assam \\
  \texttt{rik.ghosh@iitg.ac.in}
     \AND
Bibhas Chakraborty\\
Centre for Quantitative Medicine, Duke-NUS Medical School,\\
National University of Singapore, Singapore \\
and\\
Department of Statistics and Data Science, National University of Singapore, Singapore \\
and\\
Department of Biostatistics and Bioinformatics, Duke University, Durham, NC, USA\\
    \texttt{bibhas.chakraborty@duke-nus.edu.sg }
\AND
Inbal Nahum-Shani, Megan E. Patrick\\
Institute for Social Research\\
University of Michigan \\
\texttt{inbal@umich.edu, meganpat@umich.edu}
 \AND
  Palash Ghosh \\
Department of Mathematics \\
Jyoti and Bhupat Mehta School of Health Sciences and Technology\\
  Indian Institute of Technology Guwahati, Assam \\
  and \\
  Centre for Quantitative Medicine, Duke-NUS Medical School, National University of Singapore, Singapore \\
  \texttt{palash.ghosh@iitg.ac.in} \\
  \\
  Correspondence to: palash.ghosh@iitg.ac.in\\
}

\begin{document}

\maketitle

\pagebreak

\begin{abstract}
In a sequential multiple-assignment randomized trial (SMART), a sequence of treatments is given to a patient over multiple stages. In each stage, randomization may be done to allocate patients to different treatment groups. Even though SMART designs are getting popular among clinical researchers, the methodologies for adaptive randomization at different stages of a SMART are few and not sophisticated enough to handle the complexity of optimal allocation of treatments at every stage of a trial. Lack of optimal allocation methodologies can raise serious concerns about SMART designs from an ethical point of view. In this work, we develop an optimal adaptive allocation procedure to minimize the expected number of treatment failures for a SMART with a binary primary outcome. Issues related to optimal adaptive allocations are explored theoretically with supporting simulations. The applicability of the proposed methodology is demonstrated using a recently conducted SMART study named M-Bridge for developing universal and resource-efﬁcient dynamic treatment regimes (DTRs) for incoming ﬁrst-year college students as a bridge to desirable treatments to address alcohol-related risks.
	\vspace{0.5cm}
\end{abstract}


\section{Introduction}
\begin{figure}[]
     \centering
     \begin{subfigure}[b]{0.49\textwidth}
         \centering
         \includegraphics[width=\textwidth, trim=1cm 1.5cm 0cm 1.5cm]{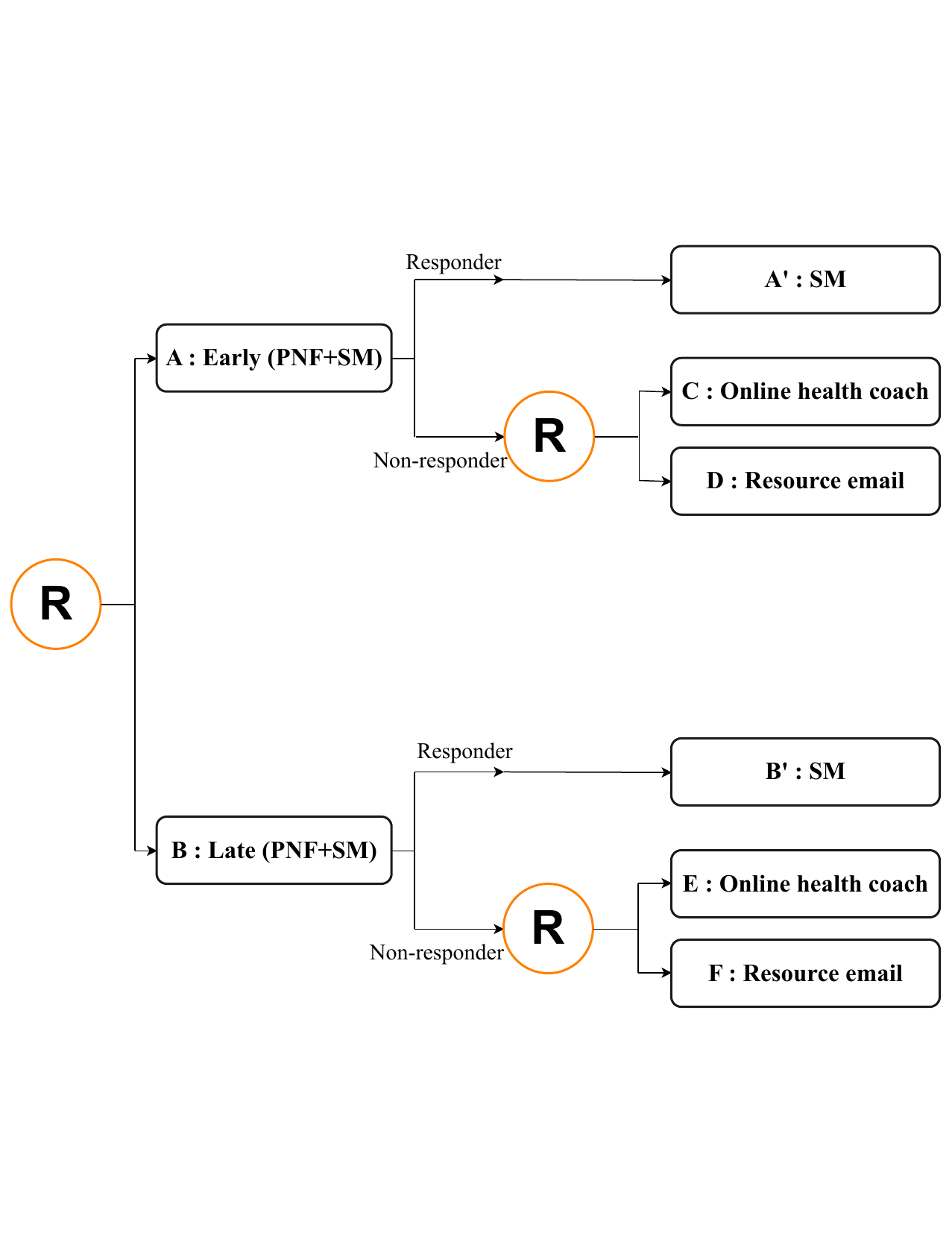}
         \caption{A schematic diagram of a 2-Stage SMART following M-Bridge study. Here $R$ represents randomization.\\ }
         \label{fig:SMART_diagram_scheme}
     \end{subfigure}
     \hfill
     \begin{subfigure}[b]{0.49\textwidth}
         \centering
         \includegraphics[width=\textwidth, height = 7cm, , trim=1cm 2cm 0cm 2cm]{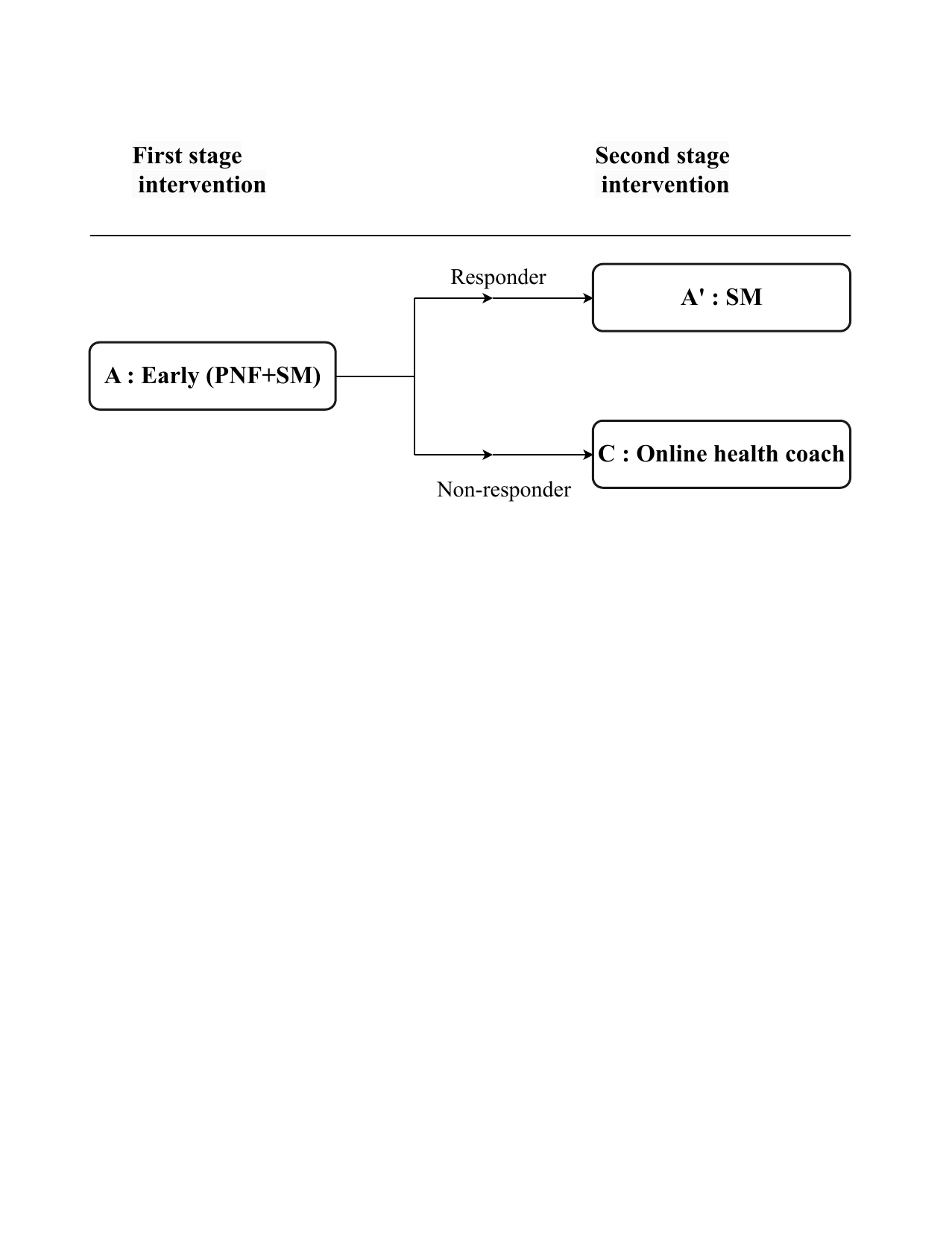}
         \caption{An example dynamic treatment regime (DTR) to reduce heavy drinking and related risks among college students from M-Bridge study. }
         \label{fig:DTR_diagram_scheme}
     \end{subfigure}
        \captionsetup{justification=centering}
        \caption{}
        \label{fig:Design diagrams}
\end{figure}

In recent times, Sequential Multiple Assignment Randomized Trials (SMARTs), an experimental design, have become popular to develop adaptive sequences of treatments (interventions) to cater to the heterogeneity in response from different treatments given to patients sequentially over a time period \citep{collins2004conceptual, Chak_book, chakraborty2014dynamic}. SMART design involves multiple stages. One patient can be randomized more than once over different stages (see Figure \ref{fig:SMART_diagram_scheme}). Using SMART, one can develop an empirically grounded dynamic treatment regime (DTR) \citep{murphy03}. A DTR is defined as a sequence of decision rules that dictates how, when, and what quantity of a specific treatment/intervention to be administered to a patient \citep{almirall2014introduction, nahum2019introduction}. DTRs are also known as adaptive interventions (AIs) \citep{ghosh2020noninferiority} or adaptive treatment strategies \citep{almirall2012smarter}. As a motivating example, consider the M-bridge SMART experimental design to reduce heavy drinking and related risks among college students \citep{patrick2020sequential}. In this study, there is an \textit{assessment only control} arm (without any treatment). We do not consider this control arm in the present work. The remaining study design of the M-Bridge is described in Figure \ref{fig:SMART_diagram_scheme}. In the first stage, all the participants (first-year college students) were administered the two cost-effective and low-burden treatments as personalized normative feedback (PNF) and self-monitoring (SM) of their alcohol use during the Fall semester, 2019. However, two distinct treatments were decided (randomized with a 1:1 ratio) based on the delivery timing of the combined universal preventive treatment (PNF + SM). The treatment `early' denotes administering PNF + SM before the start of the Fall semester. In contrast, the treatment `late' denotes giving the same (PNF + SM) during the first month of the same semester. Note that, studies have shown that PNF has the potential to reduce heavy alcohol use among college students \citep{neighbors2004targeting}.  Based on SM (high-intensity or frequent binge drinking), the participants self-identify as heavy drinkers, called non-responders. In the second stage, non-responders are then randomized (1:1 ratio) a second time to either a resource email or an invitation to have online health coaching (see Figure \ref{fig:SMART_diagram_scheme}). After receiving the first stage treatment, those who seem to be doing well, called responders, continue with the SM only treatment in the second stage. In M-Bridge, there are four embedded DTRs. In the first DTR (see Figure \ref{fig:DTR_diagram_scheme}), at the first stage, before the first semester (early treatment), participants received emailed personalized normative feedback (PNF) and were asked to self-monitoring (SM) their alcohol use during the Fall semester. In this DTR, once a student was identified as a heavy drinker (non-responder), the second stage bridging strategy was an online health coach, otherwise (responder) student would continue with SM.

Generally, in the SMART design, participants are randomized with equal probabilities to the available treatments at every stage. Even though this approach maximizes the statistical power to compare different DTRs embedded in a SMART, it raises the ethical question by ignoring intermediate information that some of the treatments are doing better than others \citep{wang2022adaptive}. In other words, despite having information about the better treatment, we assign inferior treatment to the same number of patients getting the better treatment. The use of adaptive design in SMART to change the randomization probabilities is limited as opposed to randomized controlled trials (RCTs). Here, `adaptive design' means adapting treatment decisions between patients instead of within patients \citep{cheung2015sequential}. In a non-adaptive SMART, DTRs are adapted within patients rather than between patients. Therefore, an adaptive SMART considers both within and between patients' adaption of treatments. A Q-learning-based adaption of randomization (SMART-AR) probabilities was proposed by \cite{cheung2015sequential}. However, this approach calculates empirical randomization probabilities without defining any formal optimal criteria but aims to maximize the Q-function given the history. Recently, \cite{wang2022adaptive} developed a response adaptive SMART (RA-SMART) design to skew the randomization probabilities in favor of promising treatments, using the information on treatment history and efficacy. This work assumed that the same set of treatments is available in all stages. It used information about the effectiveness of all treatments used in the first stage to calculate the randomization probabilities in the second stage. Therefore, this design cannot be used when the set of treatments in the second stage is different from that of the first stage.


In a randomized controlled trial (RCT), the randomization of patients to different treatment groups ensures that every patient has the same opportunity of receiving any of the treatments under study \citep{chow2008design}. This process creates different treatment groups, which are similar (balanced) in all the major aspects except for the treatment every group receives \citep{suresh2011overview}. RCTs mostly use a fixed randomization scheme, where the probabilities of assigning patients to any treatment groups are pre-specified and constant \citep{sim2019outcome}. However, the fixed randomization scheme in an RCT does not allow changing the allocation probabilities during the trial, even if there is accrued information that one treatment is doing better than others based on the interim analysis of the trial data \citep{rosenberger2001optimal}. Thus, the conventional RCT process allows allocating the inferior treatment (based on the interim data) and the better treatment with the same probability \citep{atkinson2013randomised}. This raises ethical concerns about RCTs \citep{harrington2000randomized}.

Adaptive designs, allowing to alter the randomization scheme during the trial, have been proposed to overcome the shortcoming of RCTs \citep{sim2019outcome}. An adaptive RCT is flexible enough to change the allocation probabilities during the trial based on the information from the interim analysis \citep{pallmann2018adaptive}. In other words, adaptive RCTs enable the allocation of more patients to better treatment in the long run, \citep{hulley2007designing}. One of the early methods used in adaptive RCTs is the play-the-winner (PW). It was first introduced for dichotomous response in clinical trials with two treatments \citep{zelen1969play}. In PW, we place a ball marked with `U' in the urn when success is obtained with treatment U or a failure with treatment V. Similarly, a ball marked with `V' is placed when a success is obtained with treatment V or a failure with treatment U \citep{wei1978randomized}. For a new patient, a ball is drawn at random from the urn without replacement to allocate the corresponding treatment; if the urn is empty, then the allocation is done by tossing an unbiased coin. However, due to delays in getting the response from patients, the PW rule is inconvenient in practical trials as most of the allocation is done by tossing an unbiased coin. Later a modified PW rule was introduced as the randomized play-the-winner (RPW) \citep{atkinson2013randomised}. In RPW, an urn consisting of two types (U and V) of balls representing two different treatments (U and V), initially with the same number. For each patient, one ball is drawn (with replacement) at random, and the corresponding treatment is administered. When the outcome of any previous patient is a success (improvement) who obtained treatment U (or V), then $\alpha$ (or $\beta$) number of U-type balls are added to the urn, and $\beta$ (or $\alpha)$ number of V-type balls are added \citep{wei1978randomized}. The exact opposite scheme is done when the outcome of that patient is a failure (no improvement). This process is repeated till the end of the trial. Note that the RPW rule is not based on any formal optimality criterion. However, in RPW, the limiting allocation is intuitively allocating the treatments using relative risk. \cite{rosenberger2001optimal} proposed an adaptive design that allocates treatments based on the formal optimal criterion for RCT with binary outcomes.

The development of DTR using SMART design is motivated by the notion that a treatment that is promising in the short term may not be beneficial in the long term \citep{butryn2011behavioral, almirall2014introduction}. Similarly, a treatment that looks not so beneficial at the first stage may be part of the best two-stage DTR among all the embedded DTRs in a SMART. Thus, adaptive SMART that alters randomization probabilities to give better treatments to more patients should be developed by considering the long-term benefits (end of the trial) of the patients. In this work, we build a novel adaptive randomization procedure that minimizes the total expected number of failures from the entire SMART with a binary primary outcome. To achieve that, we first develop the optimal allocation ratios for the second-stage randomization processes using the methodology proposed by \cite{rosenberger2001optimal} for two-arms RCT with binary outcome. Then the first-stage optimal allocation ratio is obtained recursively, passing the optimal allocation information backward from the second stage to the first stage. Derivation of the first-stage optimal allocation ratio, the corresponding adaptive allocation process and the asymptotic distribution of the optimal allocation ratio are the main contributions of the current article. 

The remaining paper proceeds as follows. In Section \ref{sec : genframe}, we develop the general framework required for this work. Section \ref{optimum_allo} describes the optimal allocation criteria and derives the optimal allocation ratios. An adaptive allocation process is developed in Section \ref{sec : adaptive alloc proc}. Section \ref{sec : testing structure} describes hypothesis testing to compare two embedded DTRs. Sections \ref{sec : simul study} and \ref{sec: Mbridge} shows the simulation study and an application to real data, respectively. Section \ref{sec:discussion} ends with a discussion.

\section{General Framework} \label{sec : genframe}

Let $Y$ be the binary primary outcome (success or failure) observed at the end of a two-stage SMART. Here, we use generic treatments termed $A, A', B, B', C, D, E$, and $F$, as shown in Figure \ref{fig:SMART_diagram_scheme}. As in Figure \ref{fig:SMART_diagram_scheme}, at the first stage, participants are randomized to the treatments $A$ (early administering of the PNF + SM) or $B$ (late administering of the PNF + SM). Let $T_1$ denote the first stage treatment, thus $T_1 \in \{A, B\}$. At the end of the first stage who started with $A$ (or $B$), based on the intermediate outcome (SM in M-bridge), responders will continue with the SM only denoted as $A'$ (or $B'$), while nonresponders will be randomized to $C$ (online health coach) or $D$ (resource email) if $T_1 = A$, and $E$ (online health coach) or $F$ (resource email) if $T_1 = B$. In this case, $C$ and $E$ (similarly $D$ and $F$) are the same treatment. However, they may be different in a SMART.  Thus, $T_2 \in \{A', C, D\}$ if $T_1 = A$ and $T_2 \in \{B', E, F\}$ if $T_1 = B$. The complete treatment sequence is expressed by $\{T_1, T_2\}$. Define $n_{T_1T_2}$ as the number of participants who obatined $\{T_1,T_2\}$.  Let $R_{T_1}, T_1 \in \{A, B\}$ be the response indicator (1: responder, 0: non-responder) who obtained treatment $T_1$ at the first stage. 


The objective of this work is to find optimal values of the three allocation ratios, $\tau_A = \frac{n_A}{n_B}$, $\tau_{AC} = \frac{n_{AC}}{n_{AD}}$ and $\tau_{BE} = \frac{n_{BE}}{n_{BF}}$, corresponding to three randomization processes in the two-stage SMART (see Figure \ref{fig:SMART_diagram_scheme}), where $n_A = n_{AA'} + n_{AC} + n_{AD} $, $n_B = n_{BB'} + n_{BE} + n_{BF}$. The optimal value of each of the three ratios is obtained for a fixed asymptotic variance (to reflect the power of the test) such that it minimizes the expected number of corresponding treatment failures \citep{rosenberger2001optimal}. Note that the second stage optimal allocation ratios ($\tau_{AC}$ and $\tau_{BE}$) can be viewed as a problem of adaptive allocation in traditional RCT (single stage) with two treatment arms. In this work, we directly apply the methodology proposed by \cite{rosenberger2001optimal} to find the second stage optimal allocation ratios. Our main contribution is to propose an adaptive sequential design that allocates patients to the first-stage treatments where both the first and the second-stage allocation ratios are optimal. The proposed method ensures the minimization of the total expected number of failures from the entire SMART by recursively passing the optimal allocation information backward (from the second stage to the first stage). In other words, the second-stage optimal allocation ratios are not dependent on the first-stage optimal allocation ratio. However, the first-stage optimal allocation ratio is obtained using the information from the second-stage optimal allocation ratios. 

Define $p_{T_1T_2}$ ($q_{T_1T_2} = 1 - p_{T_1T_2}$) as the probability of success when the (end of study) binary primary outcome $Y=1$ (success) corresponding to a patient who obtained the treatment sequence $\{T_1, T_2\}$. Similarly, $p_{T_1}$ ($q_{T_1} = 1 - p_{T_1}$) is the probability of success for a patient who obtained treatment $T_1$ at the first stage. It can be shown (see Appendix \ref{ap : first stage success probability wrt second stage}) that,
$$p_{T_1} = \gamma_{T_1} p_{T_1T_1'}+(1-\gamma_{T_1})\frac{\tau_{T_1T_2}}{1+\tau_{T_1T_2}}p_{T_1T_2}+(1-\gamma_{T_1})\frac{1}{1+\tau_{T_1T_2}}p_{T_1T_2^*},$$
where $T_2 = C, T_2^* = D$ if $T_1 = A$; $T_2 = E, T_2^* = F$ if $T_1 = B$, and $\gamma_{T_1} = P(R_{T_1} = 1)$ denotes the probability of response among the patients who obtained treatment $T_1$.

\section{Optimal Allocation Criteria}\label{optimum_allo}
In this section, we describe the general approach to find the optimal allocation ratios. The total expected number of failures from the entire SMART is minimized  considering an objective function $g(\cdot, \cdot)$ that compares two binomial success probabilities subject to a fixed asymptotic variance (\textit{avar}) of the same objective function. Note that there are multiple randomizations involved in a SMART. With respect to a specific randomization process, the expected number of failures means the expected number of patients who obtained treatments in that randomization process and then failed (in future). Among the patients who obtained $T_1$ at the first stage, the number of failures after a randomization process at the second stage is defined as 
$$F_2(\tau_{T_1T_2}) = n_{T_1T_2} q_{T_1T_2} + n_{T_1T_2^*} q_{T_1T_2^*} = \frac{n_{T_1}(1-\gamma_{T_1})}{1+\tau_{T_1T_2}} \bigg[ \tau_{T_1T_2} q_{T_1T_2} + q_{T_1T_2^*} \bigg].$$
Then the optimum allocation ratio for the second stage is
\begin{equation}
\tau_{T_1T_2}^{*} = \arg \min_{\tau_{T_1T_2}} F_2(\tau_{T_1T_2}) \mbox{ subject to }   avar\{g(\hat p_{T_1T_2}, \hat p_{T_1T_2^*})\} = \epsilon_2, \nonumber
\end{equation}
for some constant $\epsilon_2 > 0$. Similarly, the total number of failures from the entire SMART is
\begin{eqnarray}
&& F_1(\tau_{A}, \tau_{AC}, \tau_{BE}) = (n_{AA'}q_{AA'}+n_{AC}q_{AC} + n_{AD}q_{AD}) + (n_{BB'}q_{BB'}+n_{BE}q_{BE} + n_{BF} q_{BF}) \nonumber \\ 
&&= \frac{n}{1+\tau_A} \bigg [ \tau_A \gamma_Aq_{AA'} + \tau_A \frac{(1-\gamma_A)}{1+\tau_{AC}} \bigg( \tau_{AC}q_{AC} + q_{AD} \bigg) 
+  \gamma_Bq_{BB'} + \frac{(1-\gamma_B)}{1+\tau_{BE}} \bigg( \tau_{BE}q_{BE} + q_{BF} \bigg) \bigg ].   \nonumber
\end{eqnarray}
The optimum allocation ratio for the first stage is
\begin{equation}
\tau_{A}^{*} = \arg \min_{\tau_{A}} F_1(\tau_{A}, \tau_{AC}^{*}, \tau_{BE}^{*})  \mbox{ subject to }   avar\{g(\hat p_{A}, \hat p_{B})\} = \epsilon_1, \nonumber
\end{equation}
for some constant $\epsilon_1 > 0$. The common choices for the objective function $g(\cdot, \cdot)$ could be the simple difference (e.g. : $p_{AC} - p_{AD}$), the relative risk (e.g. : $q_{AC}/q_{AD}$) and the odds-ratio (e.g. : $p_{AC}q_{AD}/p_{AD}q_{AC}$).

\subsection{Optimal Allocation Ratios} \label{ssec : optimal}
Let the objective function be the simple difference between the success probabilities. Thus, for the second stage randomization to allocate patients to available treatments who started with treatment $A$ at the first stage and became non-responder, the objective function is $g(p_{AC}, p_{AD}) = p_{AC} - p_{AD}$. Similarly, the same is $g(p_{BE}, p_{BF}) = p_{BE} - p_{BF}$ among those who started with treatment $B$. For the first stage of randomization, $g(p_{A}, p_{B}) = p_{A} - p_{B}$. Using the optimum allocation criteria stated in Section \ref{optimum_allo}, the optimum allocation ratios are
\begin{equation*}
\tau_{AC}^*=\sqrt{\frac{p_{AC}}{p_{AD}}}, \hspace{0.5cm} \tau_{BE}^*=\sqrt{\frac{p_{BE}}{p_{BF}}}, \hspace{0.25cm} \mbox{and } 
\end{equation*}
\begin{equation*}
    \tau_A^*=\sqrt{\frac{(1+\tau_{BE}^*)(\gamma_A p_{AA'}(1+\tau_{AC}^*)+(1-\gamma_A) (\tau_{AC}^* p_{AC}+p_{AD}))}{(1+\tau_{AC}^*)(\gamma_B p_{BB'}(1+\tau_{BE}^*)+(1-\gamma_B) (\tau_{BE}^* p_{BE}+p_{BF}))}}.
\end{equation*}
See Appendix \ref{ap : second stage optimal alloc for diff} and \ref{ap : first stage optimal alloc for diff} for details. 
Similarly, considering the objective function as the odds-ratio, the optimum allocation ratios are
\begin{equation*}
\tau_{AC}^* = \left(\sqrt{\frac{p_{AD}}{p_{AC}}}\right)\left(\frac{q_{AD}}{q_{AC}}\right), \hspace{0.5cm} 
\tau_{BE}^* = \left(\sqrt{\frac{p_{BF}}{p_{BE}}}\right)\left(\frac{q_{BF}}{q_{BE}}\right) \hspace{0.25cm} \mbox{and } 
\end{equation*}

\begin{align*}
    \tau_A^* = & \left(\frac{1+\tau_{AC}^*}{1+\tau_{BE}^*}\right)^{\frac{3}{2}}\left(\frac{(\gamma_B p_{BB'}(1+\tau_{BE}^*)+(1-\gamma_B) (\tau_{BE}^* p_{BE}+p_{BF}))}{(\gamma_A p_{AA'}(1+\tau_{AC}^*)+(1-\gamma_A) (\tau_{AC}^* p_{AC}+p_{AD}))}\right)^{\frac{1}{2}}\\ 
    & \hspace{1.75cm} \times \left(\frac{(\gamma_B q_{BB'}(1+\tau_{BE}^*)+(1-\gamma_B) (\tau_{BE}^* q_{BE}+q_{BF}))}{(\gamma_A q_{AA'}(1+\tau_{AC}^*)+(1-\gamma_A) (\tau_{AC}^* q_{AC}+q_{AD}))}\right).
\end{align*}
See Appendix \ref{ap : second stage optimal alloc for odds ratio} and \ref{ap : first stage optimal alloc for odds ratio} for details. 
Similarly, considering the objective function as the relative-risk, the optimum allocation ratios are
\begin{equation*}
\tau_{AC}^* = \left(\sqrt{\frac{p_{AC}}{p_{AD}}}\right)\left(\frac{q_{AD}}{q_{AC}}\right), \hspace{0.5cm} 
\tau_{BE}^* = \left(\sqrt{\frac{p_{BE}}{p_{BF}}}\right)\left(\frac{q_{BF}}{q_{BE}}\right) \hspace{0.25cm} \mbox{and }
\end{equation*}

\begin{align*}
    \tau_A^* = & \left(\frac{1+\tau_{AC}^*}{1+\tau_{BE}^*}\right)^{\frac{1}{2}}\left(\frac{(\gamma_A p_{AA'}(1+\tau_{AC}^*)+(1-\gamma_A) (\tau_{AC}^* p_{AC}+p_{AD}))}{(\gamma_B p_{BB'}(1+\tau_{BE}^*)+(1-\gamma_B) (\tau_{BE}^* p_{BE}+p_{BF}))}\right)^{\frac{1}{2}}\\ 
    & \hspace{1.75cm} \times \left(\frac{(\gamma_B q_{BB'}(1+\tau_{BE}^*)+(1-\gamma_B) (\tau_{BE}^* q_{BE}+q_{BF}))}{(\gamma_A q_{AA'}(1+\tau_{AC}^*)+(1-\gamma_A) (\tau_{AC}^* q_{AC}+q_{AD}))}\right).
\end{align*}
See Appendix \ref{ap : second stage optimal alloc for relative risk} and \ref{ap : first stage optimal alloc for relative risk} for details. 

Note that the optimal allocation ratios ($\tau_{AC}^*$ and $\tau_{BE}^*$) corresponding to the second stage randomizations are functions of related success probabilities ( \{$p_{AC}, p_{AD}$\} or \{$p_{BE}, p_{BF}$\}) only. However, the first stage optimal allocation ratio, $\tau_A^*$, is a function of all the success probabilities in the entire SMART, the second stage optimal allocation ratios, and the response probabilities ($\gamma_A$ and $\gamma_B$). In other words, $\tau_A^*$ takes into account optimal allocations at the second stage. For high values of $\gamma_A$ and $\gamma_B$ (close to 1), the expression for $\tau_A^*$ approximately becomes
$$\sqrt{\frac{p_{AA'}}{p_{BB'}}},  \left(\sqrt{\frac{p_{BB'}}{p_{AA'}}}\right)\left(\frac{q_{BB'}}{q_{AA'}}\right) \mbox{ and } \left(\sqrt{\frac{p_{AA'}}{p_{BB'}}}\right)\left(\frac{q_{BB'}}{q_{AA'}}\right),$$
corresponding to the choice of the objective function as, simple difference, odds-ratio and relative-risk, respectively. In other words, when $\gamma_A$ and $\gamma_B$ are close to 1, the SMART design described here essentially becomes a two-arm RCT. 

\section{Adaptive Allocation Process} \label{sec : adaptive alloc proc}
We need to develop an adaptive allocation (randomization) procedure that ensures patients are allocated to available treatments in accordance with the optimal allocation ratios derived in Section \ref{ssec : optimal}. However, the true optimal allocation ratios  are dependent on unknown parameters. Hence, for implementation in a real SMART, we should develop a sequential design that will approximate the optimal allocation ratios. It is also desirable that the proposed sequential design should ensure the convergence of sample allocation ratios to the corresponding optimal allocation ratios. Let $Y_i$ be the binary primary outcome variable, which can take two values, $1$ for success and $0$ for failure, corresponding to the $i^{th}$ patient; $i=1, \cdots, n$. $T_{1i}$ and $T_{2i}$ denote the assigned first and second stage treatments to the $i^{th}$ patient, respectively. 
Thus, we get the total number of patients getting treatment $(T_1,T_2)$ out of $k$ patients, as $n_{T_1T_2,k} = \sum_{i=1}^k I(T_{1i}=T_1, T_{2i}=T_2)$, where $T_1$ and $T_2$ takes the values as stated in Section \ref{sec : genframe}, and $n_{T_1T_2,n} = n_{T_1T_2}$. The corresponding estimates of the success probabilities are obtained as $\hat{p}_{T_1T_2,k} = \left(\sum_{i=1}^k I(T_{1i}=T_1, T_{2i}=T_2)Y_i\right)/n_{T_1T_2,k}$. Define $\mathscr{F}_i = \{Y_1,Y_2,...,Y_i, T_{11},T_{12},...,T_{1i}, T_{21},T_{22},...,T_{2i} \}$ as the history of primary outcome variables, first and second stage allocated treatments for the first $i$ patients. Let $E_i(\cdot) = E(\cdot|\mathscr{F}_i)$ denote the conditional expectation. 

Now, taking the objective function as the simple difference of the success probabilities, the second stage allocation process can be explained with the help of the above expectation as,
\begin{equation}\label{2nd stage allo process}
    E_{i-1}(I(T_{2i}=t_2|T_{1i}=t_1,R_{T_{1i}}=0)) = \frac{\sqrt{\hat{p}_{t_1t_2,i-1}}}{\sqrt{\hat{p}_{t_1t_2,i-1}}+\sqrt{\hat{p}_{t_1t_2^*,i-1}}},
\end{equation}
where, $t_2, t_2^* \in (C,D)$ if $t_1=A$ or $t_2, t_2^* \in (E,F)$ if $t_1 = B$ and $t_2^* \neq t_2$ and $R_{T_{1i}}$ is the same as defined in Section \ref{sec : genframe} for the $i^{th}$ patient. In words, (\ref{2nd stage allo process}) refers to the estimation of the second stage success probability $(\hat{p}_{t_1 t_2,i})$ based on the first $(i-1)$ sequentially enrolled patients to be used for the adaptive randomization of the $i^{th}$ patient.  In the same line, in the first stage randomization, the allocation process can be expressed as,
\begin{equation}\label{1st stage allo process}
    E_{i-1}(T_{1i}) = \frac{\sqrt{l_{i-1}}}{\sqrt{l_{i-1}}+\sqrt{m_{i-1}}},
\end{equation}
where $l_{i-1} = (1+ \hat\tau_{BE, i})(\gamma_A \hat{p}_{AA',i-1}(1+\hat\tau_{AC, i})+(1-\gamma_A) (\hat\tau_{AC, i}\ \hat{p}_{AC,i-1}+\hat{p}_{AD,i-1}))$ and $m_{i-1} = (1+\hat\tau_{AC, i})(\gamma_B \hat{p}_{BB',i-1}(1+\hat\tau_{BE, i})+(1-\gamma_B) (\hat\tau_{BE, i}\ \hat{p}_{BE,i-1}+\hat{p}_{BF,i-1}))$. Note that, $\hat\tau_{AC, i} = \sqrt{\frac{\hat p_{AC, i-1}}{\hat p_{AD, i-1}}}$, $\hat\tau_{BE, i} = \sqrt{\frac{\hat p_{BE, i-1}}{\hat p_{BF, i-1}}}$ denote the estimated second stage allocation ratios for the $i^{th}$ patient (who obtain either $A$ or $B$ at the first stage and is a non-responder)  based on the history of $(i-1)$ patients. Similar to the explanation mentioned in Section \ref{ssec : optimal}, the first stage allocation process recursively considers the estimated allocations ($\hat\tau_{AC, i}, \hat\tau_{BE, i}$) at the second stage. It can be seen that the above allocation process replaces the unknown success probabilities in the optimal allocation ratios obtained in Section \ref{ssec : optimal} by the current estimates of the success probabilities of each treatment sequence ($\hat{p}_{t_1t_2,i-1})$.

In practice, an investigator allocates patients to available treatments using the allocation process described in (\ref{2nd stage allo process}) and (\ref{1st stage allo process}) for the second and the first stages, respectively. Therefore, it is desirable that limiting allocation (using (\ref{2nd stage allo process}) and (\ref{1st stage allo process})) is optimal. Using results from \citet{rosenberger2001optimal, melfi2001adaptive}, for the second stage randomization with the objective function as the simple difference between the success probabilities, 
\begin{eqnarray}
\hat\tau_{AC, n} & \xrightarrow{a.s} \sqrt{\frac{p_{AC}}{p_{AD}}}, \mbox{ and } \hat\tau_{BE, n} & \xrightarrow{a.s} \sqrt{\frac{p_{BE}}{p_{BF}}},
\label{eq : Asymptotics of second stages}
\end{eqnarray}
where $a.s$ denotes the almost sure convergence for a large value of $n$. 
Similarly, as shown in (\ref{eq : asymp conv of first stage optimal alloc for diff}) of Appendix \ref{ap : first stage optimal alloc for diff}, 
\begin{eqnarray}
\hat\tau_{A,n} \xrightarrow{a.s} \sqrt{\frac{\left[\sqrt{p_{BE}}+\sqrt{p_{BF}}\right]\left[\gamma_Ap_{AA'}(\sqrt{p_{AD}}+\sqrt{p_{AC}})+(1-\gamma_A)\left((p_{AC})^\frac{3}{2}+(p_{AD})^\frac{3}{2}\right)\right]}{\left[\sqrt{p_{AC}}+\sqrt{p_{AD}}\right]\left[\gamma_Bp_{BB'}(\sqrt{p_{BE}}+\sqrt{p_{BF}})+(1-\gamma_B)\left((p_{BE})^\frac{3}{2}+(p_{BF})^\frac{3}{2}\right)\right]}},
\label{eq : Asymptotics of first stages}
\end{eqnarray}
where $\hat\tau_{A,n}$ denotes the estimated first stage allocation ratio for the $n^{th}$ patient based on the history of $(n-1)$ patients. The asymptotic distributions of estimated optimum allocation ratios are given by
\begin{eqnarray}
\sqrt{n}(\hat \tau_{AC}-\tau_{AC}) &\xrightarrow[]{d} & N\left(0,\frac{1}{4}\left(\frac{v_{AC}^{-1}}{p_{AC}p_{AD}}+\frac{v_{AD}^{-1}p_{AC}}{p_{AD}^3}\right)\right),\nonumber \\ 
\sqrt{n}(\hat \tau_{BE}-\tau_{BE}) &\xrightarrow[]{d}& N\left(0,\frac{1}{4}\left(\frac{v_{BE}^{-1}}{p_{BE}p_{BF}}+\frac{v_{BF}^{-1}p_{BE}}{p_{BF}^3}\right)\right), \nonumber \\
\sqrt{n}(\hat \tau_{A}-\tau_{A}) &\xrightarrow[]{d}& N\left(0, \sigma_{\tau_{A}}^2\right), \nonumber
\end{eqnarray}
where the expressions of $v_{AC}^{-1}, v_{AD}^{-1}, v_{BE}^{-1}, v_{BF}^{-1}$ and $\sigma_{\tau_{A}}^2$ are given in \ref{ap : Variance of success prob for diff} along with their derivations in detail. The notation \textit{d} over arrow denotes the convergence in distribution. The adaptive allocation processes corresponding to the two other objective functions are derived in Section \ref{ap :  Variance second stage alloc odds} and \ref{ap :  Variance first stage alloc odds} (for odds-ratio);  and \ref{ap :  Variance second stage alloc relative risk} and \ref{ap :  Variance first stage alloc relative risk} (for relative risk) of the Appendix.

\section{Hypothesis Testing} \label{sec : testing structure}
The total expected number of failures from the entire SMART can be minimized using the developed adaptive procedure. However, testing (at the end of the trial) whether two given treatment sequences (embedded DTRs) have the same or different efficacy is also essential. This inference can be made by using the Wald-type statistic \citep{rosenberger2001optimal}.  In Figure \ref{fig:SMART_diagram_scheme}, there are four embedded DTRs, denoted as $d_1 : (A,A'^RC^{1-R})$, $d_2 : (A,A'^RD^{1-R})$, $d_3 : (B,B'^RE^{1-R})$, and $d_4 : (B,B'^RF^{1-R})$. A patient whose treatment sequence is consistent with $d_1$ implies that they will start with treatment $A$ at the first stage, then continue with $A$ if they are doing well (responder) or switch to treatment $C$ otherwise (non-responder). 
Now, we compare two DTRs. For any pair of DTRs, say $(d_i, d_j)$, the proportion of success is represented by $p_{d_i}$ and $p_{d_j}$, respectively. Hence, we consider the test,
\begin{eqnarray}
    H_0:p_{d_i}=p_{d_j} \ \  vs \  \ H_1: p_{d_i}\not =p_{d_j} \ \ \  \text{where} \ \ i, j \ \in \{1,2, 3,4\}, i \neq j. \nonumber
\end{eqnarray}
The probability of success of an embedded DTR can be expressed as $p_{d_i} = E(\bar{Y}_{d_i})=\gamma_{T_1}p_{T_1T_1}+(1-\gamma_{T_1})p_{T_1T_2}$.
Thus, the Wald-type test statistic is,
\begin{eqnarray}
    Z = \frac{\widehat{p}_{d_i}-\widehat{p}_{d_j}}{\sqrt{Var(\widehat{p}_{d_i}-\widehat{p}_{d_j})}}. \nonumber
\end{eqnarray}

\section{Simulation Study} \label{sec : simul study}
Simulations are conducted to evaluate the performance of the optimal adaptive allocation process developed in Section \ref{sec : adaptive alloc proc}. The main aims of this section are to check (a) empirical convergence of the estimated allocation ratios $(\hat{\tau})$ to the proposed optimal allocation ratios $(\tau)$, (b) empirically showing that the total expected number of failures is lowered using optimal adaptive SMART compared to a non-adaptive SMART, and (c) the number of patients allocated to the dynamic treatment regimes (DTRs) are in synchronization with the performance of the corresponding DTRs. 

Here, we consider a two-stage SMART as described in Figure \ref{fig:SMART_diagram_scheme} with a binary primary outcome having a sample size 500. Similar simulations with sample sizes 1000 and 2000 are conducted in the Appendix (see Tables \ref{tab : Tau with other statistics for diff function with 1000} and \ref{tab : Tau with other statistics for diff function with 2000}
of the Appendix). In all the simulations, response probabilities $\gamma_A$, and $\gamma_B$ are assumed to be constant at $0.40$, and $0.30$, respectively. The second column of Table \ref{tab:Tau-difference} shows the considered success probabilities $p_{AA'}, p_{AC}, p_{AD}, p_{BB'}, p_{BE}$ and $p_{BF}$, corresponding to six feasible combinations of the first and second stages treatments $\{T_1, T_2\}$, (see Figure \ref{fig:SMART_diagram_scheme}). However, these success probabilities are unknown to the investigator and must be estimated based on the interim data from the SMART to implement the adaptive allocation (randomization) procedure described in Section \ref{sec : adaptive alloc proc}. Therefore, the initial 30 patients (any other number can also be taken) sequentially enrolled in the SMART are randomized with equal probabilities at both stages. The initial estimates of success probabilities ($\hat{p}_{t_1 t_2, i}$) are based on the observed $Y_i, i = 1, \cdots, 30$. Using the estimated success probabilities, the allocation ratios corresponding to the first ($\hat\tau_{A, i}$) and second stages ($\hat\tau_{AC, i}; \hat\tau_{BE, i}$) are estimated for 31st$(i=31)$ patient. Thus, the 31st patient is randomized using $\hat\tau_{A, 31}$ at the first stage. Now, if the 31st patient is a non-responder, then, depending on the treatment assigned at the first stage, the same patient is randomized using  $\hat\tau_{AC, 31}$ or $\hat\tau_{BE, 31}$. The same process (re-estimation of success probabilities and then allocation ratios) is repeated for subsequent patients till the end of the trial. 

Table \ref{tab:Tau-difference} shows the simulation study results. The success probabilities (in the second column) are chosen in such ways that the true values of the optimum allocation ratios are around 0.5, 1, or 2 in different scenarios. In Table \ref{tab:Tau-difference}, in rows 1 to 3, the true values of the optimal allocation ratio $\tau_{A}$ are 0.521, 1.002, and 2.025, respectively. Similarly, in rows 4 to 6, the true values of the optimal allocation ratio $\tau_{AC}$ are 0.5, 1, and 2, respectively; and, in rows 7 to 9, the true values of the optimal allocation ratio $\tau_{BE}$ are 0.5, 1, and 2, respectively. Rows 10 to 16 show the performance of the adaptive allocation procedure when some (or all) of the success probabilities are very high or low. From Table \ref{tab:Tau-difference}, we observe that $\hat\tau_{A}$ and $\tau_{A}$ are close to each other in all the scenarios. The $\hat\tau_{AC}$ and $\tau_{AC}$ (similarly, $\hat\tau_{BE}$ and $\tau_{BE}$ ) are close for the value of $\tau_{AC}$ ($\tau_{BE}$) near 0.5 or 1. However, when the value of $\tau_{AC}$ or $\tau_{BE}$ are 2 or more, we observe that the corresponding estimates $\hat\tau_{AC}$ and $\hat\tau_{BE}$ are overestimating the respective true quantities. Note that the high value of the optimal allocation ratio is seen when one or more success probabilities are very low or high (failure probability is low). The overestimation of the optimal allocation ratio is because of incurred bias in the estimation of low probabilities. We also observe that the sample standard error (SSE) and asymptotic standard error (ASE) are close in most scenarios. The estimated coverage probabilities (CP) correspond to three optimal allocation ratios, mostly near 0.95, except for a few scenarios when at least one success probability is low. The estimated CP improved considerably when the total sample size of the SMART increased from 500 to 1000 and then to 2000 (see Appendix material). The last two columns of Table \ref{tab:Tau-difference} show the total expected number of failures from the entire SMART at the end of the trial when the proposed adaptive (randomization) allocation procedure (denoted as `Optimal' in Table \ref{tab:Tau-difference}) or non-adaptive randomization (probability is fixed at 0.5 and denoted as `Equal' in Table \ref{tab:Tau-difference}) are followed, respectively. It is evident from the results that the proposed adaptive procedure can reduce the total expected number of failures from the entire SMART by a considerable number. The reduced number of failures in rows 3, 14, and 15 are 53 (10.6\%), 85 (17\%), and 85 (17\%) out of 500 patients. From an ethical point of view, we can prefer the adaptive randomization procedure to the non-adaptive one in a SMART, as fewer patients experience failure in the trial. Also note that when all optimal allocation ratios are 1 (in row 10 to 14), the total expected number of failures from the entire SMART at the end of the trial are equal for both the `Optimal' and `Equal' randomization processes. In other words, when the optimal allocation ratio is 1, adaptive and non-adaptive randomizations are the same, as expected.

\begin{table}[hbt!]
\caption{Estimated first stage ($\hat\tau_A$) and second stage ($\hat\tau_{AC},\hat\tau_{BE}$) allocation ratios along with corresponding SSE, ASE and CP based on 5000 simulations. $\tau_A$, $\tau_{AC}$, and $\tau_{BE}$ denote true values of optimum allocation ratios. Here, $\gamma_A = 0.4,\ and\ \gamma_B=0.3$, and the sample size is 500 using objective function as Simple Difference. It also shows the total expected number of failures at the end of SMART using optimal allocation (proposed method) and equal randomization. \\}
\resizebox{\columnwidth}{!}{%
\renewcommand{\arraystretch}{2}%
\begin{tabular}{|c|c|c|c|c|c|c|}
\hline
No. & $(p_{AA'},p_{AC},p_{AD})$ & \multirow{2}{*}{$\tau_{A}\ (\hat{\tau}_A,\ SSE,\ ASE,\ \widehat{CP})$} & \multirow{2}{*}{$\tau_{AC}\ (\hat{\tau}_{AC},\ SSE,\ ASE,\ \widehat{CP})$} & \multirow{2}{*}{$\tau_{BE}\ (\hat{\tau}_{BE},\ SSE,\ ASE,\ \widehat{CP})$} & \multicolumn{2}{c|}{Expected number of failures} \\ \cline{6-7} 
& $(p_{BB'},p_{BE},p_{BF})$ & & & & Optimal & Equal \\ \hline


1 & (0.20, 0.15, 0.15) & \multirow{2}{*}{0.521 (0.516, 0.046, 0.046, 0.947)} & \multirow{2}{*}{1.000 (1.099, 0.589, 0.803, 0.929)} & \multirow{2}{*}{0.931 (0.931, 0.043, 0.041, 0.941)} & \multirow{2}{*}{267} & \multirow{2}{*}{302} \\ 
& (0.45, 0.65, 0.75) &  & & & &  \\ \hline

2 & (0.30, 0.80, 0.20) & \multirow{2}{*}{1.002 (1.008, 0.043, 0.043, 0.953)} & \multirow{2}{*}{2.000 (2.254, 1.057, 1.651, 0.949)} & \multirow{2}{*}{1.044 (1.047, 0.073, 0.070, 0.946)} & \multirow{2}{*}{260} & \multirow{2}{*}{277} \\ 
& (0.25, 0.60, 0.55) &  & & & &  \\ \hline

3 & (0.80, 0.95, 0.85) & \multirow{2}{*}{2.025 (2.049, 0.160, 0.161, 0.957)} & \multirow{2}{*}{1.057 (1.058, 0.026, 0.026, 0.945)} & \multirow{2}{*}{1.000 (1.078, 0.503, 0.632, 0.925)} & \multirow{2}{*}{179} & \multirow{2}{*}{232} \\
& (0.35, 0.15, 0.15) & & & & & \\ \hline

4 & (0.30, 0.20, 0.80) & \multirow{2}{*}{1.109 (1.109, 0.054, 0.053, 0.946)} & \multirow{2}{*}{0.500 (0.481, 0.094, 0.074, 0.914)} & \multirow{2}{*}{0.500 (0.473, 0.112, 0.088, 0.900)} & \multirow{2}{*}{278} & \multirow{2}{*}{311} \\
& (0.25, 0.15, 0.60) & & & & & \\ \hline

5 & (0.30, 0.20, 0.20) & \multirow{2}{*}{0.686 (0.681, 0.047, 0.046, 0.948)} & \multirow{2}{*}{1.000 (1.041, 0.376, 0.449, 0.939)} & \multirow{2}{*}{0.500 (0.479, 0.096, 0.077, 0.913)} & \multirow{2}{*}{297} & \multirow{2}{*}{325} \\ 
& (0.65, 0.15, 0.60) & & & & & \\ \hline

6 & (0.30, 0.80, 0.20) & \multirow{2}{*}{0.985 (0.991, 0.042, 0.041, 0.951)} & \multirow{2}{*}{2.000 (2.266, 1.055, 1.651, 0.950)} & \multirow{2}{*}{1.000 (1.002, 0.065, 0.063, 0.949)} & \multirow{2}{*}{256} & \multirow{2}{*}{273}\\ 
& (0.25, 0.60, 0.60) & & & & & \\ \hline

7 & (0.30, 0.80, 0.80) & \multirow{2}{*}{1.085 (1.080, 0.041, 0.041, 0.946)} & \multirow{2}{*}{1.000 (1.002, 0.041, 0.041, 0.954)} & \multirow{2}{*}{0.500 (0.475, 0.108, 0.087, 0.906)} & \multirow{2}{*}{220} & \multirow{2}{*}{235} \\ 
& (0.65, 0.15, 0.60)  & & & & & \\ \hline

8 & (0.30, 0.80, 0.80) & \multirow{2}{*}{1.414 (1.420, 0.073, 0.073, 0.951)} & \multirow{2}{*}{1.000 (1.001, 0.038, 0.038, 0.954)} & \multirow{2}{*}{1.000 (1.050, 0.377, 0.409, 0.944)} & \multirow{2}{*}{262} & \multirow{2}{*}{274} \\ 
& (0.65, 0.15, 0.15) & & & & &\\ \hline

9 & (0.30, 0.20, 0.20) & \multirow{2}{*}{0.686 (0.681, 0.047, 0.046, 0.947)} & \multirow{2}{*}{1.000 (1.036, 0.368, 0.433, 0.938)} & \multirow{2}{*}{2.000 (2.227, 0.863, 1.112, 0.953)} & \multirow{2}{*}{297} & \multirow{2}{*}{325} \\ 
& (0.65, 0.60, 0.15)  & & & & & \\ \hline 

\multicolumn{3}{|c}{Very high/low success probability values} & \multicolumn{4}{c|}{} \\ \hline 
10 & (0.10, 0.10, 0.10) & \multirow{2}{*}{1.000 (1.011, 0.147, 0.144, 0.951)} & \multirow{2}{*}{1.000 (1.061, 0.437, 0.489, 0.927)} & \multirow{2}{*}{1.000 (1.058, 0.402, 0.425, 0.939)} & \multirow{2}{*}{450} & \multirow{2}{*}{450} \\
& (0.10, 0.10, 0.10) & & & & & \\ \hline

11 & (0.05, 0.05, 0.05) & \multirow{2}{*}{1.000 (1.026, 0.223, 0.226, 0.963)} & \multirow{2}{*}{1.000 (1.113, 0.539, 0.718, 0.957)} & \multirow{2}{*}{1.000 (1.093, 0.486, 0.609, 0.957)} & \multirow{2}{*}{475} & \multirow{2}{*}{475} \\ 
& (0.05, 0.05, 0.05) & & & & &\\ \hline

12 & (0.90, 0.90, 0.90) & \multirow{2}{*}{1.000 (1.000, 0.015, 0.015, 0.948)} & \multirow{2}{*}{1.000 (1.000, 0.028, 0.028, 0.951)} & \multirow{2}{*}{1.000 (1.001, 0.025, 0.026, 0.956)} & \multirow{2}{*}{50} & \multirow{2}{*}{50} \\
& (0.90, 0.90, 0.90) & &  & & &\\ \hline

13 & (0.95, 0.95, 0.95) & \multirow{2}{*}{1.000 (1.000, 0.010, 0.010, 0.947)} & \multirow{2}{*}{1.000 (1.000, 0.019, 0.019, 0.967)} & \multirow{2}{*}{1.000 (1.000, 0.018, 0.018, 0.962)} & \multirow{2}{*}{25} & \multirow{2}{*}{25} \\ 
& (0.95, 0.95, 0.95) & & & & &\\ \hline

14 & (0.35, 0.95, 0.05) & \multirow{2}{*}{0.943 (0.946, 0.037, 0.044, 0.974)} & \multirow{2}{*}{4.359 (6.402, 2.901, 8.560, 0.937)} & \multirow{2}{*}{3.000 (4.063, 2.236, 4.198, 0.944)} & \multirow{2}{*}{169} & \multirow{2}{*}{254} \\ 
& (0.65, 0.90, 0.10) & & & & &\\ \hline

15 & (0.45, 0.05, 0.95) & \multirow{2}{*}{1.072 (1.071, 0.046, 0.051, 0.968)} & \multirow{2}{*}{0.229 (0.196, 0.088, 0.110, 0.983)} & \multirow{2}{*}{3.000 (4.114, 2.306, 4.483, 0.950)} & \multirow{2}{*}{190} & \multirow{2}{*}{275} \\ 
& (0.25, 0.90, 0.10) & & & & &\\ \hline

16 & (0.95, 0.95, 0.05) & \multirow{2}{*}{1.057 (1.057, 0.031, 0.036, 0.974)} & \multirow{2}{*}{4.359 (6.362, 2.882, 8.246, 0.934)} & \multirow{2}{*}{0.333 (0.298, 0.102, 0.088, 0.980)} & \multirow{2}{*}{89} & \multirow{2}{*}{174} \\ 
& (0.90, 0.10, 0.90) & & & & &\\ \hline
\end{tabular}
\label{tab:Tau-difference}%
}
\end{table}

\begin{figure}[ht]
     \centering
     \begin{subfigure}[b]{0.3\textwidth}
         \centering
         \includegraphics[width=\textwidth]{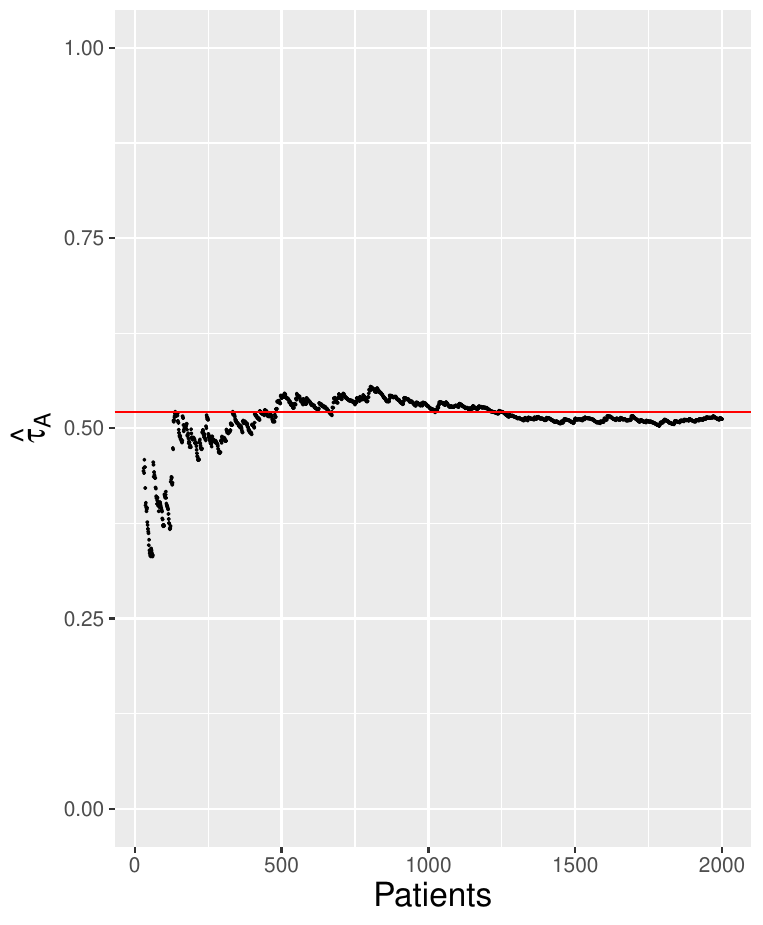}
         \caption{$\tau_{A}=0.521$}
         \label{fig:tau_A = 0.5}
     \end{subfigure}
     \hfill
     \begin{subfigure}[b]{0.3\textwidth}
         \centering
         \includegraphics[width=\textwidth]{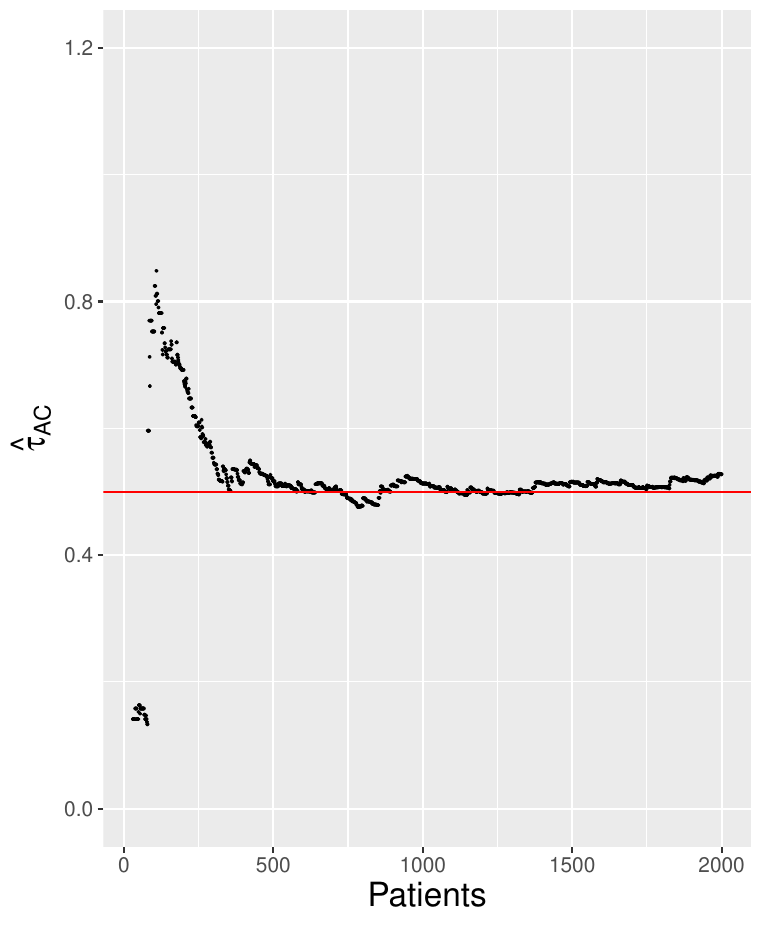}
         \caption{$\tau_{AC}=0.5$}
         \label{fig:tau_{AC} = 0.5}
     \end{subfigure}
     \hfill
     \begin{subfigure}[b]{0.3\textwidth}
         \centering
         \includegraphics[width=\textwidth]{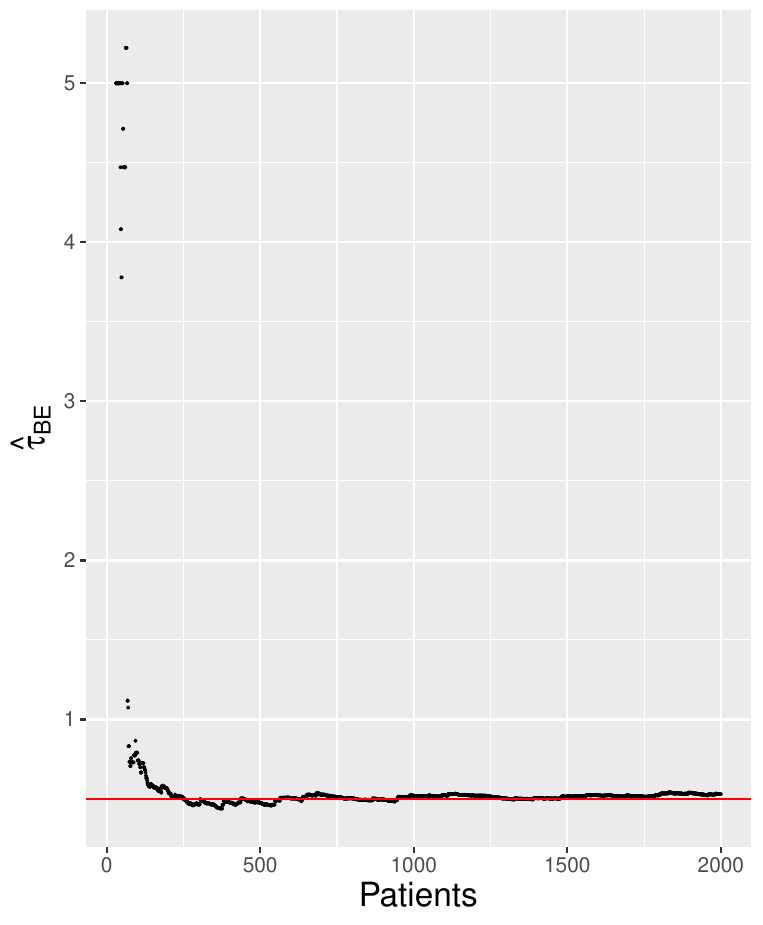}
         \caption{$\tau_{BE}=0.5$}
         \label{fig:tau_{BE} = 0.5}
     \end{subfigure}
        \captionsetup{justification=centering}
        \caption{Convergence patterns of estimated (black dots) optimal allocation ratios ($\hat \tau_{A}$, $\hat \tau_{AC}$, and $\hat \tau_{BE}$) to corresponding true values $(\tau_{A}, \tau_{AC}$,  and $\tau_{BE})$ indicated by red lines.}
        \label{fig:convergence graphs}
\end{figure}

Figure \ref{fig:convergence graphs} shows the convergence patterns of the estimated (black dots) optimal allocation ratios to the corresponding true values (red lines) as the sample size increases. In this specific instance (one of 5000 simulations), we observe that $\hat \tau_{A}$ and $\hat \tau_{AC}$ both narrow down the gap between their values and corresponding true values after inclusion of 250 or more patients in the SMART. On the other hand, $\hat \tau_{BE}$ started to close down the same gap a little earlier. In summary, Figure \ref{fig:convergence graphs} empirically ensures the convergence property of the estimated optimal allocation ratios following the adaptive procedure described in Section \ref{sec : adaptive alloc proc}.

\begin{figure}[h]
     \centering
     \begin{subfigure}[b]{0.3\textwidth}
         \centering
         \includegraphics[width=\textwidth]{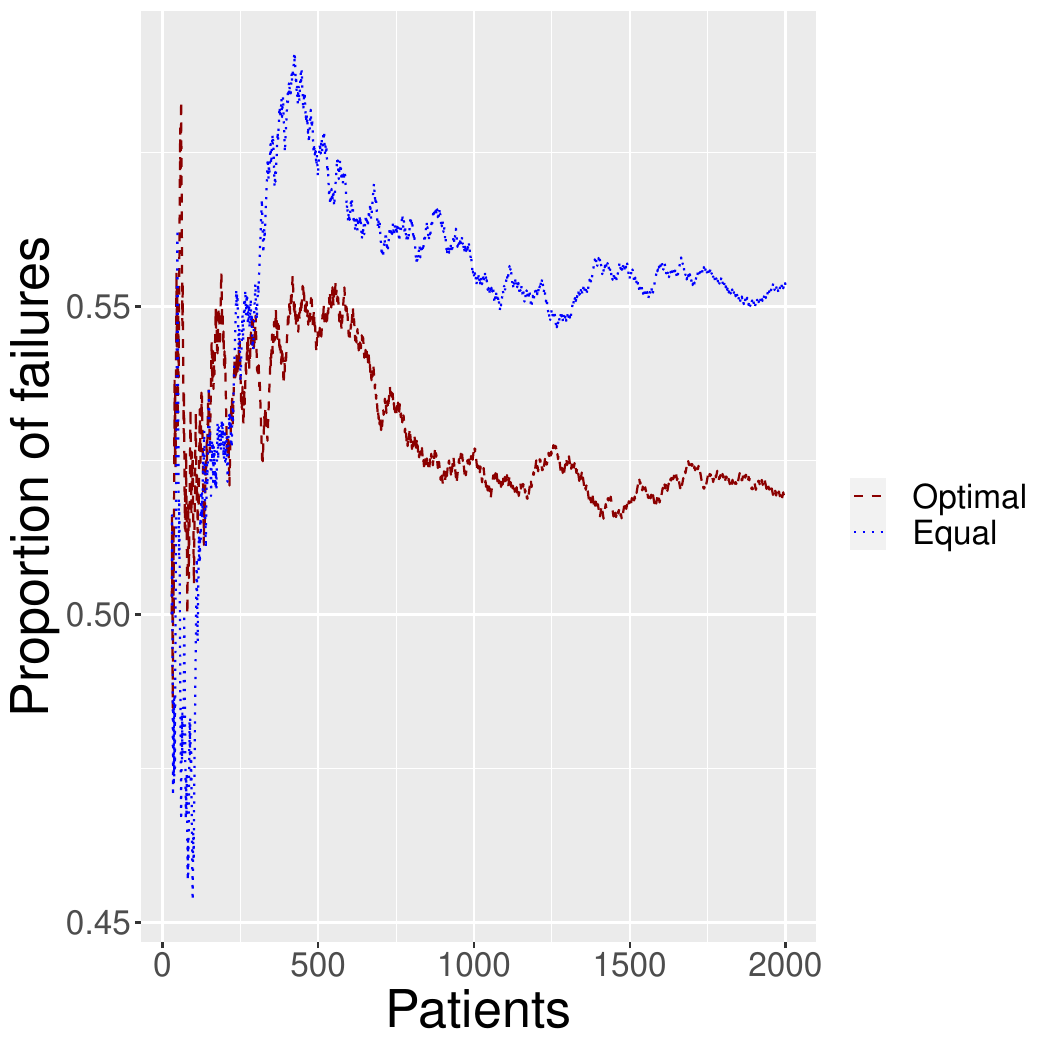}
         \caption{Comparison of proportion of failures for Optimal and Equal allocations}
         \label{fig: failure prop}
     \end{subfigure}
     \hfill
     \begin{subfigure}[b]{0.3\textwidth}
         \centering
         \includegraphics[width=\textwidth]{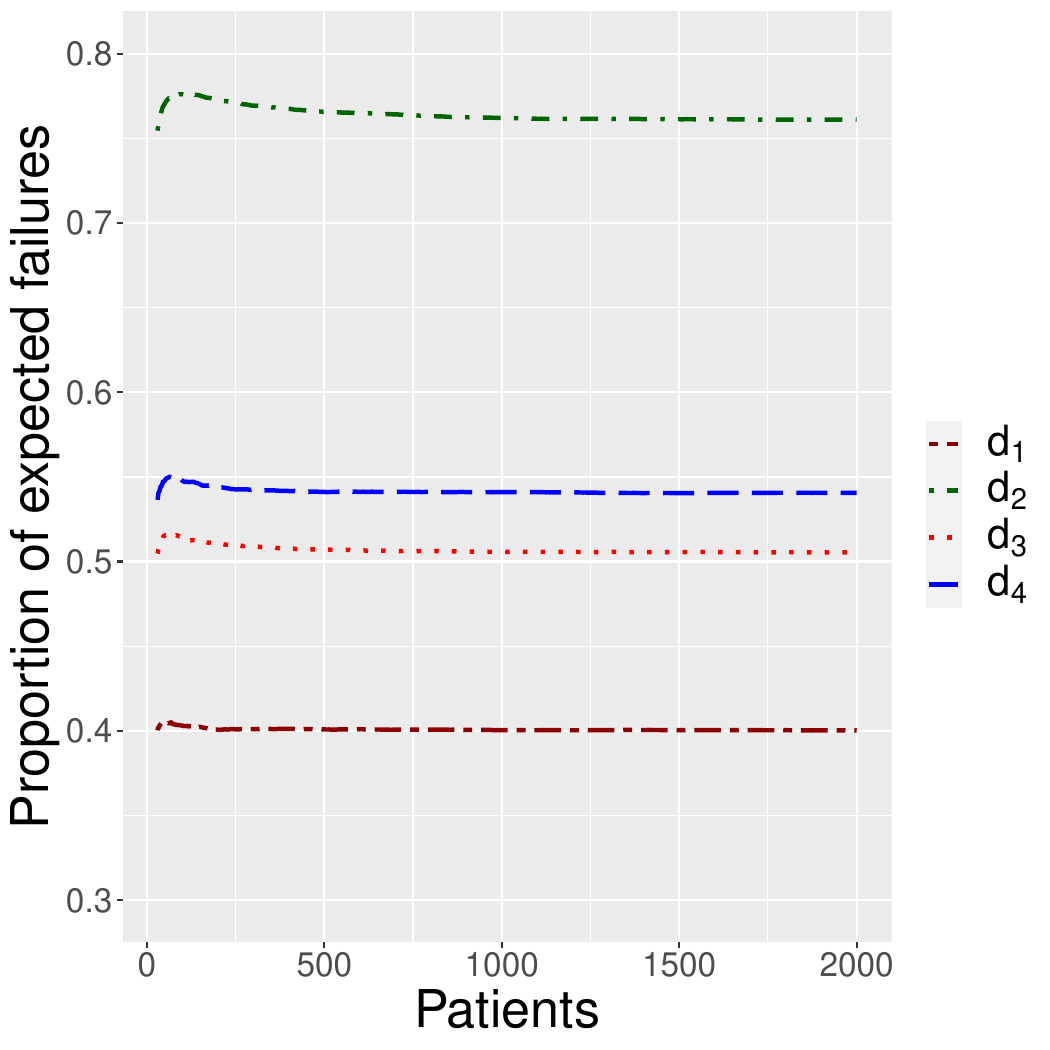}
         \caption{Proportions of expected failures along embedded DTRs}
         \label{fig:dtr alloc prop}
     \end{subfigure}     
     \hfill
     \begin{subfigure}[b]{0.3\textwidth}
         \centering
         \includegraphics[width=\textwidth]{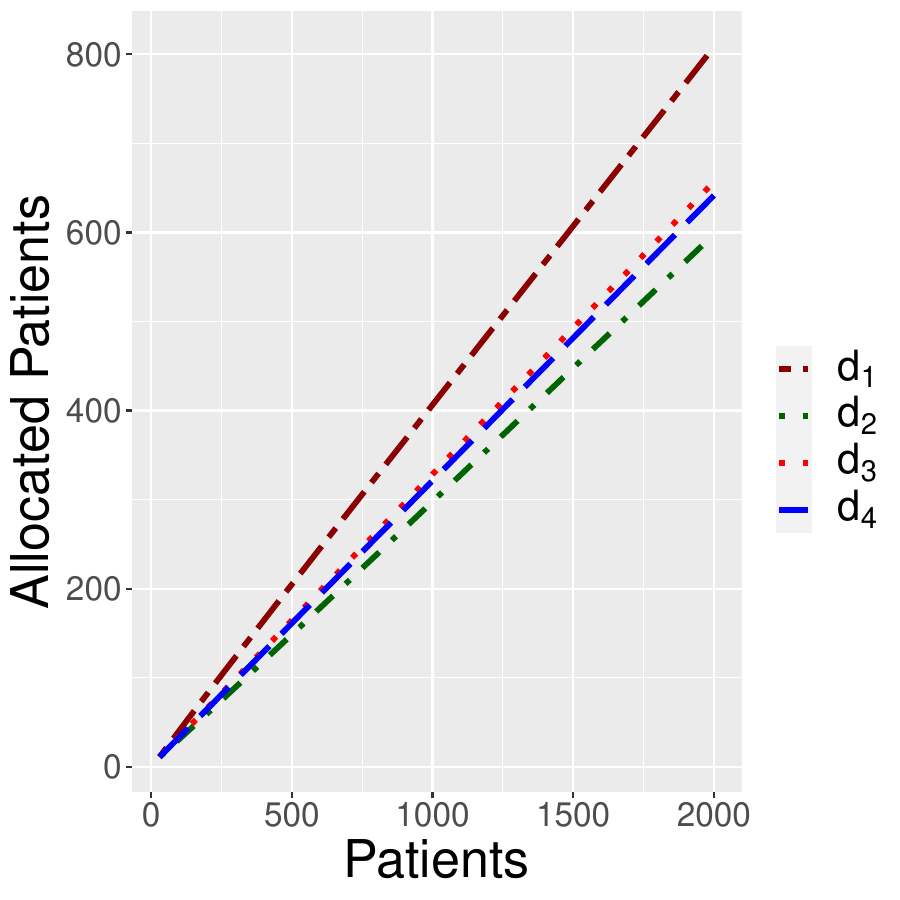}
         \caption{Allocated patients along four embedded DTRs}
         \label{fig:dtr alloc}
     \end{subfigure}
        \caption{}
        \label{fig:comparison graphs}
\end{figure}

We have seen that the proposed adaptive randomization procedure minimizes the total expected number of failures (compared to a non-adaptive SMART) at the end of the SMART using the last two columns of Table \ref{tab:Tau-difference}. Figure \ref{fig: failure prop} shows the trajectories of the proportion of failures (one of the 5000 simulations) following `Optimal' (dashed brown line) and `Equal' (dotted blue line) allocation processes over the sequential enrollment of 2000 patients. Notice that the proportion of failures is initially higher in the `Optimal' allocation process than in the `Equal' allocation. However, as the number of enrolled patients increases, the proportion of failures in `Optimal' become lower compared to `Equal' allocation. Interesting to observe that, after 500 patients, the vertical distance between (and the patterns of the two graphs) the two lines is almost the same till the end. Approximately after 1000 patients' enrollment, the `Optimal' and `Equal' graphs stabilize around 0.52 and 0.56 (with some variations), respectively. The considered values of success probabilities for all three graphs in Figure \ref{fig:comparison graphs} are $p_{AA'} = 0.3,\ p_{AC} = 0.3,\ p_{AD} = 0.4,\ p_{BB'} = 0.65,\ p_{BE} = 0.6$, and $p_{BF} = 0.65$; along with $\gamma_A=0.4$, $\gamma_B=0.3$.

Figure \ref{fig:dtr alloc prop} and \ref{fig:dtr alloc} compare the embedded DTRs with respect to  the proportions of the expected number of failures and the number of allocated patients averaged over 5000 simulations. We observed that the order (lowest to the highest proportion of expected failures) of embedded DTRs in Figure \ref{fig:dtr alloc prop} is $d_1, d_3, d_4$, and $d_2$, which is the exact same order (highest to lowest number of allocated patients) of the embedded DTRs in Figure \ref{fig:dtr alloc}. Thus, from the ethical point of view, we can claim that the proposed methodology assigns more patients to better DTRs (having a lower proportion of failures). At the end of the SMART, we compare two distinct path DTRs, $d_1$ and $d_3$ using the hypothesis testing procedure described in Section \ref{sec : testing structure}. The test concludes that $d_1$ is significantly better than $d_3$ with p-value $< 0.01$ and the test statistic, $Z_{obs} = 42.90$. This conclusion is in the same line as we observed from Figures \ref{fig:dtr alloc prop} and \ref{fig:dtr alloc}.

\section{Application to M-Bridge Data}\label{sec: Mbridge}
In this section, we demonstrate an application of the developed adaptive allocation (randomization) procedure using M-Bridge data \citep{patrick2021main}. The binary primary outcome is based on the frequency of consuming 4/5+ drinks by the participants within a two-hour period in the past 30 days in any of the three follow-ups at the end of the study. If the frequency is one or more, then the binary outcome is 0 (failure); otherwise, 1 (success). Note that the M-Bridge study used equal randomizations both at the first and the second stage randomization processes. Here, the objective is to show what benefit would have happened if the adaptive allocation procedure had been used instead of equal randomization during the allotment of the participants to different treatments in the M-Bridge study. Specifically, we show that the developed procedure would have resulted in fewer failures, and more participants would have got better DTRs (having a higher chance of success). To retrospectively apply the developed adaptive allocation procedure in the M-Bridge study, first, we arrange (in increasing order) all the 521 participants according to their entry date and time in the study. Note that 70 participants who did not appear in any follow-up studies or were administered both treatment options, namely online health coach and resource email at the second stage, were removed from the current analysis. For illustration purposes, let us consider a version of the M-Bridge study where the recruitment of participants can be done on a rolling basis as opposed to one-time recruitment. We also assume that the binary primary outcome for a participant is available before the entry of the next participant. Here, we consider the first 60 participants without adaptive randomization (randomization using the M-Bridge protocol with a 1:1 ratio) to obtain the initial estimates of the success probabilities. After that, each participant is allocated to a treatment following the developed adaptive allocation procedure (see Section \ref{sec : adaptive alloc proc}) using the simple difference of the success probabilities as the objective function $g(\cdot, \cdot)$. During the retrospective adaptive randomization (say 60 participants are already allocated), if the allocated treatments at the first stage and the second stage for the $61^{st}$ participant are $A$ and $C$, respectively, then we will pick that participant who was first given the treatment sequence $\{A, C\}$ from the remaining arranged participants. The selected participant may be higher ranked than 61 in the arranged list of participants. 

\begin{figure}[!hbt]
     \centering
     \begin{subfigure}[b]{0.3\textwidth}
         \centering
         \includegraphics[width=\textwidth]{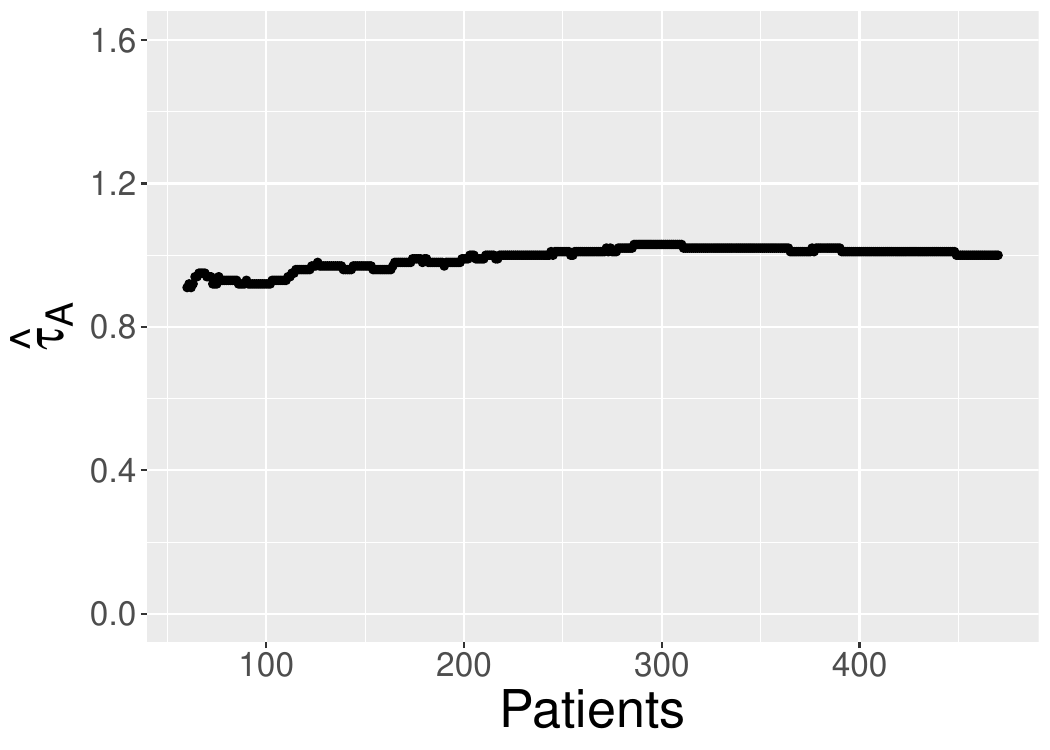}
         \caption{$\hat \tau_{A}$: The estimated first stage optimal allocation ratio.\\}
         \label{fig: mbridge tau a}
     \end{subfigure}
     \hfill
     \begin{subfigure}[b]{0.3\textwidth}
         \centering
         \includegraphics[width=\textwidth]{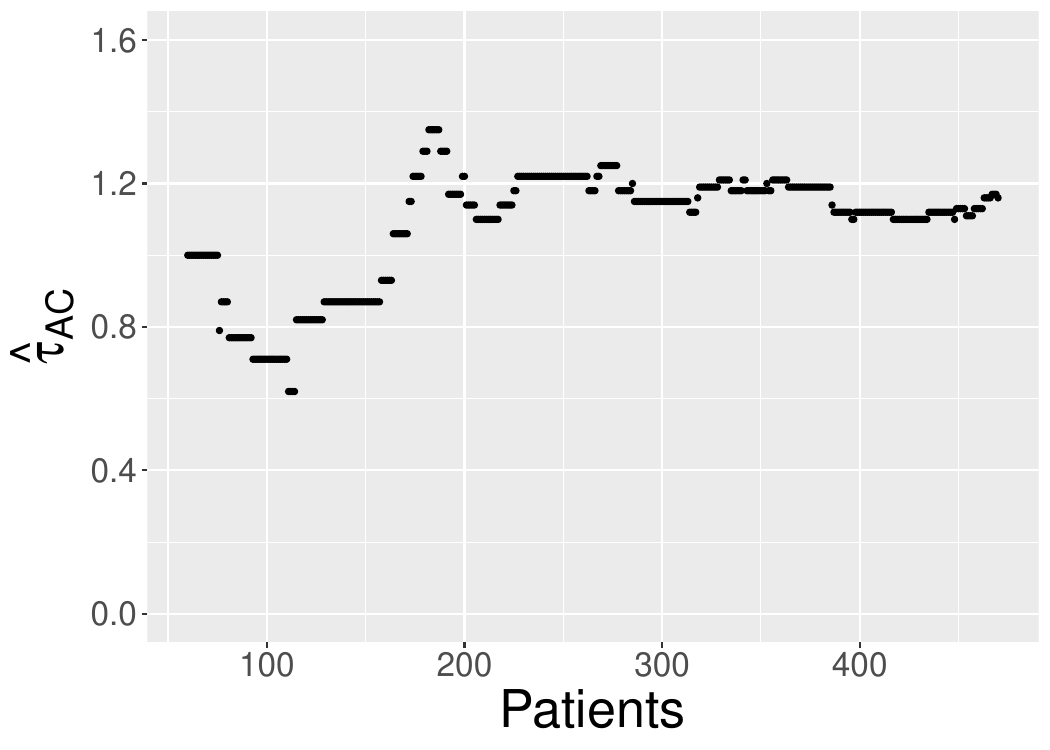}
         \caption{$\hat \tau_{AC}$: The estimated second stage optimal allocation ratio for non-responders who obtained $A$ at the first stage.}
         \label{fig:mbridge tau ac}
     \end{subfigure}     
     \hfill
     \begin{subfigure}[b]{0.3\textwidth}
         \centering
         \includegraphics[width=\textwidth]{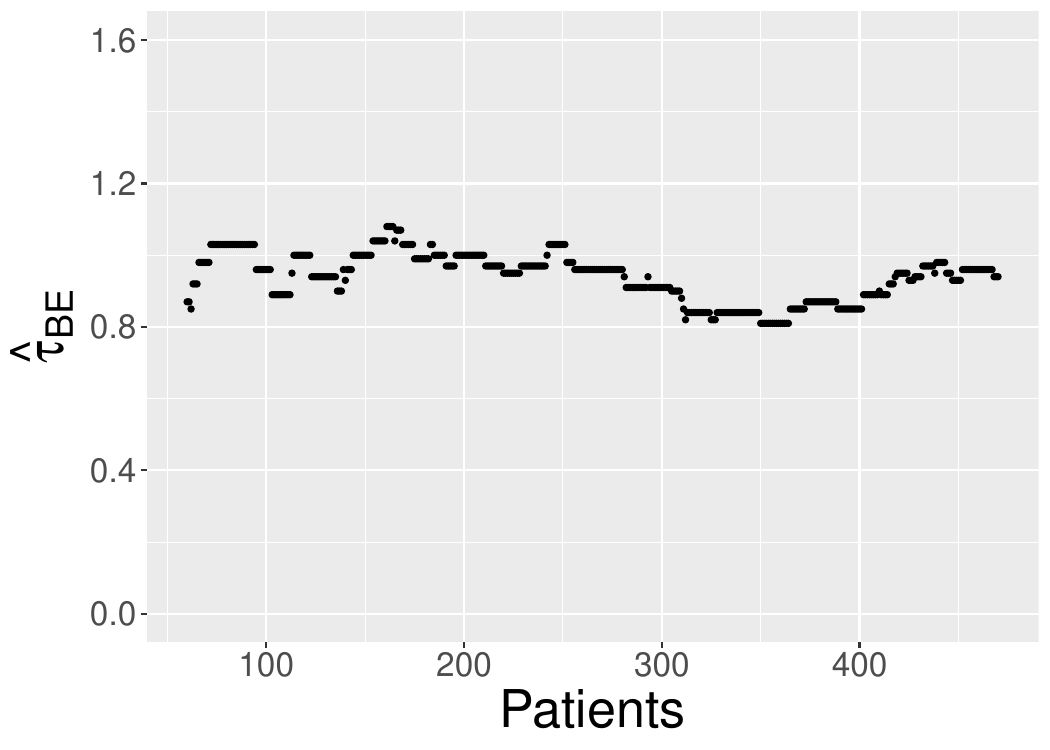}
         \caption{$\hat \tau_{BE}$: The estimated second stage optimal allocation ratio for non-responders who obtained $B$ at the first stage.}
         \label{fig:mbridge tau be}
     \end{subfigure}
        \caption{Convergence patterns of estimated (black dots) optimal allocation ratios $\hat \tau_{A}$, $\hat \tau_{AC}$, and $\hat \tau_{BE}$ in the M-Bridge study.}
        \label{fig:convergence of all tau for mbridge}
\end{figure}

Figure \ref{fig:convergence of all tau for mbridge} presents the convergence patterns of the estimated optimal allocation ratios $\hat \tau_{A}$, $\hat \tau_{AC}$, and $\hat \tau_{BE}$ in the M-Bridge study. From Figure \ref{fig: mbridge tau a}, we observe that $\hat \tau_{A}$ converges around 1, with the last estimated value of $\hat \tau_{A}$ is $1.00$ with ASE $0.023$. On the other hand, $\hat \tau_{AC}$ converges to a value just below of 1.2, with the last estimated value of $\hat \tau_{AC}$ is $1.16$ with ASE $0.169$ (see Figure \ref{fig:mbridge tau ac}), and $\hat \tau_{BE}$ takes the values between 0.8 and 1 at the end of the study, with the last estimated value of $\hat \tau_{BE}$ is $0.94$ with ASE $0.136$ (see Figure \ref{fig:mbridge tau be}). Notice that the convergence pattern of $\hat \tau_{A}$ is more convincing than the other two optimal allocation ratios at the second stage of M-Bridge. This is because $\hat \tau_{A}$ is based on all the samples (470, see Table \ref{tab : dtr alloc for mbridge study}) of M-Bridge, but $\hat \tau_{AC}$ and $\hat \tau_{BE}$ are based on much fewer samples which are consistent with the corresponding treatment options.

	\begin{table}[!ht]
\caption{Allocated participants and proportion of failures (in parentheses) following optimal adaptive allocation (OAA) and 1:1 allocation in M-Bridge study. The simple difference of the success probabilities is used as the objective function $g(\cdot, \cdot)$. The OAA has to stop after 470 participants as treatment sequence $\{A, C\}$ of the M-Bridge SMART has no available participants. The proportion of failures for $d_i$ is $q_{d_{i}}$, whereas the proportion of failure (in the last row) is the ratio of the total number of failures to total participants.   \\}
\resizebox{\columnwidth}{!}{%
\renewcommand{\arraystretch}{2}%
\begin{tabular}{|c|c|c|c|c|c|}
\hline
\multirow{2}{*}{$DTR$} & \multicolumn{5}{c|}{Responder (R) + Non-Responder (NR) = Total (Proportion of failures) }\\
\cline{2-6} 
 & \multicolumn{2}{c|}{Optimal Adaptive Allocation (OAA)} & \multicolumn{2}{c|}{M-Bridge Allocation} & M-Bridge Study allocation (end of study) \\ \cline{2-6} 
& Participants with OAA  & Remaining participants & Till 470 participants & Remaining 51 participants & All participants \\ \cline{2-6} 
\hline
$d_1$ & 174 + 37 = 211 (0.233) & 13 + 0 = 13 (0.795) & 167 + 34 = 201 (0.237) & 20 + 3 = 23 (0.798) & 187 + 37 = 224 (0.235) \\ \hline
$d_2$ & 174 + 22 = 196 (0.268) & 13 + 16 = 29 (0.676) & 167 + 35 = 202 (0.274) & 20 + 3 = 23 (0.607) & 187 + 38 = 225 (0.284) \\ \hline
$d_3$ & 166 + 35 = 201 (0.259) & 10 + 1 = 11 (0.544) & 160 + 30 = 190 (0.284) & 16 + 6 = 22 (0.808) & 176 + 36 = 212 (0.266) \\ \hline
$d_4$ & 166 + 36 = 202 (0.244) & 10 + 11 = 21 (0.660) & 160 + 44 = 204 (0.249) & 16 + 3 = 19 (0.702) & 176 + 47 = 223 (0.252) \\ \hline
$Total$ & 470 (0.236) & 51 (0.471) & 470 (0.260) & 51 (0.255) & 521 (0.259) \\ \hline
\end{tabular}
\label{tab : dtr alloc for mbridge study}%
}
\end{table}

Table \ref{tab : dtr alloc for mbridge study} compares optimal adaptive allocation (OAA) with the original (1:1) randomized allocation with respect to allocated participants and the proportion of failures. The comparison is made among the four embedded DTRs $d_1, d_2, d_3, d_4$ using the total number of participants and the proportion of failures in those four embedded DTRs. Note that the proportion of failures for $d_i$ is $q_{d_{i}} = 1- p_{d_{i}}, i = 1, \cdots, 4$, whereas the proportion of failure (in the last row of Table \ref{tab : dtr alloc for mbridge study}) is the ratio of the total number of failures to total participants. In OAA, we have used 470 participants as the treatment sequence $\{A, C\}$ of M-Bridge SMART has no further available participants. The last row of Table \ref{tab : dtr alloc for mbridge study} confirms that the proportion of failure using OAA is 0.236, which is lower than the proportion of failure of 0.260 in the original M-Bridge study considering the same number of 470 participants. It supports our claim that the developed procedure OAA would have resulted in fewer failures. In the original M-Bridge study, the DTR $d_2$ had the highest proportion of failures (0.284), but has the highest (225) participants allocated to it using a 1:1 randomization scheme at both stages. 
However, the developed OAA procedure rightfully allocated a maximum number of participants (211) to the best performing DTR $d_1$ having the lowest failure proportion of 0.233. Similarly, OAA allocates the lowest number of participants (196) to the DTR $d_2$ having the highest proportion of failures (0.268). Therefore, from the second column of Table \ref{tab : dtr alloc for mbridge study}, we observe that the allocated number of participants is decreasing in DTRs $d_1, d_4, d_3$ and $d_2$, respectively, which is consistent with the increasing order of proportion of failures in DTRs $d_1, d_4, d_3$ and $d_2$, respectively. However, we found no statistically significant evidence in the pairwise comparison of embedded DTRs $\{d_i, d_j\}, i = 1,2; j = 3,4$, following the hypothesis testing procedure described in Section \ref{sec : testing structure}. This result is consistent with the finding as reported in \cite{patrick2021main}.

\section{Discussion}\label{sec:discussion}
In this work, we have developed an optimal adaptive allocation (randomization) procedure that minimizes the total expected number of failures in a SMART and gives better DTRs to more participants. Adaptive randomization is frequently used in traditional (two-arm and single-stage) randomized controlled trials. Using simulation and artificially infusing adaptive randomization in the M-Bridge study, we have shown how a SMART design can incorporate adaptive randomization to make it more popular among clinicians by better addressing ethical concerns. The developed procedure needs only the response rates (as constants) and the estimation of success probabilities for the implementation. Retrospective implementation in the M-Bridge study endorses the feasibility of adaptive randomization in a SMART with a sample size of less than 500. Due to the unavailability of participants in the treatment sequence $\{A, C\}$ of the M-Bridge study, the adaptive procedure had to stop after including 470 participants. If we were able to use all the 521 participants of M-Bridge, the convergence of the second-stage optimal allocation ratios would have been better. Notice that, in Figures \ref{fig:mbridge tau ac} and \ref{fig:mbridge tau be}, the graphs for $\hat\tau_{AC}$ and $\hat\tau_{BE}$ are more like step functions whereas the graph for $\hat\tau_{A}$ in Figure \ref{fig: mbridge tau a} is more like a continuous function. This is because the value of the $\hat\tau_{A}$ is updated for each new participant, but $\hat\tau_{AC}$ or $\hat\tau_{BE}$ are updated for a new participant only if the earlier participant had a treatment sequence that is consistent with those second stage optimal ratios. Thus, it is possible $\hat\tau_{AC}$ or $\hat\tau_{BE}$ may remain constant for few consecutive participants. 

Note that, till date, no SMART has been conducted using an adaptive allocation (randomization) procedure. The original M-bridge study was specifically designed around the timing of the academic year. The first randomization started early (before the start of the Fall semester) or late (during the first month of the same semester). The original M-bridge study design did not allow for waiting to see how the treatment went for the first set of participants. Therefore, to motivate researchers/clinicians about the importance and advantages of the optimal adaptive SMART, we 
retrospectively applied the developed adaptive allocation procedure in the M-Bridge study assuming the recruitment of participants can be done on a rolling basis as opposed to one-time recruitment. In other words, using the application to M-Bridge data in Section \ref{sec: Mbridge}, we have shown that the developed adaptive allocation (randomization) procedure could maximize the benefit of treatments by providing better treatment sequences to more number of participants compared to the 1:1 randomization used in the original study. In the future, an adaptive version of the M-bridge study can be developed where participants can be recruited over a few semesters. In that scenario, the optimal allocation ratios can be updated (to be used for the upcoming semester) at the end of each semester by observing the performances of the participants.

The developed procedure can be extended for similar types of SMART with more than two stages. In that case, the developed optimal allocation ratios will include extra response rate parameters from the added stages. Irrespective of the number of stages in a SMART, one main limitation of the adaptive randomization procedure is the requirement for earlier participants' primary outcomes before considering the next participant \citep{robertson2023response}. Therefore, adaptive randomization may not be feasible if the primary outcome of the trial is available after a long duration. Here, we have developed a procedure that updates the optimal allocation ratios for each new participant. It may be practically challenging in some SMARTs. \cite{wu2023interim} developed a group sequential approach to SMARTs for interim monitoring using the multivariate chi-square distribution. The developed optimal adaptive allocation procedure can be implemented in those SMARTs where interim monitoring is possible. Also, an application of group sequential design in SMART will be helpful in this situation where the optimal allocation ratios can be updated after the entry of a batch/group of participants. We plan to persue this as a future work. 

In simulation studies and in the application to the MBridge study, we have used 
the initial period of equal randomization, which is referred to as the ``burn-in'' or ``warm-up'' or ``warm-start'' period \citep{du2018comparing}. In the adaptive randomization procedure, the burn-in period is necessary to get the initial estimates of the unknown success probabilities that are required to calculate the optimal allocation ratios. How much of such initial exploration (equal randomization) is necessary? In this work, we have decided the length of the burn-in period as an ad-hoc choice. There is no optimality claim associated with the initial period of equal randomization we are employing in our simulations and in the application to the MBridge study. This issue also arises in online reinforcement learning / contextual bandit algorithms literature in computer science, where they call it the ``exploration-exploitation dilemma'' \citep{sutton2018reinforcement}. Future work that decides the length of the burn-in period based on some optimal criteria will be useful in the context of SMART. 

In this work, we have focused on developing an optimal adaptive allocation procedure for the binary primary outcome variable. However, the primary outcome variable can be continuous too \citep{zhao2011reinforcement}. The optimal adaptive allocation procedure for the continuous primary outcome variable will be studied in future work. Furthermore, it will be interesting work to adjust the optimal adaptive allocation ratios with respect to covariates in SMART \citep{moore2009covariate}.

\section*{Acknowledgments}
Bibhas Chakraborty would like to acknowledge the grant MOE-T2EP20122-0013 from the Ministry of Education, Singapore, and the start-up grant from the Duke-NUS Medical School, Singapore. Megan Patrick would like to acknowledge funding support by the National Institute on Alcohol Abuse and Alcoholism (NIAAA) (R01 AA026574). Palash Ghosh would like to acknowledge partial support from the IDEAS-Technology Innovation HUB, ISI Kolkata (grant no: OO/ISI/IDEAS-TIH/2023-24/59), and ICMR Centre for Excellence (grant no. 5/3/8/20/2019-ITR).

\newpage

\setlength{\tabcolsep}{5.5pt} 
\renewcommand{\arraystretch}{1.35}

\bibliographystyle{chicago} 
\bibliography{ref}

\newpage 

\appendix

\section{\centering \textbf{Appendix : Technical Details}}

\subsection{\centering
    \textbf{Derivation of the first stage success probability $(p_{T_1})$}\label{ap : first stage success probability wrt second stage}
}
$p_{T_1},\ q_{T_1}(=1-p_{T_1})$ is the first stage probability of success and failure for a patient receiving treatment $T_1\ (T_1 \in (A,B))$ at the first stage, respectively. The $p_{T_1T_2}$ is the success probability (observed at the end of study) for the patient who obtained the treatment sequence $\{T_1,T_2\}$, where $T_2 \in \{A', C, D\}$ when $T_1 = A$ and $T_2 \in \{B', E, F\}$ for $T_1 = B$. Then the first stage probability can be expressed as,
\begin{align}
	p_{T_1} & =  \frac{Number\ of\ successes\ for\ participants\ started\ with\ T_1}{n_{T_1}}   \nonumber \\
	& =  \frac{n_{T_1T_1^{'}}p_{T_1T_1^{'}}+n_{T_1T_2}p_{T_1T_2}+n_{T_1T_2^{*}}p_{T_1T_2^{*}}}{n_{T_1}}, \mbox{ where } T_2 = C, T_2^* = D \mbox{ if } T_1 = A; T_2 = E, T_2^* = F \mbox{ if } T_1 = B \nonumber \\
	& =  \frac{n_{T_1}\gamma_{T_1}p_{T_1T_1^{'}}+n_{T_1}(1-\gamma_{T_1})\frac{\tau_{T_1T_2}}{1+\tau_{T_1T_2}}p_{T_1T_2}+n_{T_1}(1-\gamma_{T_1})\frac{1}{1+\tau_{T_1T_2}}p_{T_1T_2^{*}}}{n_{T_1}} \nonumber \\
	& =  \gamma_{T_1}p_{T_1T_1^{'}}+(1-\gamma_{T_1})\frac{\tau_{T_1T_2}}{1+\tau_{T_1T_2}}p_{T_1T_2}+(1-\gamma_{T_1})\frac{1}{1+\tau_{T_1T_2}}p_{T_1T_2^{*}}.
\label{succ_prob_p_T_1}
\end{align}

Note that the sum of the coefficients in (\ref{succ_prob_p_T_1}) of above expression is,
\begin{align}
	& \gamma_{T_1}+(1-\gamma_{T_1})\frac{\tau_{T_1T_2}}{1+\tau_{T_1T_2}}+(1-\gamma_{T_1})\frac{1}{1+\tau_{T_1T_2}} \nonumber \\ 
 & = \gamma_{T_1}+(1-\gamma_{T_1})(\frac{\tau_{T_1T_2}}{1+\tau_{T_1T_2}}+\frac{1}{1+\tau_{T_1T_2}}) \nonumber \\
	& = \gamma_{T_1}+(1-\gamma_{T_1}) \nonumber \\
	& = 1. \label{coeff_sum_1}
	\end{align}
	Thus, using (\ref{succ_prob_p_T_1}) and (\ref{coeff_sum_1}), we have 
\begin{align*}
    q_{T_1} & = 1-p_{T_1} \\ 
    & = 1- \gamma_{T_1}p_{T_1T_1^{'}}-(1-\gamma_{T_1})\frac{\tau_{T_1T_2}}{1+\tau_{T_1T_2}}p_{T_1T_2}-(1-\gamma_{T_1})\frac{1}{1+\tau_{T_1T_2}}p_{T_1T_2^{*}} \\ 
	& = \gamma_{T_1}+(1-\gamma_{T_1})\frac{\tau_{T_1T_2}}{1+\tau_{T_1T_2}}+(1-\gamma_{T_1})\frac{1}{1+\tau_{T_1T_2}}-\gamma_{T_1}p_{T_1T_1^{'}}-(1-\gamma_{T_1})\frac{\tau_{T_1T_2}}{1+\tau_{T_1T_2}}p_{T_1T_2}-(1-\gamma_{T_1})\frac{1}{1+\tau_{T_1T_2}}p_{T_1T_2^{*}} \\
	& = \gamma_{T_1}q_{T_1T_1^{'}}+(1-\gamma_{T_1})\frac{\tau_{T_1T_2}}{1+\tau_{T_1T_2}}q_{T_1T_2}+(1-\gamma_{T_1})\frac{1}{1+\tau_{T_1T_2}}q_{T_1T_2^{*}}.
\end{align*}


\subsection{\centering
    \textbf{Derivation of Second Stage Optimal Allocation Ratio for Simple Difference}\label{ap : second stage optimal alloc for diff}
    }
We consider the objective function $g(\cdot, \cdot)$ as introduced in Section \ref{optimum_allo} to be the simple difference. The objective function of simple difference for comparing the two-second stage probabilities $(p_{T_1T_2}$, and $p_{T_1T_2^*})$ is given by $p_{T_1T_2}-p_{T_1T_2^*}$, where $T_2 = C, T_2^* = D$ if $T_1 = A$; $T_2 = E, T_2^* = F$ if $T_1 = B$. The optimality criterion (as defined in Section \ref{optimum_allo} of main paper) using the asymptotic variance of the objective function $avar(g(\cdot,\cdot))$ can be expressed as,
	\begin{equation*}
	\frac{p_{T_1T_2} q_{T_1T_2}}{n_{T_1T_2}}+\frac{p_{T_1T_2^*} q_{T_1T_2^*}}{n_{T_1T_2^*}} = \epsilon_2, \mbox{ for some constant } \epsilon_2 > 0.
	\end{equation*}
 
Note that, $\tau_{T_1T_2}=\frac{n_{T_1T_2}}{n_{T_1T_2^*}}$. The $n_{T_1T_2}$ and $n_{T_1T_2^*}$ can be written as,
	\begin{equation*}
	n_{T_1T_2}=n_{T_1}^{NR} \left(\frac{\tau_{T_1T_2}}{1+\tau_{T_1T_2}}\right), 	n_{T_1T_2^*}=\frac{n_{T_1}^{NR}}{1+\tau_{T_1T_2}},
	\end{equation*}
where $n_{T_1}^{NR} = n_{T_1T_2}+n_{T_1T_2^*}$, is the total number of patients who obtained treatment $T_1$ at the first stage and become non-responders at the end of the first stage. Substituting the expressions of $n_{T_1T_2}$ and $n_{T_1T_2^*}$ in asymptotic variance expression obtained earlier,
	\noindent 
	\begin{eqnarray}
	\frac{p_{T_1T_2}q_{T_1T_2}}{\frac{n_{T_1}^{NR}\tau_{T_1T_2}}{1+\tau_{T_1T_2}}} + \frac{p_{T_1T_2^*}q_{T_1T_2^*}}{\frac{n_{T_1}^{NR}}{1+\tau_{T_1T_2}}} =\epsilon_2. \nonumber \\ 
 \implies   n_{T_1}^{NR} = \frac{\left(1+\tau_{T_1T_2}\right) \left(p_{T_1T_2}q_{T_1T_2}+\tau_{T_1T_2}p_{T_1T_2^*}q_{T_1T_2^*}\right)}{\epsilon_2\tau_{T_1T_2}}.
     \label{eq : n for second stage}
 \end{eqnarray}

From Section \ref{optimum_allo},  the second stage optimal allocation ratio is obtained as,
\begin{equation}
\tau_{T_1T_2}^{*} = \arg \min_{\tau_{T_1T_2}} F_2(\tau_{T_1T_2}) \mbox{ subject to }   \frac{p_{T_1T_2} q_{T_1T_2}}{n_{T_1T_2}}+\frac{p_{T_1T_2^*} q_{T_1T_2^*}}{n_{T_1T_2^*}} = \epsilon_2. \nonumber
\end{equation}

Among the patients who obtained $T_1$ at the first stage, the number of failures after a randomization process at the second stage is (see Section \ref{optimum_allo})
\begin{align*}
F_2(\tau_{T_1T_2}) & = n_{T_1T_2}q_{T_1T_2}+n_{T_1T_2^*}q_{T_1T_2^*} \\
& =  n_{T_1}^{NR} \left(\frac{\tau_{T_1T_2}}{1+\tau_{T_1T_2}}\right) q_{T_1T_2}+\frac{n_{T_1}^{NR}}{1+\tau_{T_1T_2}}q_{T_1T_2^*} \\
& = \left(p_{T_1T_2}q_{T_1T_2}+\tau_{T_1T_2}p_{T_1T_2^*}q_{T_1T_2^*}\right) \frac{q_{T_1T_2}}{\epsilon_2} + \frac{\left(p_{T_1T_2} q_{T_1T_2}+\tau_{T_1T_2}p_{T_1T_2^*}q_{T_1T_2^*}\right)q_{T_1T_2^*}}{\epsilon_2 \tau_{T_1T_2}}.
\end{align*}

Now, we have
\begin{align}
\frac{\partial F_2(\tau_{T_1T_2})}{\partial\tau_{T_1T_2}} & = p_{T_1T_2^*}q_{T_1T_2^*}\frac{q_{T_1T_2}}{\epsilon_2}-\frac{p_{T_1T_2}q_{T_1T_2}q_{T_1T_2^*}}{\epsilon_2\tau_{T_1T_2}^2}. \nonumber 
\end{align}

Equating the above expression with $0$ gives the optimal value of $\tau_{T_1T_2}$ as,
\begin{eqnarray}
\tau_{T_1T_2}^* = \sqrt{\frac{p_{T_1T_2}}{p_{T_1T_2^*}}}.
\label{eq : second stage optimal alloc prop for diff}
\end{eqnarray}

Now, we show that the estimated second stage allocation ratio for the $n^{th}$ patient (see Section \ref{sec : adaptive alloc proc})
$$\hat\tau_{T_1T_2, n} \xrightarrow{a.s} \tau_{T_1T_2}^*.$$
Define \textbf{Lemma 1} (same as \textbf{Lemma 1.2.6} in \citet{sokol2013advanced}) : $\{X_n\}$ be a sequence of random variables, and let $X$ be some other random variable. Let $f : R \rightarrow R$ be a continuous function. If $X_n$ converges almost surely (\textit{a.s}) to $X$, then $f(X_n)$
converges almost surely to $f(X)$. If $X_n$ converges in probability to $X$, then $f(X_n)$ converges	in probability to $f(X)$.

Note that $\hat p_{T_1T_2, n} \xrightarrow{a.s} p_{T_1T_2}$; and $\left(\hat{p}_{AA'},\hat{p}_{AC},\hat{p}_{AD},\hat{p}_{BB'},\hat{p}_{BE},\hat{p}_{BF}\right) \in  (0,1)$.
Now consider $f(x) = \frac{1}{x}$, $x\  \in (0,1)$. Then the $f(x)$ is a continuous function in the support space.
Thus, using Lemma 1, 
\begin{eqnarray}
    \hat\tau_{T_1T_2, n} \xrightarrow{a.s} \sqrt{\frac{p_{T_1T_2}}{p_{T_1T_2^*}}} \implies \hat\tau_{T_1T_2, n} \xrightarrow{a.s} \tau_{T_1T_2}^*.
\label{eq : asymp conv of second stage optimal alloc}
\end{eqnarray}

\subsection{\centering
    \textbf{Derivation of First Stage Optimal Allocation Ratio for Simple Difference}\label{ap : first stage optimal alloc for diff}
    }
We consider the objective function $g(\cdot, \cdot)$ as introduced in Section \ref{optimum_allo} to be a simple difference. The objective function of simple difference for comparing the two first stage probabilities $(p_A$, and $p_B)$ is given by $p_A-p_B$. The optimality criterion (as defined in Section \ref{optimum_allo} of main paper) for the first stage allocation ratio using the asymptotic variance of the objective function $avar(g(\cdot,\cdot))$ can be expressed as
\begin{equation*}
\frac{p_A q_A}{n_A}+\frac{p_B q_B}{n_B} = \epsilon_1.
\end{equation*}
 
Using the expression for first stage success probability $(p_{T_1})$ and failure probability $(q_{T_1})$ as obtained in \ref{ap : first stage success probability wrt second stage}, in above equation, we get,
\begin{multline*}
\frac{\left(\gamma_Ap_{AA'}+(1-\gamma_A)\frac{\tau_{AC}}{1+\tau_{AC}}p_{AC}+(1-\gamma_A)\frac{1}{1+\tau_{AC}}p_{AD}\right)\left(\gamma_Aq_{AA'}+(1-\gamma_A)\frac{\tau_{AC}}{1+\tau_{AC}}q_{AC}+(1-\gamma_A)\frac{1}{1+\tau_{AC}}q_{AD}\right)}{n_A} \\  +\frac{\left(\gamma_Bp_{BB'}+(1-\gamma_B)\frac{\tau_{BE}}{1+\tau_{BE}}p_{BE}+(1-\gamma_B)\frac{1}{1+\tau_{BE}}p_{BF}\right)\left(\gamma_Bq_{BB'}+(1-\gamma_B)\frac{\tau_{BE}}{1+\tau_{BE}}q_{BE}+(1-\gamma_B)\frac{1}{1+\tau_{BE}}q_{BF}\right)}{n_B} \\ =\epsilon_1.
\end{multline*}
	
Since, $\tau_A=\frac{n_A}{n_B}$, $n_A$, and $n_B$ can be written as,
\begin{equation*}
n_A=n \left(\frac{\tau_A}{1+\tau_A}\right), \mbox{ } 	n_B=\frac{n}{1+\tau_A}.
\end{equation*}
	
Substituting the expression of $n_A$ and $n_B$ in asymptotic variance expression obtained earlier,
\noindent 
\begin{multline*}
\frac{\left(\gamma_Ap_{AA'}+(1-\gamma_A)\frac{\tau_{AC}}{1+\tau_{AC}}p_{AC}+(1-\gamma_A)\frac{1}{1+\tau_{AC}}p_{AD}\right)\left(\gamma_Aq_{AA'}+(1-\gamma_A)\frac{\tau_{AC}}{1+\tau_{AC}}q_{AC}+(1-\gamma_A)\frac{1}{1+\tau_{AC}}q_{AD}\right)}{\frac{n\tau_A}{1+\tau_A}} \\  +\frac{\left(\gamma_Bp_{BB'}+(1-\gamma_B)\frac{\tau_{BE}}{1+\tau_{BE}}p_{BE}+(1-\gamma_B)\frac{1}{1+\tau_{BE}}p_{BF}\right)\left(\gamma_Bq_{BB'}+(1-\gamma_B)\frac{\tau_{BE}}{1+\tau_{BE}}q_{BE}+(1-\gamma_B)\frac{1}{1+\tau_{BE}}q_{BF}\right)}{\frac{n}{1+\tau_A}} \\ =\epsilon_1.
\end{multline*}
	
Let,
\begin{equation*}
\gamma_Ap_{AA'}+\frac{(1-\gamma_A)\tau_{AC}}{1+\tau_{AC}}p_{AC}+\frac{(1-\gamma_A)}{1+\tau_{AC}}p_{AD}=t_1,
\end{equation*}
\begin{equation*}
\gamma_Aq_{AA'}+\frac{(1-\gamma_A)\tau_{AC}}{1+\tau_{AC}}q_{AC}+\frac{(1-\gamma_A)}{1+\tau_{AC}}q_{AD}=t_2,
\end{equation*}
\begin{equation*}
\gamma_Bp_{BB'}+\frac{(1-\gamma_B)\tau_{BE}}{1+\tau_{BE}}p_{BE}+\frac{(1-\gamma_B)}{1+\tau_{BE}}p_{BF}=l_1,
\end{equation*}
\begin{equation*}
\gamma_Bq_{BB'}+\frac{(1-\gamma_B)\tau_{BE}}{1+\tau_{BE}}q_{BE}+\frac{(1-\gamma_B)}{1+\tau_{BE}}q_{BF}=l_2.
\end{equation*}

Further using the expressions of $t_1,t_2,l_1$ and $l_2$ in the asymptotic variance expression we get,
\begin{equation}
\frac{t_1t_2}{\frac{n\tau_A}{1+\tau_A}}+\frac{l_1l_2}{\frac{n}{1+\tau_A}}=\epsilon_1.
\end{equation}

The total number of failures after the completion of SMART can be expressed as,
\begin{align*}
F_1(\tau_A,\tau_{AC},\tau_{BE}) & = n_{AA'}q_{AA'}+n_{AC}q_{AC}+n_{AD}q_{AD}+n_{BB'}q_{BB'}+n_{BE}q_{BE}+n_{BF}q_{BF} \\
& = \frac{n\tau_A}{1+\tau_A}\gamma_Aq_{AA'} + \frac{n\tau_A(1-\gamma_A)}{1+\tau_A}\frac{\tau_{AC}}{1+\tau_{AC}}q_{AC} +\frac{n\tau_A}{1+\tau_A}(1-\gamma_A)\frac{1}{1+\tau_{AC}}q_{AD} \\
& +\frac{n}{1+\tau_A}\gamma_Bq_{BB'} +\frac{n(1-\gamma_B)}{1+\tau_A}\frac{\tau_{BE}}{1+\tau_{BE}}q_{BE} +\frac{n(1-\gamma_B)}{1+\tau_A}\frac{1}{1+\tau_{BE}}q_{BF} \\
& = \frac{t_1t_2+\tau_Al_1l_2}{\epsilon_1}\gamma_Aq_{AA'} + \frac{t_1t_2+\tau_Al_1l_2}{\epsilon_1} \frac{1-\gamma_A}{1+\tau_{AC}} (\tau_{AC}q_{AC}+q_{AD}) \\ & +\frac{t_1t_2+\tau_Al_1l_2}{\tau_A\epsilon_1}\gamma_Bq_{BB'} + \frac{t_1t_2+\tau_Al_1l_2}{\tau_A\epsilon_1} \frac{1-\gamma_B}{1+\tau_{BE}} (\tau_{BE}q_{BE}+q_{BF}). 
\end{align*}

To obtain the first stage optimal allocation ratio, the above expression is differentiated with respect to the allocation ratio $(\tau_A)$ and is to be equated with $0$. Note that the two second-stage allocation ratios are replaced by corresponding optimal allocation ratios from (\ref{eq : asymp conv of second stage optimal alloc}) (see Section \ref{optimum_allo}). Hence,
\begin{align}
\frac{\partial F_1(\tau_A,\tau_{AC}^*,\tau_{BE}^*)}{\partial\tau_A} & = \frac{l_1l_2}{\epsilon_1}\gamma_Aq_{AA'}+\frac{l_1l_2}{\epsilon_1} \frac{1-\gamma_A}{1+\tau_{AC}^*}(\tau_{AC}^* q_{AC}+q_{AD}) \nonumber \\ & -\frac{t_1t_2}{\epsilon_1\tau_A^2}\gamma_Bq_{BB'}-\frac{t_1t_2}{\epsilon_1\tau_A^2}\frac{1-\gamma_B}{1+\tau_{BE}^*}(\tau_{BE}^* q_{BE}+q_{BF}).
\label{eq : expression of second stage total patient number for diff}
\end{align}

 Equating (\ref{eq : expression of second stage total patient number for diff}) with $0$ gives us the value of $\tau_A$ to be,
	\begin{align}
	\tau_A^* & =\sqrt{\frac{t_1t_2\gamma_Bq_{BB'}+t_1t_2\frac{1-\gamma_B}{1+\tau_{BE}^*}(\tau_{BE}^*q_{BE}+q_{BF})}{l_1l_2\gamma_Aq_{AA'}+l_1l_2\frac{1-\gamma_A}{1+\tau_{AC}^*}(\tau_{AC}^*q_{AC}+q_{AD})}} \nonumber \\
	& = \sqrt{\frac{(1+\tau_{AC}^*)t_1t_2(\gamma_B q_{BB'}(1+\tau_{BE}^*)+(1-\gamma_B) (\tau_{BE}^* q_{BE}+q_{BF}))}{(1+\tau_{BE}^*)l_1l_2(\gamma_A q_{AA'}(1+\tau_{AC}^*)+(1-\gamma_A) (\tau_{AC}^* q_{AC}+q_{AD}))}} \nonumber \\ 
	& = \sqrt{\frac{(1+\tau_{BE}^*)(\gamma_A p_{AA'}(1+\tau_{AC}^*)+(1-\gamma_A) (\tau_{AC}^* p_{AC}+p_{AD}))}{(1+\tau_{AC}^*)(\gamma_B p_{BB'}(1+\tau_{BE}^*)+(1-\gamma_B) (\tau_{BE}^* p_{BE}+p_{BF}))}}.
    \label{eq : first stage optimal alloc prop for diff}
	\end{align}
	
\textbf{Lemma 2}	(same as \textbf{Lemma 1.2.10} in \citet{sokol2013advanced}) states that, let $\{X_n\}$ and $\{Y_n\}$ be sequences of random variables, and let $X$ and $Y$ be two other random variables. If $X_n$ converges in probability to $X$ and $Y_n$ converges in probability to $Y$, then $X_n + Y_n$ converges in probability to $X + Y$, and $X_nY_n$ converges in probability to $XY$. Also, if $X_n$ converges almost surely to $X$ and $Y_n$ converges almost surely to $Y$, then $X_n + Y_n$ converges almost surely to $X + Y$, and $X_nY_n$ converges almost surely to $XY$. Thus, using Lemma 2, we have (see (\ref{eq : asymp conv of second stage optimal alloc})),
\begin{equation*}
\gamma_A\hat{p}_{AA',n}(1+\hat{\tau}_{AC,n}) \xrightarrow{a.s} \gamma_Ap_{AA'}\left(\frac{\sqrt{p_{AC}}+\sqrt{p_{AD}}}{\sqrt{p_{AD}}}\right)
\end{equation*}
\begin{equation*}
(1-\gamma_A)(\hat{\tau}_{AC,n} \hat{p}_{AC,n} + \hat{p}_{AD,n}) \xrightarrow{a.s} (1-\gamma_A)\left(\frac{(p_{AC})^{\frac{3}{2}}+(p_{AD})^{\frac{3}{2}}}{\sqrt{p_{AD}}}\right).
\end{equation*}
Using Lemma 2 on equation (\ref{eq : asymp conv of second stage optimal alloc}) and the above two asymptotic expressions, we obtain that,
\begin{eqnarray}
    & (1+\hat{\tau}_{BE,n})(\gamma_A\hat{p}_{AA',n}(1+\hat{\tau}_{AC,n})+(1-\gamma_A)(\hat{\tau}_{AC,n}\hat{p}_{AC,n}+\hat{p}_{AD,n})) \nonumber \\ & \xrightarrow{a.s}  \left(\frac{\sqrt{p_{BF}}+\sqrt{p_{BE}}}{\sqrt{p_{BF}}}\right) \left(\gamma_Ap_{AA'}\left(\frac{\sqrt{p_{AC}}+\sqrt{p_{AD}}}{\sqrt{p_{AD}}}\right)+(1-\gamma_A)\left(\frac{(p_{AC})^{\frac{3}{2}}+(p_{AD})^{\frac{3}{2}}}{\sqrt{p_{AD}}}\right)\right).
\label{eq : asymp conv of first stage numerator for diff}
\end{eqnarray}
Similarly, using Lemma 2, we have,
\begin{eqnarray}
    \gamma_B\hat{p}_{BB',n}(1+\hat{\tau}_{BE,n}) \xrightarrow{a.s} \gamma_Bp_{BB'}\left(\frac{\sqrt{p_{BE}}+\sqrt{p_{BF}}}{\sqrt{p_{BF}}}\right) \nonumber
\end{eqnarray}
\begin{eqnarray}
(1-\gamma_B)(\hat{\tau}_{BE,n}\hat{p}_{BE,n}+\hat{p}_{BF,n}) \xrightarrow{a.s} (1-\gamma_B)\left(\frac{(p_{BE})^{\frac{3}{2}}+(p_{BF})^{\frac{3}{2}}}{\sqrt{p_{BF}}}\right). \nonumber
\end{eqnarray}
Again using Lemma 2, on the above two obtained expressions, we get,
\begin{eqnarray}
& (1+\hat{\tau}_{AC,n})(\gamma_B\hat{p}_{BB',n}(1+\hat{\tau}_{BE,n})+(1-\gamma_B)(\hat{\tau}_{BE,n}\hat{p}_{BE,n}+\hat{p}_{BF,n})) \nonumber \\ & \xrightarrow{a.s}  \left(\frac{\sqrt{p_{AC}}+\sqrt{p_{AD}}}{\sqrt{p_{AD}}}\right) \left(\gamma_Bp_{BB'}\left(\frac{\sqrt{p_{BE}}+\sqrt{p_{BF}}}{\sqrt{p_{BF}}}\right)+(1-\gamma_B)\left(\frac{(p_{BE})^{\frac{3}{2}}+(p_{BF})^{\frac{3}{2}}}{\sqrt{p_{BF}}}\right)\right).
\label{eq : asymp conv of first stage denominator for diff}
\end{eqnarray}
	
Thus, using Lemma 1, Lemma 2, and equations (\ref{eq : first stage optimal alloc prop for diff}), (\ref{eq : asymp conv of first stage numerator for diff}) and (\ref{eq : asymp conv of first stage denominator for diff}), we have
\begin{eqnarray}
\hat{\tau}_{A,n} \hspace{0.5cm} \xrightarrow{a.s} && \sqrt{\frac{\left[\sqrt{p_{BE}}+\sqrt{p_{BF}}\right]\left[\gamma_Ap_{AA'}(\sqrt{p_{AD}}+\sqrt{p_{AC}})+(1-\gamma_A)\left((p_{AC})^\frac{3}{2}+(p_{AD})^\frac{3}{2}\right)\right]}{\left[\sqrt{p_{AC}}+\sqrt{p_{AD}}\right]\left[\gamma_Bp_{BB'}(\sqrt{p_{BE}}+\sqrt{p_{BF}})+(1-\gamma_B)\left((p_{BE})^\frac{3}{2}+(p_{BF})^\frac{3}{2}\right)\right]}} \nonumber\\
&=& \sqrt{\frac{(1+\tau_{BE}^*)(\gamma_A p_{AA'}(1+\tau_{AC}^*)+(1-\gamma_A) (\tau_{AC}^* p_{AC}+p_{AD}))}{(1+\tau_{AC}^*)(\gamma_B p_{BB'}(1+\tau_{BE}^*)+(1-\gamma_B) (\tau_{BE}^* p_{BE}+p_{BF}))}}.
\label{eq : asymp conv of first stage optimal alloc for diff}
\end{eqnarray}

\subsection{\centering
\textbf{Derivation of Asymptotic Variance of Success Probabilities:  $\hat p_{T_1T_2}$}\label{ap : Variance of success prob for diff}
}
Let us consider $\{Y_1,Y_2,...,Y_n\}$ be the binary primary outcome variables, where,
\begin{equation*}
    Y_i=
    \begin{cases}
      1, & \text{if $i^{th}$ patient observes success at the end of study,} \\
      0, & \text{otherwise.}
    \end{cases}
\end{equation*}
$T_{1i}$ and $T_{2i}$ denote the assigned first and second stage treatments to the $i^{th}$ patient, respectively. $T_{1i}$ can take values $A$ and $B$; $T_{2i} \in \{A', C, D\}$ where $T_{1i} = A$ and $T_{2i} \in \{B', E, F\}$ where $T_{1i} = B$. As defined in Section \ref{sec : adaptive alloc proc}, $\mathscr{F}_i = \{Y_1,Y_2,...,Y_i, T_{11},T_{12},...,T_{1i}, T_{21},T_{22},...,T_{2i} \}$ as the history of primary outcome variables, first and second stage allocated treatments for the first $i$ patients. Also, the conditional expectation as, $E_{i-1}(\cdot)=E(\cdot|\mathscr{F}_{i-1})$. Let, $T_1^n=(T_{11},T_{12},...,T_{1n})$, and $T_2^n=(T_{21},T_{22},...,T_{2n})$ be the history of treatment assignment of the first and second stage allocated treatments, respectively; and $Y^n=(Y_1,Y_2,...,Y_n)$ be the history of primary outcome variables. The likelihood function from the data is \citep{rosenberger1997asymptotic},
\begin{align*}
    \mathscr{L}_n & \equiv \{Y^n,T_1^n,T_2^n\}\\
    & \equiv \mathscr{L}(Y_n,T_{1n},T_{2n}|\mathscr{F}_{n-1})\mathscr{L}_{n-1},
\end{align*}
where $\mathscr{L}(Y_n,T_{1n},T_{2n}|\mathscr{F}_{n-1})$ is the likelihood contribution from the $n^{th}$ patient given the history of earlier patients. Now, 
\begin{equation}
    \mathscr{L}_n=\prod_{i=1}^n\mathscr{L}(Y_i,T_{1i},T_{2i}|\mathscr{F}_{i-1}),
    \label{eq : crude likelihood}
\end{equation}
with $\mathscr{L}_0 = 1$, where,
\begin{align}
    & \mathscr{L}(Y_i,T_{1i},T_{2i}|\mathscr{F}_{i-1}) \nonumber \\ 
    & = \mathscr{L}(Y_i|T_{1i},T_{2i},\mathscr{F}_{i-1})\mathscr{L}(T_{1i},T_{2i}|\mathscr{F}_{i-1}) = \mathscr{L}(Y_i|T_{1i},T_{2i},\mathscr{F}_{i-1})\mathscr{L}(T_{2i}|T_{1i},\mathscr{F}_{i-1})\mathscr{L}(T_{1i}|\mathscr{F}_{i-1}) \nonumber \\ & = p_{AC}^{Y_iI(T_{1i}=A,T_{2i} = C)}(1-p_{AC})^{(1-Y_i)I(T_{1i}=A,T_{2i} = C)} \times \nonumber \\ & p_{AD}^{Y_iI(T_{1i}=A,T_{2i} = D)}(1-p_{AD})^{(1-Y_i)I(T_{1i}=A,T_{2i} = D)} \times \nonumber \\ & p_{AA'}^{Y_iI(T_{1i}=A,T_{2i} = A')}(1-p_{AA'})^{(1-Y_i)I(T_{1i}=A,T_{2i} = A')} \times \nonumber \\ & p_{BE}^{Y_iI(T_{1i}=B,T_{2i} = E)}(1-p_{BE})^{(1-Y_i)I(T_{1i}=B,T_{2i} = E)} \times \nonumber \\ & p_{BF}^{Y_iI(T_{1i}=B,T_{2i} = F)}(1-p_{BF})^{(1-Y_i)I(T_{1i}=B,T_{2i} = F)} \times \nonumber \\ 
    & p_{BB'}^{Y_iI(T_{1i}=B,T_{2i} = B')}(1-p_{BB'})^{(1-Y_i)I(T_{1i}=B,T_{2i} = B')} \times \nonumber \\
    & \{E_{i-1}(I(T_{2i}=A'|T_{1i}=A))\}^{I(T_{1i}=A,T_{2i}=A')} \times \nonumber \\
    & \left[\{E_{i-1}(I(T_{2i}=C|T_{1i}=A))\}^{I(T_{1i}=A,T_{2i}=C)}\{E_{i-1}(1-I(T_{2i}=C|T_{1i}=A))\}^{I(T_{1i}=A,T_{2i}=D)}\right]^{1-I(T_{1i}=A,T_{2i}=A')} \times \nonumber \\
    & \{E_{i-1}(I(T_{2i}=B'|T_{1i}=B))\}^{I(T_{1i}=B,T_{2i}=B')} \times \nonumber \\
    & \left[\{E_{i-1}(I(T_{2i}=E|T_{1i}=B))\}^{I(T_{1i}=B,T_{2i}=E)}\{E_{i-1}(1-I(T_{2i}=E|T_{1i}=B))\}^{I(T_{1i}=B,T_{2i}=F)}\right]^{1-I(T_{1i}=B,T_{2i}=B')} \times \nonumber \\ 
    & \{E_{i-1}(I(T_{1i}=A))\}^{I(T_{1i}=A)}\{E_{i-1}(1-I(T_{1i}=A))\}^{I(T_{1i}=B)}. 
    \label{likelihood_for_i_patient}
\end{align}

Thus, using (\ref{likelihood_for_i_patient}), the equation (\ref{eq : crude likelihood}) can be expressed as,
\begin{align}
    & \mathscr{L}_n \nonumber \\
    & = \left(p_{AC}^{\sum_i Y_i I(T_{1i}=A,T_{2i} = C)}(1-p_{AC})^{\sum_i (1-Y_i) I(T_{1i}=A,T_{2i} = C)}\right) \nonumber \\ 
    & \left(p_{AD}^{\sum_i Y_i I(T_{1i}=A,T_{2i} = D)}(1-p_{AD})^{\sum_i (1-Y_i) I(T_{1i}=A,T_{2i} = D)}\right) \times \nonumber \\ 
    & \left(p^{\sum_i Y_i I(T_{1i}=A,T_{2i} = A')}_{AA'}(1-p_{AA'})^{\sum_i (1-Y_i) I(T_{1i}=A,T_{2i} = A')}\right) \times \nonumber \\ 
    & \left(\prod_{i=1}^n \{E_{i-1}(I(T_{2i}=A'|T_{1i}=A))\}^{I(T_{1i}=A,T_{2i}=A')} \right) \times \nonumber \\
    & \left(\prod_{i=1}^n\left[\{E_{i-1}(I(T_{2i}=C|T_{1i}=A))\}^{I(T_{1i}=A,T_{2i}=C)}\{E_{i-1}(1-I(T_{2i}=C|T_{1i}=A))\}^{I(T_{1i}=A,T_{2i}=D)}\right]^{1-I(T_{1i}=A,T_{2i}=A')}\right) \times \nonumber \\
    & \left(p_{BE}^{\sum_i Y_i I(T_{1i}=B,T_{2i} = E)}(1-p_{BE})^{\sum_i (1-Y_i) I(T_{1i}=B,T_{2i} = E)}\right) \times \nonumber \\ 
    & \left(p_{BF}^{\sum_i Y_i I(T_{1i}=B,T_{2i} = F)}(1-p_{BF})^{\sum_i (1-Y_i) I(T_{1i}=B,T_{2i} = F)}\right) \times \nonumber \\ 
    & \left(p^{\sum_i Y_i I(T_{1i}=B,T_{2i} = B')}_{BB'}(1-p_{BB'})^{\sum_i (1-Y_i) I(T_{1i}=B,T_{2i} = B')}\right) \times \nonumber \\
    & \left(\prod_{i=1}^n \{E_{i-1}(I(T_{2i}=B'|T_{1i}=B))\}^{I(T_{1i}=B,T_{2i}=B')} \right) \times \nonumber \\
    & \left(\prod_{i=1}^n\left[\{E_{i-1}(I(T_{2i}=E|T_{1i}=B))\}^{I(T_{1i}=B,T_{2i}=E)}\{E_{i-1}(1-I(T_{2i}=E|T_{1i}=B))\}^{I(T_{1i}=B,T_{2i}=F)}\right]^{1-I(T_{1i}=B,T_{2i}=B')}\right) \times \nonumber\\ 
    & \left(\prod_{i=1}^n(E_{i-1}(I(T_{1i}=A))^{I(T_{1i}=A)}(1-E_{i-1}(I(T_{1i}=A)))^{I(T_{1i}=B)}\right).
    \label{likelihood_for_all_n_patient}
\end{align}

Now, using the equation $(A3)$ of \citet{rosenberger1997asymptotic}, $-n^{-1}\sum_{i=1}^{n}E_{i-1}\{\frac{\partial^2 log \mathscr{L}_i}{\partial p_{AC}^2}\}$ from (\ref{likelihood_for_i_patient}) becomes
\begin{align}
& n^{-1}\sum_{i=1}^n\left[p_{AC}^{-2}E_{i-1}(Y_i I(T_{1i}=A,T_{2i} = C))+(1-p_{AC})^{-2}E_{i-1}((1-Y_i) I(T_{1i}=A,T_{2i} = C))\right] \nonumber \\
& = n^{-1}\sum_{i=1}^n\left(p_{AC}^{-1}+(1-p_{AC})^{-1}\right)E_{i-1}(I(T_{1i}=A,T_{2i} = C)) \nonumber \\
& \xrightarrow[]{a.s} (1-\gamma_A)(p_{AC}^{-1}+(1-p_{AC})^{-1})\left(\frac{\tau_{AC}}{1+\tau_{AC}}\right)\left(\frac{\tau_A}{1+\tau_A}\right) \equiv v_{AC}.
\label{Variance of prob along (A,C)}
\end{align}
The last step of the above is done using the results from the Appendix of \cite{rosenberger2001optimal}. Thus, the variance of $\hat p_{AC} (\equiv \hat p_{AC, n})$ is $\frac{1}{n} v_{AC}^{-1}$. Similarly, 
\begin{align}
& n^{-1}\sum_{i=1}^n\left[p_{AD}^{-2}E_{i-1}(Y_i I(T_{1i}=A,T_{2i} = D))+(1-p_{AD})^{-2}E_{i-1}((1-Y_i) I(T_{1i}=A,T_{2i} = D))\right] \nonumber \\
& = n^{-1}\sum_{i=1}^n\left(p_{AD}^{-1}+(1-p_{AD})^{-1}\right)E_{i-1}(I(T_{1i}=A,T_{2i} = D)) \nonumber \\
& \xrightarrow[]{a.s} (1-\gamma_A)(p_{AD}^{-1}+(1-p_{AD})^{-1})\left(\frac{1}{1+\tau_{AC}}\right)\left(\frac{\tau_A}{1+\tau_A}\right) \equiv v_{AD}.
\label{Variance of prob along (A,D)}
\end{align}
Thus, the variance of $\hat p_{AD} (\equiv \hat p_{AD, n})$ is $\frac{1}{n} v_{AD}^{-1}$.
Now, 
\begin{align}
    & n^{-1}\sum_{i=1}^n\left[p_{AA'}^{-2}E_{i-1}(Y_i I(T_{1i}=A,T_{2i} = A'))+(1-p_{AA'})^{-2}E_{i-1}((1-Y_i) I(T_{1i}=A,T_{2i} = A'))\right] \nonumber \\
    & = n^{-1}\sum_{i=1}^n\left(p_{AA'}^{-1}+(1-p_{AA'})^{-1}\right)E_{i-1}(I(T_{1i}=A,T_{2i} = A')) \nonumber \\
    & \xrightarrow[]{a.s} \gamma_A(p_{AA'}^{-1}+(1-p_{AA'})^{-1})\left(\frac{\tau_{A}}{1+\tau_{A}}\right) \equiv v_{AA'}.
    \label{Variance of prob along (A,A)}
\end{align}
So, the variance of $\hat p_{AA'} (\equiv \hat p_{AA', n})$ is $\frac{1}{n} v_{AA'}^{-1}$.
Similarly, 
\begin{align}
    & n^{-1}\sum_{i=1}^n\left[p_{BE}^{-2}E_{i-1}(Y_i I(T_{1i}=B,T_{2i} = E))+(1-p_{BE})^{-2}E_{i-1}((1-Y_i) I(T_{1i}=B,T_{2i} = E))\right] \nonumber \\
    & = n^{-1}\sum_{i=1}^n\left(p_{BE}^{-1}+(1-p_{BE})^{-1}\right)E_{i-1}(I(T_{1i}=B,T_{2i} = E)) \nonumber \\
    & \xrightarrow[]{a.s} (1-\gamma_B)(p_{BE}^{-1}+(1-p_{BE})^{-1})\left(\frac{\tau_{BE}}{1+\tau_{BE}}\right)\left(\frac{1}{1+\tau_{A}}\right) \equiv v_{BE}.
    \label{Variance of prob along (B,E)}
\end{align}
Thus, the variance of $\hat p_{BE} (\equiv \hat p_{BE, n})$ is $\frac{1}{n} v_{BE}^{-1}$. Similarly, 
\begin{align}
    & n^{-1}\sum_{i=1}^n\left[p_{BF}^{-2}E_{i-1}(Y_i I(T_{1i}=B,T_{2i} = F))+(1-p_{BF})^{-2}E_{i-1}((1-Y_i) I(T_{1i}=B,T_{2i} = F))\right] \nonumber \\
    & = n^{-1}\sum_{i=1}^n\left(p_{BF}^{-1}+(1-p_{BF})^{-1}\right)E_{i-1}(I(T_{1i}=B,T_{2i} = F)) \nonumber \\
    & \xrightarrow[]{a.s} (1-\gamma_B)(p_{BF}^{-1}+(1-p_{BF})^{-1})\left(\frac{1}{1+\tau_{BE}}\right)\left(\frac{1}{1+\tau_{A}}\right) \equiv v_{BF}.
    \label{Variance of prob along (B,F)}
\end{align}
So, the variance of $\hat p_{BF} (\equiv \hat p_{BF, n})$ is $\frac{1}{n} v_{BF}^{-1}$. Similarly,
\begin{align}
    & n^{-1}\sum_{i=1}^n\left[p_{BB'}^{-2}E_{i-1}(Y_i I(T_{1i}=B,T_{2i} = B'))+(1-p_{BB'})^{-2}E_{i-1}((1-Y_i) I(T_{1i}=B,T_{2i} = B'))\right] \nonumber \\
    & = n^{-1}\sum_{i=1}^n\left(p_{BB'}^{-1}+(1-p_{BB'})^{-1}\right)E_{i-1}(I(T_{1i}=B,T_{2i} = B')) \nonumber \\
    & \xrightarrow[]{a.s} \gamma_B(p_{BB'}^{-1}+(1-p_{BB'})^{-1})\left(\frac{1}{1+\tau_{A}}\right) \equiv v_{BB'}.
    \label{Variance of prob along (B,B)}
\end{align}
Thus, the variance of $\hat p_{BB'} (\equiv \hat p_{BB', n})$ is $\frac{1}{n} v_{BB'}^{-1}$.

\subsubsection{\centering
    \textbf{Derivation of Asymptotic Variance of Second Stage Allocation Ratio} \label{ap : Variance second stage alloc diff}
}
Using equations (\ref{Variance of prob along (A,C)}), (\ref{Variance of prob along (A,D)}), (\ref{Variance of prob along (B,E)}), and (\ref{Variance of prob along (B,F)}) from Appendix \ref{ap : Variance of success prob for diff}, we have (See \cite{rosenberger1997asymptotic}, \cite{rosenberger2001optimal}),
\begin{eqnarray}
\sqrt{n}(\hat{p}_{AC,n}-p_{AC}) & \xrightarrow[]{d} N(0,v_{AC}^{-1}), \nonumber \\ 
\sqrt{n}(\hat{p}_{AD,n}-p_{AD}) & \xrightarrow[]{d} N(0,v_{AD}^{-1}), \nonumber \\
\sqrt{n}(\hat{p}_{BE,n}-p_{BE}) & \xrightarrow[]{d}N(0,v_{BE}^{-1}), \nonumber \\
\sqrt{n}(\hat{p}_{BF,n}-p_{BF}) & \xrightarrow[]{d}N(0,v_{BF}^{-1}), \nonumber
\end{eqnarray}
and are asymptotically independent. Using Slutsky's theorem,

\begin{eqnarray}
    \sqrt{n}\left(\begin{bmatrix}
\hat{p}_{AC,n}-p_{AC}\\
\hat{p}_{AD,n}-p_{AD}
\end{bmatrix}\right) & \xrightarrow[]{d} N\left(\begin{bmatrix}
0\\
0
\end{bmatrix},
\begin{bmatrix}
v_{AC}^{-1} & 0\\
0 & v_{AD}^{-1}
\end{bmatrix}\right), \nonumber \\
\sqrt{n}\left(\begin{bmatrix}
\hat{p}_{BE,n}-p_{BE}\\
\hat{p}_{BF,n}-p_{BF}
\end{bmatrix}\right) & \xrightarrow[]{d} N\left(\begin{bmatrix}
0\\
0
\end{bmatrix},\begin{bmatrix}
v_{BE}^{-1} & 0\\
    0 & v_{BF}^{-1}
\end{bmatrix}\right). \nonumber
\end{eqnarray}

Note that, $\hat{\tau}_{AC,n}=\sqrt{\frac{\hat{p}_{AC,n}}{\hat{p}_{AD,n}}}$, and $\hat{\tau}_{BE,n}=\sqrt{\frac{\hat{p}_{BE,n}}{\hat{p}_{BF,n}}}$. Now using Delta Method (with function, $h(x,y)=\sqrt{\frac{x}{y}}$), we have,
\begin{eqnarray}
\sqrt{n}(\hat \tau_{AC,n}-\tau_{AC}^*) \xrightarrow[]{d} N\left(0,\frac{1}{4}\left(\frac{v_{AC}^{-1}}{p_{AC}p_{AD}}+\frac{v_{AD}^{-1}p_{AC}}{p_{AD}^3}\right)\right), \nonumber \\ 
\sqrt{n}(\hat \tau_{BE,n}-\tau_{BE}^*) \xrightarrow[]{d} N\left(0,\frac{1}{4}\left(\frac{v_{BE}^{-1}}{p_{BE}p_{BF}}+\frac{v_{BF}^{-1}p_{BE}}{p_{BF}^3}\right)\right). \nonumber
\end{eqnarray}

Finally, the asymptotic variances of the second stage optimal allocation ratios are obtained as,
\begin{eqnarray}
Var(\hat \tau_{AC,n}) = \frac{1}{4n}\left(\frac{v_{AC}^{-1}}{p_{AC}p_{AD}}+\frac{v_{AD}^{-1}p_{AC}}{p_{AD}^3}\right), \nonumber \\ 
Var(\hat \tau_{BE,n}) = \frac{1}{4n}\left(\frac{v_{BE}^{-1}}{p_{BE}p_{BF}}+\frac{v_{BF}^{-1}p_{BE}}{p_{BF}^3}\right). \nonumber
\end{eqnarray}

\subsubsection{\centering
    \textbf{Derivation of Asymptotic Variance of First Stage Allocation Ratio}\label{ap : Variance first stage alloc diff}
}

Using equations (\ref{Variance of prob along (A,C)}), (\ref{Variance of prob along (A,D)}), (\ref{Variance of prob along (A,A)}), (\ref{Variance of prob along (B,E)}), (\ref{Variance of prob along (B,F)}), and (\ref{Variance of prob along (B,B)}) from Appendix \textbf{\ref{ap : Variance of success prob for diff}}, we have,
\begin{eqnarray}
    \sqrt{n}(\hat{p}_{AA',n}-p_{AA'}) & \xrightarrow[]{d} N(0,v_{AA'}^{-1}), \nonumber \\
    \sqrt{n}(\hat{p}_{AC,n}-p_{AC}) & \xrightarrow[]{d} N(0,v_{AC}^{-1}), \nonumber \\ 
    \sqrt{n}(\hat{p}_{AD,n}-p_{AD}) & \xrightarrow[]{d} N(0,v_{AD}^{-1}), \nonumber \\
    \sqrt{n}(\hat{p}_{BB',n}-p_{BB'}) & \xrightarrow[]{d} N(0,v_{BB'}^{-1}), \nonumber \\
    \sqrt{n}(\hat{p}_{BE,n}-p_{BE}) & \xrightarrow[]{d}N(0,v_{BE}^{-1}), \nonumber \\
    \sqrt{n}(\hat{p}_{BF,n}-p_{BF}) & \xrightarrow[]{d}N(0,v_{BF}^{-1}), \nonumber
\end{eqnarray}
and are asymptotically independent. Using Slutsky's theorem,

\begin{eqnarray}
        \sqrt{n}\left(\begin{bmatrix}
\hat{p}_{AA',n}-p_{AA'}\\
\hat{p}_{AC,n}-p_{AC}\\
\hat{p}_{AD,n}-p_{AD}
\end{bmatrix}\right)  \xrightarrow[]{d} N\left(\begin{bmatrix}
0\\
0\\
0
\end{bmatrix},\begin{bmatrix}
v^{-1}_{AA'} & 0 & 0 \\
0 & v^{-1}_{AC} & 0 \\
0 & 0 & v^{-1}_{AD}
\end{bmatrix}\right). \nonumber 
\end{eqnarray}
Similarly, 

\begin{eqnarray}
        \sqrt{n}\left(\begin{bmatrix}
\hat{p}_{BB',n}-p_{BB'}\\
\hat{p}_{BE,n}-p_{BE}\\
\hat{p}_{BF,n}-p_{BF}
\end{bmatrix}\right)  \xrightarrow[]{d} N\left(\begin{bmatrix}
0\\
0\\
0
\end{bmatrix},\begin{bmatrix}
v^{-1}_{BB'} & 0 & 0 \\
0 & v^{-1}_{BE} & 0 \\
0 & 0 & v^{-1}_{BF}
\end{bmatrix}\right). \nonumber 
\end{eqnarray}

Now, we use the Delta method to derive the asymptotic distribution of the estimated first stage success probability from \ref{ap : first stage success probability wrt second stage}. The chosen functional form for the Delta method is $h_1(x,y,z) = \gamma_{T_1}x+(1-\gamma_{T_1})\left[\frac{y^{\frac{3}{2}}+z^{\frac{3}{2}}}{y^{\frac{1}{2}}+z^{\frac{1}{2}}}\right]$. Now, we have,
\begin{align*}
    Var(\sqrt{n} \hat p_{T_1,n}) & = \gamma_{T_1}^2 \times v_{T_1T_1^{'}}^{-1} \\ & +\left(\frac{\left(1-\gamma_{T_1}\right)\left(2\left(p_{T_1T_2}\right)^{\frac{3}{2}}+3p_{T_1T_2}(p_{T_1T_2^*})^{\frac{1}{2}}-\left(p_{T_1T_2^*}\right)^{\frac{3}{2}}\right)}{2\left(p_{T_1T_2}\right)^{\frac{1}{2}}\left(p_{T_1T_2}^{\frac{1}{2}}+p_{T_1T_2^*}^{\frac{1}{2}}\right)^{2}}\right)^2 \times v_{T_1T_2}^{-1} \\ & +\left(\frac{\left(1-\gamma_{T_1}\right)\left(2\left(p_{T_1T_2^*}\right)^{\frac{3}{2}}+3p_{T_1T_2^*}(p_{T_1T_2})^{\frac{1}{2}}-\left(p_{T_1T_2}\right)^{\frac{3}{2}}\right)}{2\left(p_{T_1T_2^*}\right)^{\frac{1}{2}}\left(p_{T_1T_2}^{\frac{1}{2}}+p_{T_1T_2^*}^{\frac{1}{2}}\right)^{2}}\right)^2 \times v_{T_1T_2^{*}}^{-1} \\
    & \equiv v_{T_1}.
\end{align*}

Thus, 
\begin{eqnarray}
    \sqrt{n}\left(\hat p_{T_1,n}-p_{T_1}\right) & \xrightarrow[]{d} N(0,v_{T_1}). \nonumber
\end{eqnarray}
Using the above equation, we have,
\begin{eqnarray}
        \sqrt{n}\left(\begin{bmatrix}
\hat{p}_{A}-p_{A}\\
\hat{p}_{B}-p_{B}
\end{bmatrix}\right)  \xrightarrow[]{d} N\left(\begin{bmatrix}
0\\
0
\end{bmatrix},\begin{bmatrix}
v_{A} & 0 \\
0 & v_{B}
\end{bmatrix}\right). \nonumber 
\end{eqnarray}

Now, using the same approach as in Appendix \ref{ap : Variance second stage alloc diff}, the variance of the first stage optimal allocation ratio is obtained as,

\begin{eqnarray}
        \sigma^2_{\tau_A} = Var(\hat \tau_{A,n}) = \left(\frac{1}{4n}\right) \left( \frac{v_{A}}{\left(\gamma_Ap_{AA'}+\left(1-\gamma_A\right)\left(\frac{p_{AC}^{\frac{3}{2}}+p_{AD}^{\frac{3}{2}}}{p_{AC}^{\frac{1}{2}}+p_{AD}^{\frac{1}{2}}}\right)\right)\left(\gamma_Bp_{BB'}+\left(1-\gamma_B\right)\left(\frac{p_{BE}^{\frac{3}{2}}+p_{BF}^{\frac{3}{2}}}{p_{BE}^{\frac{1}{2}}+p_{BF}^{\frac{1}{2}}}\right)\right)}\right) \nonumber \\ + \left(\frac{1}{4n}\right)\left(\frac{v_{B}\left(\gamma_Ap_{AA'}+\left(1-\gamma_A\right)\left(\frac{p_{AC}^{\frac{3}{2}}+p_{AD}^{\frac{3}{2}}}{p_{AC}^{\frac{1}{2}}+p_{AD}^{\frac{1}{2}}}\right)\right)}{\left(\gamma_Bp_{BB'}+\left(1-\gamma_B\right)\left(\frac{p_{BE}^{\frac{3}{2}}+p_{BF}^{\frac{3}{2}}}{p_{BE}^{\frac{1}{2}}+p_{BF}^{\frac{1}{2}}}\right)\right)^3}\right). \nonumber
\end{eqnarray}

\subsection{\centering
    \textbf{Derivation of Second Stage Optimal Allocation Ratio for Odds Ratio}\label{ap : second stage optimal alloc for odds ratio}
    }
Let us consider the objective function $g(\cdot, \cdot)$ as introduced in Section \ref{optimum_allo} to be odds ratio. We consider the same setup as in Appendix \ref{ap : second stage optimal alloc for diff}. The odds ratio of corresponding to treatment sequences $\{T_1, T_2\}$ and $\{T_1, T_2^*\}$ is defined as, $\frac{p_{T_1T_2}(1-p_{T_1T_2^*})}{p_{T_1T_2^*}(1-p_{T_1T_2})}$. The optimality criterion (as defined in Section \ref{optimum_allo} of main paper) using the asymptotic variance of the objective function $avar(g(\hat p_{T_1T_2},\hat p_{T_1T_2^*}))$ can be expressed,
\begin{equation*}
\left(\frac{p_{T_1T_2}q_{T_1T_2^*}}{p_{T_1T_2^*}q_{T_1T_2}}\right)^2 \left(\frac{1}{n_{T_1T_2}p_{T_1T_2}}+\frac{1}{n_{T_1T_2}q_{T_1T_2}}+\frac{1}{n_{T_1T_2^*}p_{T_1T_2^*}}+\frac{1}{n_{T_1T_2^*}q_{T_1T_2^*}}\right) = \epsilon_2, \mbox{ for some constant } \epsilon_2 > 0.
\end{equation*}
 
Note that, $\tau_{T_1T_2}=\frac{n_{T_1T_2}}{n_{T_1T_2^*}}$. The  $n_{T_1T_2}$, and $n_{T_1T_2^*}$ can be written as,
\begin{equation*}
n_{T_1T_2}=n_{T_1}^{NR} \left(\frac{\tau_{T_1T_2}}{1+\tau_{T_1T_2}}\right),
n_{T_1T_2^*}=\frac{n_{T_1}^{NR}}{1+\tau_{T_1T_2}},
\end{equation*}
where $n_{T_1}^{NR} = n_{T_1T_2}+n_{T_1T_2^*}$, is the total number of patients who obtained treatment $T_1$ at the first stage and become non-responders at the end of the first stage. Substituting the expressions of $n_{T_1T_2}$ and $n_{T_1T_2^*}$ in asymptotic variance expression obtained earlier,
\noindent 
\begin{equation*}
\frac{1}{\frac{n_{T_1}^{NR}\tau_{T_1T_2}}{1+\tau_{T_1T_2}}p_{T_1T_2}q_{T_1T_2}} + \frac{1}{\frac{n_{T_1}^{NR}}{1+\tau_{T_1T_2}}p_{T_1T_2^*}q_{T_1T_2^*}} =\epsilon_2 \left(\frac{p_{T_1T_2^*}q_{T_1T_2}}{p_{T_1T_2}q_{T_1T_2^*}}\right)^2 .
\end{equation*}
Solving for $n_{T_1}^{NR}$, gives,
\begin{eqnarray}
 n_{T_1}^{NR} = \frac{\left(1+\tau_{T_1T_2}\right) \left(p_{T_1T_2^*}q_{T_1T_2^*}+\tau_{T_1T_2}p_{T_1T_2}q_{T_1T_2}\right)}{\epsilon_2\tau_{T_1T_2}p_{T_1T_2}q_{T_1T_2}p_{T_1T_2^*}q_{T_1T_2^*}} \left(\frac{p_{T_1T_2}q_{T_1T_2^*}}{p_{T_1T_2^*}q_{T_1T_2}}\right)^2.
 \label{eq : n for second stage for odds ratio}
\end{eqnarray}
 

From Section \ref{optimum_allo},  the second stage optimal allocation ratio is obtained as,
\begin{eqnarray}
    \tau_{T_1T_2}^{*} = \arg \min_{\tau_{T_1T_2}} F_2(\tau_{T_1T_2}) & \mbox{ subject to } \nonumber \\ &  \left(\frac{p_{T_1T_2}q_{T_1T_2^*}}{p_{T_1T_2^*}q_{T_1T_2}}\right)^2 \left(\frac{1}{n_{T_1T_2}p_{T_1T_2}}+\frac{1}{n_{T_1T_2}q_{T_1T_2}}+\frac{1}{n_{T_1T_2^*}p_{T_1T_2^*}}+\frac{1}{n_{T_1T_2^*}q_{T_1T_2^*}}\right) = \epsilon_2. \nonumber
\end{eqnarray} 

Thus using the above optimality criterion, the optimal value of $\tau_{T_1T_2}$ is,
\begin{eqnarray}
    \tau_{T_1T_2}^* = \left(\sqrt{\frac{p_{T_1T_2^*}}{p_{T_1T_2}}}\right)\left(\frac{q_{T_1T_2^*}}{q_{T_1T_2}}\right).
\label{eq : second stage optimal alloc prop for odds ratio}
\end{eqnarray}
	
As derived in Section \ref{ap : second stage optimal alloc for diff} using Lemma 1, on equation (\ref{eq : second stage optimal alloc prop for odds ratio}) we have,
\begin{eqnarray}
    \hat{\tau}_{T_1T_2,n} \xrightarrow{a.s} \left(\sqrt{\frac{p_{T_1T_2^*}}{p_{T_1T_2}}}\right)\left(\frac{q_{T_1T_2^*}}{q_{T_1T_2}}\right).
\label{eq : asymp conv of second stage optimal alloc for odds}
\end{eqnarray}

 \subsubsection{\centering Allocation Procedure}
Similar to the procedure of Section \ref{sec : adaptive alloc proc} (as given by (\ref{2nd stage allo process}) for simple difference), the adaptive allocation process for the second stage optimal allocation ratio using odds ratio as the objective function is,
 \begin{equation}\label{2nd stage allo process for odds ratio}
    E_{i-1}(I(T_{2i}=t_2|T_{1i}=t_1,R_{T_{1i}}=0)) = \frac{\sqrt{\hat{p}_{t_1t_2^*,i-1}}\hat{q}_{t_1t_2^*,i-1}}{\sqrt{\hat{p}_{t_1t_2,i-1}}\hat{q}_{t_1t_2,i-1}+\sqrt{\hat{p}_{t_1t_2^*,i-1}}\hat{q}_{t_1t_2^*,i-1}}.
\end{equation}

\subsubsection{\centering Asymptotic Variance}\label{ap :  Variance second stage alloc odds}

Following the same procedure, as mentioned in Section \ref{ap : Variance second stage alloc diff}, using Delta method with the function $h(x,y)$ as $\sqrt{\frac{y}{x}} \frac{1-y}{1-x}$, the variance of the two second stage optimal allocation ratios $\hat{\tau}_{AC,n}$, and $\hat{\tau}_{BE,n}$ are,
\begin{eqnarray}
    Var(\hat{\tau}_{AC,n}) = \frac{1}{4n} \left(\frac{v_{AC}^{-1} \left(1-3p_{AC}\right)^2 \left(1-p_{AD}\right)^2 p_{AD}}{p_{AC}^{3} \left(1-p_{AC}\right)^4}+\frac{v_{AD}^{-1} \left(1-3p_{AD}\right)^2}{p_{AC} \left(1-p_{AC}\right)^2 p_{AD}}\right), \nonumber
\end{eqnarray}
\begin{eqnarray}
    Var(\hat{\tau}_{BE,n}) = \frac{1}{4n} \left(\frac{v_{BE}^{-1} \left(1-3p_{BE}\right)^2 \left(1-p_{BF}\right)^2 p_{BF}}{p_{BE}^{3} \left(1-p_{BE}\right)^4}+\frac{v_{BF}^{-1} \left(1-3p_{BF}\right)^2}{p_{BE} \left(1-p_{BE}\right)^2 p_{BF}}\right), \nonumber
\end{eqnarray}
where $v_{AC},v_{AD},v_{BE},$ and $v_{BF}$ are from Section \ref{ap : Variance of success prob for diff}. Thus, the asymptotic distributions of the estimated second stage optimum allocation ratios are, 
\begin{eqnarray}
    \sqrt{n}(\hat \tau_{AC,n}-\tau_{AC}^*) &\xrightarrow[]{d} & N\left(0,\frac{1}{4} \left(\frac{v_{AC}^{-1} \left(1-3p_{AC}\right)^2 \left(1-p_{AD}\right)^2 p_{AD}}{p_{AC}^{3} \left(1-p_{AC}\right)^4}+\frac{v_{AD}^{-1} \left(1-3p_{AD}\right)^2}{p_{AC} \left(1-p_{AC}\right)^2 p_{AD}}\right)\right),\nonumber \\ 
    \sqrt{n}(\hat \tau_{BE,n}-\tau_{BE}^*) &\xrightarrow[]{d}& N\left(0,\frac{1}{4} \left(\frac{v_{BE}^{-1} \left(1-3p_{BE}\right)^2 \left(1-p_{BF}\right)^2 p_{BF}}{p_{BE}^{3} \left(1-p_{BE}\right)^4}+\frac{v_{BF}^{-1} \left(1-3p_{BF}\right)^2}{p_{BE} \left(1-p_{BE}\right)^2 p_{BF}}\right)\right). \nonumber
\end{eqnarray}

\subsection{\centering
    \textbf{Derivation of First Stage Optimal Allocation Ratio for Odds Ratio}\label{ap : first stage optimal alloc for odds ratio}
    }
We consider the objective function $g(\cdot, \cdot)$ as introduced in Section \ref{optimum_allo} to be the odds ratio. The objective function of odds ratio for comparing the two first stage probabilities $(p_A$, and $p_B)$ is given by $\frac{p_{A}(1-p_{B})}{p_{B}(1-p_{A})}$. The optimality criterion (as defined in Section \ref{optimum_allo} of main paper) for the first stage allocation ratio using the asymptotic variance of the objective function $avar(g(\hat p_A,\hat p_B))$ can be expressed as
\begin{equation*}
\left(\frac{p_Aq_B}{p_Bq_A}\right)^2\left(\frac{1}{n_Ap_A}+\frac{1}{n_Aq_A}+\frac{1}{n_Bp_B}+\frac{1}{n_Bq_B}\right) = \epsilon_1.
\end{equation*}
 
Using the expression for first stage success probability $(p_{T_1})$ and failure probability $(q_{T_1})$ as obtained in \ref{ap : first stage success probability wrt second stage}, in above equation, we get,
\begin{eqnarray}
    & \left(\frac{\left(\gamma_{A}p_{AA'}+(1-\gamma_{A})\frac{\tau_{AC}}{1+\tau_{AC}}p_{AC}+(1-\gamma_{A})\frac{1}{1+\tau_{AC}}p_{AD}\right)\left(\gamma_{B}q_{BB'}+(1-\gamma_{B})\frac{\tau_{BE}}{1+\tau_{BE}}q_{BE}+(1-\gamma_{B})\frac{1}{1+\tau_{BE}}q_{BF}\right)}{\left(\gamma_{B}p_{BB'}+(1-\gamma_{B})\frac{\tau_{BE}}{1+\tau_{BE}}p_{BE}+(1-\gamma_{B})\frac{1}{1+\tau_{BE}}p_{BF}\right)\left(\gamma_{A}q_{AA'}+(1-\gamma_{A})\frac{\tau_{AC}}{1+\tau_{AC}}q_{AC}+(1-\gamma_{A})\frac{1}{1+\tau_{AC}}q_{AD}\right)}\right)^2 \nonumber \\ & \left[\frac{1}{n_Ap_Aq_A}+\frac{1}{n_Bp_Bq_B}\right] = \epsilon_1. \nonumber
\end{eqnarray}
	
Since, $\tau_A=\frac{n_A}{n_B}$, $n_A$, and $n_B$ can be written as,
\begin{equation*}
n_A=n \left(\frac{\tau_A}{1+\tau_A}\right), 
n_B=\frac{n}{1+\tau_A}.
\end{equation*}

The total number of failures that is obtained after completion of SMART can be expressed as,
\begin{align*}
F_1(\tau_A,\tau_{AC},\tau_{BE}) & = n_{AA'}q_{AA'}+n_{AC}q_{AC}+n_{AD}q_{AD}+n_{BB'}q_{BB'}+n_{BE}q_{BE}+n_{BF}q_{BF} 
\end{align*}

Now, from Section \ref{optimum_allo}, we have,
\begin{eqnarray}
\tau_{A}^{*} = & \arg \min_{\tau_{A}} F_1(\tau_{A}, \tau_{AC}^{*}, \tau_{BE}^{*}) \mbox{ subject to } \nonumber \\ 
    & \left(\frac{\left(\gamma_{A}p_{AA'}+(1-\gamma_{A})\frac{\tau_{AC}}{1+\tau_{AC}}p_{AC}+(1-\gamma_{A})\frac{1}{1+\tau_{AC}}p_{AD}\right)\left(\gamma_{B}q_{BB'}+(1-\gamma_{B})\frac{\tau_{BE}}{1+\tau_{BE}}q_{BE}+(1-\gamma_{B})\frac{1}{1+\tau_{BE}}q_{BF}\right)}{\left(\gamma_{B}p_{BB'}+(1-\gamma_{B})\frac{\tau_{BE}}{1+\tau_{BE}}p_{BE}+(1-\gamma_{B})\frac{1}{1+\tau_{BE}}p_{BF}\right)\left(\gamma_{A}q_{AA'}+(1-\gamma_{A})\frac{\tau_{AC}}{1+\tau_{AC}}q_{AC}+(1-\gamma_{A})\frac{1}{1+\tau_{AC}}q_{AD}\right)}\right)^2 \nonumber \\ 
    & \left[\frac{1}{n_Ap_Aq_A}+\frac{1}{n_Bp_Bq_B}\right] = \epsilon_1. \nonumber
\end{eqnarray}

Using the above optimality criterion and the expression of $n_A$, $n_B$, we get the first stage optimal allocation ratio as,
\begin{align}
\tau_A^* = & \left(\frac{1+\tau_{AC}^*}{1+\tau_{BE}^*}\right)^{\frac{3}{2}}\left(\frac{(\gamma_B p_{BB'}(1+\tau_{BE}^*)+(1-\gamma_B) (\tau_{BE}^* p_{BE}+p_{BF}))}{(\gamma_A p_{AA'}(1+\tau_{AC}^*)+(1-\gamma_A) (\tau_{AC}^* p_{AC}+p_{AD}))}\right)^{\frac{1}{2}} \nonumber \\ 
& \hspace{1.75cm} \times \left(\frac{(\gamma_B q_{BB'}(1+\tau_{BE}^*)+(1-\gamma_B) (\tau_{BE}^* q_{BE}+q_{BF}))}{(\gamma_A q_{AA'}(1+\tau_{AC}^*)+(1-\gamma_A) (\tau_{AC}^* q_{AC}+q_{AD}))}\right).
\label{eq : optimal first stage proportion for odds}
\end{align}

Similar to the Section \ref{ap : first stage optimal alloc for diff}, using Lemma 1, and Lemma 2 on equation (\ref{eq : optimal first stage proportion for odds}), we get,
\begin{align}
    \hat{\tau}_{A,n} & \xrightarrow{a.s} \left(\frac{\left(\sqrt{p_{AC}}q_{AC}+\sqrt{p_{AD}}q_{AD}\right)}{\left(\sqrt{p_{BE}}q_{BE}+\sqrt{p_{BF}}q_{BF}\right)}\right)^{\frac{3}{2}} \nonumber \\ & \hspace{1.75cm} \times \left(\frac{(\gamma_B p_{BB'}(\sqrt{p_{BE}}q_{BE}+\sqrt{p_{BF}}q_{BF})+(1-\gamma_B) (\sqrt{p_{BF}}q_{BF}p_{BE}+\sqrt{p_{BE}}q_{BE}p_{BF}))}{(\gamma_A p_{AA'}(\sqrt{p_{AC}}q_{AC}+\sqrt{p_{AD}}q_{AD})+(1-\gamma_A) (\sqrt{p_{AD}}q_{AD}p_{AC}+\sqrt{p_{AC}}q_{AC}p_{AD}))}\right)^{\frac{1}{2}} \nonumber \\ 
& \hspace{1.75cm} \times \left(\frac{(\gamma_B q_{BB'}(\sqrt{p_{BE}}q_{BE}+\sqrt{p_{BF}}q_{BF})+(1-\gamma_B) (\sqrt{p_{BF}}q_{BF}q_{BE}+\sqrt{p_{BE}}q_{BE}q_{BF}))}{(\gamma_A q_{AA'}(\sqrt{p_{AC}}q_{AC}+\sqrt{p_{AD}}q_{AD})+(1-\gamma_A) (\sqrt{p_{AD}}q_{AD}q_{AC}+\sqrt{p_{AC}}q_{AC}q_{AD}))}\right) \nonumber \\
& \hspace{1cm} = \tau_A^*.
\label{eq : asymp conv of first stage optimal alloc for odds}
\end{align}

\subsubsection{\centering Allocation Procedure}
Similar to the procedure of Section \ref{sec : adaptive alloc proc} (as given by (\ref{2nd stage allo process}) for simple difference), the adaptive allocation process for the first stage optimal allocation ratio using odds ratio as the objective function is,
\begin{equation}\label{1st stage allo process for odds}
E_{i-1}(T_{1i}) = \frac{\sqrt{l_{i-1}}}{\sqrt{l_{i-1}}+\sqrt{m_{i-1}}},
\end{equation}
where
\begin{eqnarray}
l_{i-1} = \left(1+\hat{\tau}_{AC,i}\right)^{\frac{3}{2}}\left((\gamma_B \hat{p}_{BB',i-1}(1+\hat{\tau}_{BE,i})+(1-\gamma_B) (\hat{\tau}_{BE,i} \hat{p}_{BE,i-1}+\hat{p}_{BF,i-1}))\right)^{\frac{1}{2}} \nonumber \\ \times \left((\gamma_B \hat{q}_{BB',i-1}(1+\hat{\tau}_{BE,i})+(1-\gamma_B) (\hat{\tau}_{BE,i} \hat{q}_{BE,i-1}+\hat{q}_{BF,i-1}))\right), \nonumber \\
m_{i-1} = \left(1+\hat{\tau}_{BE,i}\right)^{\frac{3}{2}}\left((\gamma_A \hat{p}_{AA',i-1}(1+\hat{\tau}_{AC,i})+(1-\gamma_A) (\hat{\tau}_{AC,i} \hat{p}_{AC,i-1}+\hat{p}_{AD,i-1}))\right)^{\frac{1}{2}} \nonumber \\ \times \left((\gamma_A \hat{q}_{AA',i-1}(1+\hat{\tau}_{AC,i})+(1-\gamma_A) (\hat{\tau}_{AC,i} \hat{q}_{AC,i-1}+\hat{q}_{AD,i-1}))\right). \nonumber
\end{eqnarray}

\subsubsection{\centering Asymptotic Variance}\label{ap :  Variance first stage alloc odds}

Following the same procedure, as mentioned in Section \ref{ap : Variance first stage alloc diff}, and \ref{ap : Variance second stage alloc diff}, using Delta method with the function $h(x,y)$ as $\sqrt{\frac{y}{x}}\frac{1-y}{1-x}$, the variance of the first stage optimal allocation ratio $(\hat{\tau}_{A,n})$ is obtain as,
\begin{eqnarray}
Var(\hat{\tau}_{A,n}) = \frac{1}{4n} \left(\frac{v_{A} \left(1-3p_{A}\right)^2 \left(1-p_{B}\right)^2 p_{B}}{p_{A}^{3} \left(1-p_{A}\right)^4}+\frac{v_{B} \left(1-3p_{B}\right)^2}{p_{A} \left(1-p_{A}\right)^2 p_{B}}\right), \nonumber
\end{eqnarray}
where the expressions of $v_{A},$ and $v_{B}$ are from Section \ref{ap : Variance first stage alloc diff}, and $p_{A}$, and $p_{B}$ are from Section \ref{ap : first stage success probability wrt second stage}. 

Thus, the asymptotic distribution of the estimated first stage optimum allocation ratio is, 
\begin{eqnarray}
\sqrt{n}(\hat \tau_{A,n}-\tau_{A}^*) &\xrightarrow[]{d} & N\left(0,\frac{1}{4} \left(\frac{v_{A} \left(1-3p_{A}\right)^2 \left(1-p_{B}\right)^2 p_{B}}{p_{A}^{3} \left(1-p_{A}\right)^4}+\frac{v_{B} \left(1-3p_{B}\right)^2}{p_{A} \left(1-p_{A}\right)^2 p_{B}}\right)\right).\nonumber
\end{eqnarray}

\subsection{\centering
\textbf{Derivation of Second Stage Optimal Allocation Ratio for Relative Risk}\label{ap : second stage optimal alloc for relative risk}
}
We consider the objective function $g(\cdot, \cdot)$ as introduced in Section \ref{optimum_allo} to be the relative risk. The objective function of relative risk for comparing the two second stage probabilities $(p_{T_1T_2}$, and $p_{T_1T_2^*})$ is given by $\frac{1-p_{T_1T_2^*}}{1-p_{T_1T_2}}$. The optimality criterion (as defined in Section \ref{optimum_allo} of main paper) for the first stage allocation ratio using the asymptotic variance of the objective function $avar(g(\hat p_{T_1T_2},\hat p_{T_1T_2}^*))$ can be expressed as
\begin{equation*}
\frac{1}{(1-p_{T_1T_2})^2}\left(\frac{p_{T_1T_2^*}q_{T_1T_2^*}}{n_{T_1T_2^*}}+\frac{(1-p_{T_1T_2^*})^2p_{T_1T_2}q_{T_1T_2}}{n_{T_1T_2}(1-p_{T_1T_2})^2}\right) = \epsilon_2, \mbox{ for some constant } \epsilon_2 > 0.
\end{equation*} 

Note that, $\tau_{T_1T_2}=\frac{n_{T_1T_2}}{n_{T_1T_2^*}}$. Then $n_{T_1T_2}$, and $n_{T_1T_2^*}$ can be written as,
\begin{equation*}
n_{T_1T_2}=n_{T_1}^{NR} \left(\frac{\tau_{T_1T_2}}{1+\tau_{T_1T_2}}\right),
n_{T_1T_2^*}=\frac{n_{T_1}^{NR}}{1+\tau_{T_1T_2}},
\end{equation*}
where $n_{T_1}^{NR} = n_{T_1T_2}+n_{T_1T_2^*}$, is the total number of patients who obtained treatment $T_1$ at the first stage and become non-responders at the end of the first stage. Substituting the expressions of $n_{T_1T_2}$ and $n_{T_1T_2^*}$ in asymptotic variance expression obtained earlier,

\begin{equation*}
\frac{1}{(1-p_{T_1T_2})^2}\left(\frac{p_{T_1T_2^*}q_{T_1T_2^*}}{\frac{n_{T_1}^{NR}}{1+\tau_{T_1T_2}}}+\frac{(1-p_{T_1T_2^*})^2p_{T_1T_2}q_{T_1T_2}}{n_{T_1}^{NR} \left(\frac{\tau_{T_1T_2}}{1+\tau_{T_1T_2}}\right)(1-p_{T_1T_2})^2}\right) = \epsilon_2.
\end{equation*} 

From Section \ref{optimum_allo},  the second stage optimal allocation ratio is obtained as,
\begin{eqnarray}
    \tau_{T_1T_2}^{*} = \arg \min_{\tau_{T_1T_2}} F_2(\tau_{T_1T_2}) & \mbox{ subject to } \nonumber \\ &  \frac{1}{(1-p_{T_1T_2})^2}\left(\frac{p_{T_1T_2^*}q_{T_1T_2^*}}{\frac{n_{T_1}^{NR}}{1+\tau_{T_1T_2}}}+\frac{(1-p_{T_1T_2^*})^2p_{T_1T_2}q_{T_1T_2}}{n_{T_1}^{NR} \left(\frac{\tau_{T_1T_2}}{1+\tau_{T_1T_2}}\right)(1-p_{T_1T_2})^2}\right) = \epsilon_2. \nonumber
\end{eqnarray} 

Thus using the above optimality criterion, the optimal value of $\tau_{T_1T_2}$ is,
\begin{eqnarray}
    \tau_{T_1T_2}^* = \left(\sqrt{\frac{p_{T_1T_2}}{p_{T_1T_2^*}}}\right)\left(\frac{q_{T_1T_2^*}}{q_{T_1T_2}}\right).
\label{eq : second stage optimal alloc prop for relative risk}
\end{eqnarray}
	
As derived in Section \ref{ap : second stage optimal alloc for diff} using Lemma 1, on equation (\ref{eq : second stage optimal alloc prop for relative risk}) we have,
\begin{eqnarray}
    \hat{\tau}_{T_1T_2,n} \xrightarrow{a.s} \left(\sqrt{\frac{p_{T_1T_2}}{p_{T_1T_2^*}}}\right)\left(\frac{q_{T_1T_2^*}}{q_{T_1T_2}}\right).
\label{eq : asymp conv of second stage optimal alloc for relative risk}
\end{eqnarray}

\subsubsection{\centering Allocation Procedure}
Similar to the procedure of Section \ref{sec : adaptive alloc proc} (as given by (\ref{2nd stage allo process}) for simple difference), the adaptive allocation process for the second stage optimal allocation ratio using relative risk as the objective function is,
\begin{equation}\label{2nd stage allo process for relative risk}
E_{i-1}(I(T_{2i}=t_2|T_{1i}=t_1,R_{T_{1i}}=0)) = \frac{\sqrt{\hat{p}_{t_1t_2,i-1}}\hat{q}_{t_1t_2^*,i-1}}{\sqrt{\hat{p}_{t_1t_2,i-1}}\hat{q}_{t_1t_2^*,i-1}+\sqrt{\hat{p}_{t_1t_2^*,i-1}}\hat{q}_{t_1t_2,i-1}}.
\end{equation}

\subsubsection{\centering Asymptotic Variance}\label{ap :  Variance second stage alloc relative risk}

Following the same procedure, as mentioned in Section \ref{ap : Variance second stage alloc diff}, using Delta method with the function $h(x,y) = \sqrt{\frac{x}{y}}(\frac{1-y}{1-x})$, the variance of the two second stage optimal allocation ratios $\hat{\tau}_{AC,n}$, and $\hat{\tau}_{BE,n}$ are,
\begin{eqnarray}
Var(\hat{\tau}_{AC,n}) = \frac{1}{4n} \left(\frac{v_{AC}^{-1} \left(1+p_{AC}\right)^2 \left(1-p_{AD}\right)^2}{p_{AC} \left(1-p_{AC}\right)^4 p_{AD}}+\frac{v_{AD}^{-1} \left(1+p_{AD}\right)^2 p_{AC}}{\left(1-p_{AC}\right)^2 p_{AD}^{3}}\right), \nonumber
\end{eqnarray}
\begin{eqnarray}
Var(\hat{\tau}_{BE,n}) = \frac{1}{4n} \left(\frac{v_{BE}^{-1} \left(1+p_{BE}\right)^2 \left(1-p_{BF}\right)^2}{p_{BE} \left(1-p_{BE}\right)^4 p_{BF}}+\frac{v_{BF}^{-1} \left(1+p_{BF}\right)^2 p_{BE}}{\left(1-p_{BE}\right)^2 p_{BF}^{3}}\right), \nonumber
\end{eqnarray}
where $v_{AC},v_{AD},v_{BE},$ and $v_{BF}$ are from Section \ref{ap : Variance of success prob for diff}. Thus, the asymptotic distributions of the estimated second stage optimum allocation ratios are, 
\begin{eqnarray}
    \sqrt{n}(\hat \tau_{AC,n}-\tau_{AC}^*) &\xrightarrow[]{d} & N\left(0,\frac{1}{4} \left(\frac{v_{AC}^{-1} \left(1+p_{AC}\right)^2 \left(1-p_{AD}\right)^2}{p_{AC} \left(1-p_{AC}\right)^4 p_{AD}}+\frac{v_{AD}^{-1} \left(1+p_{AD}\right)^2 p_{AC}}{\left(1-p_{AC}\right)^2 p_{AD}^{3}}\right)\right),\nonumber \\ 
    \sqrt{n}(\hat \tau_{BE,n}-\tau_{BE}^*) &\xrightarrow[]{d}& N\left(0,\frac{1}{4} \left(\frac{v_{BE}^{-1} \left(1+p_{BE}\right)^2 \left(1-p_{BF}\right)^2}{p_{BE} \left(1-p_{BE}\right)^4 p_{BF}}+\frac{v_{BF}^{-1} \left(1+p_{BF}\right)^2 p_{BE}}{\left(1-p_{BE}\right)^2 p_{BF}^{3}}\right)\right). \nonumber
\end{eqnarray}

\subsection{\centering
    \textbf{Derivation of First Stage Optimal Allocation Ratio for Relative Risk}\label{ap : first stage optimal alloc for relative risk}
    }
We consider the objective function $g(\cdot, \cdot)$ as introduced in Section \ref{optimum_allo} to be the relative risk. The objective function of relative risk for comparing the two first stage probabilities $(p_A$, and $p_B)$ is given by $\frac{1-p_{B}}{1-p_{A}}$. The optimality criterion (as defined in Section \ref{optimum_allo} of main paper) for the first stage allocation ratio using the asymptotic variance of the objective function $avar(g(\hat p_A,\hat p_B))$ can be expressed as
\begin{equation*}
\frac{1}{q_{A}^2}\left(\frac{q_{B}^{2}p_{A}}{n_{A}q_{A}}+\frac{p_{B}q_{B}}{n_{B}}\right) = \epsilon_1, \mbox{ for some constant } \epsilon_1 > 0.
\end{equation*} 
 
Using the expression for first stage success probability $(p_{T_1})$ and failure probability $(q_{T_1})$ as obtained in \ref{ap : first stage success probability wrt second stage}, in above equation, we get,
\begin{align}
 & \frac{1}{\left(\gamma_{A}q_{AA'}+(1-\gamma_{A})\frac{\tau_{AC}}{1+\tau_{AC}}q_{AC}+(1-\gamma_{A})\frac{1}{1+\tau_{AC}}q_{AD}\right)^2} \times \nonumber \\ &  \left[
     \frac{\left(\gamma_{A}p_{AA'}+(1-\gamma_{A})\frac{\tau_{AC}}{1+\tau_{AC}}p_{AC}+(1-\gamma_{A})\frac{1}{1+\tau_{AC}}p_{AD}\right)\left(\gamma_{B}q_{BB'}+(1-\gamma_{B})\frac{\tau_{BE}}{1+\tau_{BE}}q_{BE}+(1-\gamma_{B})\frac{1}{1+\tau_{BE}}q_{BF}\right)^2}{n_{A}\left(\gamma_{A}q_{AA'}+(1-\gamma_{A})\frac{\tau_{AC}}{1+\tau_{AC}}q_{AC}+(1-\gamma_{A})\frac{1}{1+\tau_{AC}}q_{AD}\right)}   \right. \nonumber \\
     &\left. + \frac{\left(\gamma_{B}p_{BB'}+(1-\gamma_{B})\frac{\tau_{BE}}{1+\tau_{BE}}p_{BE}+(1-\gamma_{B})\frac{1}{1+\tau_{BE}}p_{BF}\right)\left(\gamma_{B}q_{BB'}+(1-\gamma_{B})\frac{\tau_{BE}}{1+\tau_{BE}}q_{BE}+(1-\gamma_{B})\frac{1}{1+\tau_{BE}}q_{BF}\right)}{n_{B}} 
     \right] = \epsilon_1. \nonumber
\end{align}

Since, $\tau_A=\frac{n_A}{n_B}$, $n_A$, and $n_B$ can be written as,
\begin{equation*}
n_A=n \left(\frac{\tau_A}{1+\tau_A}\right),
n_B=\frac{n}{1+\tau_A}.
\end{equation*}

The total number of failures that is obtained after completion of SMART is,
\begin{align*}
F_1(\tau_A,\tau_{AC},\tau_{BE}) & = n_{AA'}q_{AA'}+n_{AC}q_{AC}+n_{AD}q_{AD}+n_{BB'}q_{BB'}+n_{BE}q_{BE}+n_{BF}q_{BF}.
\end{align*}
Now, from Section \ref{optimum_allo}, we have,
\begin{eqnarray}
\tau_{A}^{*} = & \arg \min_{\tau_{A}} F_1(\tau_{A}, \tau_{AC}^{*}, \tau_{BE}^{*}) \mbox{ subject to } \nonumber \\ 
      & \frac{1}{\left(\gamma_{A}q_{AA'}+(1-\gamma_{A})\frac{\tau_{AC}}{1+\tau_{AC}}q_{AC}+(1-\gamma_{A})\frac{1}{1+\tau_{AC}}q_{AD}\right)^2} \times \nonumber \\ &  \left[
     \frac{\left(\gamma_{A}p_{AA'}+(1-\gamma_{A})\frac{\tau_{AC}}{1+\tau_{AC}}p_{AC}+(1-\gamma_{A})\frac{1}{1+\tau_{AC}}p_{AD}\right)\left(\gamma_{B}q_{BB'}+(1-\gamma_{B})\frac{\tau_{BE}}{1+\tau_{BE}}q_{BE}+(1-\gamma_{B})\frac{1}{1+\tau_{BE}}q_{BF}\right)^2}{n_{A}\left(\gamma_{A}q_{AA'}+(1-\gamma_{A})\frac{\tau_{AC}}{1+\tau_{AC}}q_{AC}+(1-\gamma_{A})\frac{1}{1+\tau_{AC}}q_{AD}\right)}   \right. \nonumber \\
     &\left. + \frac{\left(\gamma_{B}p_{BB'}+(1-\gamma_{B})\frac{\tau_{BE}}{1+\tau_{BE}}p_{BE}+(1-\gamma_{B})\frac{1}{1+\tau_{BE}}p_{BF}\right)\left(\gamma_{B}q_{BB'}+(1-\gamma_{B})\frac{\tau_{BE}}{1+\tau_{BE}}q_{BE}+(1-\gamma_{B})\frac{1}{1+\tau_{BE}}q_{BF}\right)}{n_{B}} 
     \right] = \epsilon_1. \nonumber
\end{eqnarray}

Using the above optimality criterion and the expression of $n_A$, $n_B$, we get the first stage optimal allocation ratio as,
\begin{align}
\tau_A^* = & \left(\frac{1+\tau_{AC}^*}{1+\tau_{BE}^*}\right)^{\frac{1}{2}}\left(\frac{(\gamma_A p_{AA'}(1+\tau_{AC}^*)+(1-\gamma_A) (\tau_{AC}^* p_{AC}+p_{AD}))}{(\gamma_B p_{BB'}(1+\tau_{BE}^*)+(1-\gamma_B) (\tau_{BE}^* p_{BE}+p_{BF}))}\right)^{\frac{1}{2}} \nonumber \\ 
& \hspace{1.75cm} \times \left(\frac{(\gamma_B q_{BB'}(1+\tau_{BE}^*)+(1-\gamma_B) (\tau_{BE}^* q_{BE}+q_{BF}))}{(\gamma_A q_{AA'}(1+\tau_{AC}^*)+(1-\gamma_A) (\tau_{AC}^* q_{AC}+q_{AD}))}\right).
\label{eq : optimal first stage proportion for relative risk}
\end{align}

 Similar to the Section \ref{ap : first stage optimal alloc for diff}, using Lemma 1, and Lemma 2 on equation (\ref{eq : optimal first stage proportion for relative risk}), we get,
 \begin{align}
	    \hat{\tau}_{A,n} & \xrightarrow{a.s} \left(\frac{\left(\sqrt{p_{AC}}q_{AD}+\sqrt{p_{AD}}q_{AC}\right)}{\left(\sqrt{p_{BE}}q_{BF}+\sqrt{p_{BF}}q_{BE}\right)}\right)^{\frac{1}{2}} \nonumber \\ & \hspace{1.75cm} \times \left(\frac{(\gamma_A p_{AA'}(\sqrt{p_{AC}}q_{AD}+\sqrt{p_{AD}}q_{AC})+(1-\gamma_A) (\sqrt{p_{AC}}q_{AD}p_{AC}+\sqrt{p_{AD}}q_{AC}p_{AD}))}{(\gamma_B p_{BB'}(\sqrt{p_{BE}}q_{BF}+\sqrt{p_{BF}}q_{BE})+(1-\gamma_B) (\sqrt{p_{BE}}q_{BF}p_{BE}+\sqrt{p_{BF}}q_{BE}p_{BF}))}\right)^{\frac{1}{2}} \nonumber \\ 
    & \hspace{1.75cm} \times \left(\frac{(\gamma_B q_{BB'}(\sqrt{p_{BE}}q_{BF}+\sqrt{p_{BF}}q_{BE})+(1-\gamma_B) (\sqrt{p_{BE}}q_{BF}q_{BE}+\sqrt{p_{BF}}q_{BE}q_{BF}))}{(\gamma_A q_{AA'}(\sqrt{p_{AC}}q_{AD}+\sqrt{p_{AD}}q_{AC})+(1-\gamma_A) (\sqrt{p_{AC}}q_{AD}q_{AC}+\sqrt{p_{AD}}q_{AC}q_{AD}))}\right).
	\label{eq : asymp conv of first stage optimal alloc for relative risk}
\end{align}

 \subsubsection{\centering Allocation Procedure}
Similar to the procedure of Section \ref{sec : adaptive alloc proc} (as given by (\ref{2nd stage allo process}) for simple difference), the adaptive allocation process for the first stage optimal allocation ratio using relative risk as the objective function is,
\begin{equation}\label{1st stage allo process for relative risk}
E_{i-1}(T_{1i}) = \frac{\sqrt{l_{i-1}}}{\sqrt{l_{i-1}}+\sqrt{m_{i-1}}},
\end{equation}
where
\begin{eqnarray}
    l_{i-1} = \left(1+\hat{\tau}_{AC,i}\right)^{\frac{1}{2}}\left((\gamma_A \hat{p}_{AA',i-1}(1+\hat{\tau}_{AC,i})+(1-\gamma_A) (\hat{\tau}_{AC,i} \hat{p}_{AC,i-1}+\hat{p}_{AD,i-1}))\right)^{\frac{1}{2}} \nonumber \\ \times \left((\gamma_B \hat{q}_{BB',i-1}(1+\hat{\tau}_{BE,i})+(1-\gamma_B) (\hat{\tau}_{BE,i} \hat{q}_{BE,i-1}+\hat{q}_{BF,i-1}))\right), \nonumber \\
    m_{i-1} = \left(1+\hat{\tau}_{BE,i}\right)^{\frac{1}{2}}\left((\gamma_B \hat{p}_{BB',i-1}(1+\hat{\tau}_{BE,i})+(1-\gamma_B) (\hat{\tau}_{BE,i} \hat{p}_{BE,i-1}+\hat{p}_{BF,i-1}))\right)^{\frac{1}{2}} \nonumber \\ \times \left((\gamma_A \hat{q}_{AA',i-1}(1+\hat{\tau}_{AC,i})+(1-\gamma_A) (\hat{\tau}_{AC,i} \hat{q}_{AC,i-1}+\hat{q}_{AD,i-1}))\right). \nonumber
\end{eqnarray}

\subsubsection{\centering Asymptotic Variance}\label{ap :  Variance first stage alloc relative risk}

Following the same procedure, as mentioned in Section \ref{ap : Variance first stage alloc diff}, and \ref{ap : Variance second stage alloc diff}, using Delta method with the function $h(x,y)$ as $\sqrt{\frac{x}{y}}\frac{1-y}{1-x}$, the variance of the first stage optimal allocation ratio $(\hat{\tau}_{A,n})$ is obtain as,
\begin{eqnarray}
    Var(\hat{\tau}_{A,n}) = \frac{1}{4n} \left(\frac{v_{A} \left(1+p_{A}\right)^2 \left(1-p_{B}\right)^2}{p_{A} \left(1-p_{A}\right)^4 p_{B}}+\frac{v_{B} \left(1+p_{B}\right)^2 p_{A}}{\left(1-p_{A}\right)^2 p_{B}^{3}}\right), \nonumber
\end{eqnarray}
where the expressions of $v_{A},$ and $v_{B}$ are from Section \ref{ap : Variance first stage alloc diff}, and $p_{A}$, and $p_{B}$ are from Section \ref{ap : first stage success probability wrt second stage}. 

Thus, the asymptotic distribution of the estimated first stage optimum allocation ratio is, 
\begin{eqnarray}
    \sqrt{n}(\hat \tau_{A,n}-\tau_{A}^*) &\xrightarrow[]{d} & N\left(0,\frac{1}{4} \left(\frac{v_{A} \left(1+p_{A}\right)^2 \left(1-p_{B}\right)^2}{p_{A} \left(1-p_{A}\right)^4 p_{B}}+\frac{v_{B} \left(1+p_{B}\right)^2 p_{A}}{\left(1-p_{A}\right)^2 p_{B}^{3}}\right)\right).\nonumber
\end{eqnarray}

\subsection{\centering
    \textbf{Derivation of the success probability for an dynamic treatment regime.} \label{ap : proportion of success of DTRs}
}

Note that (see Section \ref{sec : genframe}),
\begin{equation}
    E(Y|T_1=t_1,T_2=t_2)=p_{t_1t_2}.
\end{equation}

Now, the success probability of the dynamic treatment regime, $d_1$, is obtained as \citep{ghosh2020noninferiority},
\begin{align*}
    p_{d_1} & = E(\overline{Y}_{d_1}) \\
    & = E(W^{d_1}Y), \text{\hspace{1cm} where, } W^{d_1} =  \frac{I(T_{1}=A,T_{2}=A'^{R_A}C^{1-R_{A}})}{\frac{\tau_A}{1+\tau_A} \times \left(\frac{\tau_{AC}}{1+\tau_{AC}}\right)^{1-R_A}} \\
    & = E\left(E\left[\frac{I(T_{1}=A,T_{2}=A'^{R_A}C^{1-R_{A}})Y}{\frac{\tau_A}{1+\tau_A} \times \left(\frac{\tau_{AC}}{1+\tau_{AC}}\right)^{1-R_A}}\;\middle|\;T_{1},T_{2}\right]\right) \\
    & = P(T_{1}=A,T_{2}=A')E\left[\frac{I(T_{1}=A,T_{2}=A')Y}{\frac{\tau_A}{1+\tau_A}}\;\middle|\;T_{1}=A,T_{2}=A'\right]\\ & +P(T_{1}=A,T_{2}=C)E\left[\frac{I(T_{1}=A,T_{2}=C)Y}{\frac{\tau_A}{1+\tau_A} \frac{\tau_{AC}}{1+\tau_{AC}}}\;\middle|\;T_{1}=A,T_{2}=C\right] \\ 
    & = P(T_1=A)P(T_2=A'|T_1=A) E\left[\frac{Y}{\frac{\tau_A}{1+\tau_A}}\;\middle|\;T_{1}=A,T_{2}=A'\right]\\ & +P(T_1=A)P(T_2=C|T_1=A) E\left[\frac{Y}{\frac{\tau_A}{1+\tau_A} \frac{\tau_{AC}}{1+\tau_{AC}}}\;\middle|\;T_{1}=A,T_{2}=C\right] \\ 
    & = \left(\frac{\tau_A}{1+\tau_A}\right)\gamma_A\left(\frac{1}{\frac{\tau_A}{1+\tau_A}}\right)p_{AA'}+\left(\frac{\tau_A}{1+\tau_A}\right)(1-\gamma_A)\left(\frac{\tau_{AC}}{1+\tau_{AC}}\right)\left(\frac{1}{\frac{\tau_A}{1+\tau_A}\frac{\tau_{AC}}{1+\tau_{AC}}}\right)p_{AC} \\
    & = \gamma_Ap_{AA'}+(1-\gamma_A)p_{AC}.
\end{align*}


Similarly, the success probabilities of other dynamic treatment regimes are
$p_{d_2}=\gamma_Ap_{AA'}+(1-\gamma_A)p_{AD};\ p_{d_3}=\gamma_Bp_{BB'}+(1-\gamma_B)p_{BE};\ p_{d_4}=\gamma_Bp_{BB'}+(1-\gamma_B)p_{BF}$.

\clearpage

\subsection{\centering \textbf{Simulation Study }}
In the main paper, we have shown the simulation studies with the objective function as the simple difference with a sample size of 500. Similar simulations for the same objective function with sample sizes 1000 and 2000 are shown in Tables \ref{tab : Tau with other statistics for diff function with 1000} and \ref{tab : Tau with other statistics for diff function with 2000}, respectively.

In Sections \ref{ap : second stage optimal alloc for odds ratio}, \ref{ap : first stage optimal alloc for odds ratio}, \ref{ap : second stage optimal alloc for relative risk}, and \ref{ap : first stage optimal alloc for relative risk}, we have obtained the adaptive optimal allocation ratios and the corresponding adaptive allocation processes with the objective functions as odds ratio and relative risk. We have also established the asymptotic distributions of the developed adaptive optimal allocation ratios for both the objective functions. Following the same simulation structure as defined in Section \ref{sec : simul study}, the estimates of the optimal allocation ratios are obtained. We have considered sample sizes of $500$, $1000$, and $2000$ to estimate the optimal allocation ratios, and the corresponding SSE, ASE, and CP values in Tables \ref{tab : Tau with other statistics for odds ratio function with 500}, \ref{tab : Tau with other statistics for odds ratio function with 1000}, \ref{tab : Tau with other statistics for odds ratio function with 2000} for objective function odds ratio, and \ref{tab : Tau with other statistics for relative risk function with 500}, \ref{tab : Tau with other statistics for relative risk function with 1000}, and \ref{tab : Tau with other statistics for relative risk function with 2000} for relative risk as the objective function. The success probability setup has been kept exactly same as in Table \ref{tab:Tau-difference}, Section \ref{sec : simul study} of the main paper for Tables \ref{tab : Tau with other statistics for odds ratio function with 500}, \ref{tab : Tau with other statistics for odds ratio function with 1000},  \ref{tab : Tau with other statistics for odds ratio function with 2000}, \ref{tab : Tau with other statistics for relative risk function with 500}, \ref{tab : Tau with other statistics for relative risk function with 1000}, and \ref{tab : Tau with other statistics for relative risk function with 2000}.

In Tables \ref{tab : Tau with other statistics for odds ratio function with 500}, \ref{tab : Tau with other statistics for odds ratio function with 1000}, and \ref{tab : Tau with other statistics for odds ratio function with 2000}, we observe when the true values of the optimal allocation ratios ($\tau_{A}^*, \tau_{AC}^*, \tau_{BE}^*$) are close to 0.5 or 1, the estimated optimal allocation ratios ($\hat{\tau}_{A}, \hat{\tau}_{AC}, \hat{\tau}_{BE}$) are close to their respective true values.  However, when the value of $\tau_A^*$, $\tau_{AC}^*$ or $\tau_{BE}^*$ are 2 or more, we observe corresponding estimates $\hat\tau_A$, $\hat\tau_{AC}$ and $\hat\tau_{BE}$ are overestimating the respective optimal allocation ratios. Such high value of the optimal allocation ratios are seen when one or more success probabilities are very low or high (failure probability is low). It can be further observed that when the true values of optimal allocation ratios are close to $0.5$ or $1$, SSE and ASE are close to each other, and the coverage probability (CP) is near to $0.95$. However, when the true values of any of the optimal allocation ratios are more than $2$, we observe the estimated optimal allocation ratios deviate from the corresponding true values.

In Tables \ref{tab : Tau with other statistics for relative risk function with 500}, \ref{tab : Tau with other statistics for relative risk function with 1000}, and \ref{tab : Tau with other statistics for relative risk function with 2000}, we observed that for all the values (rows 1 to 9) of the estimated optimal allocation ratios are close to the corresponding true values of the same. Similarly, in each case, the SSE and the ASE are close, and the coverage probabilities (CP) are near to $0.95$. However, for very large (or small) values for some success probabilities (rows 14 to 16), the estimated optimal allocation ratios deviate from their corresponding true values. In those cases, the values of SSE and ASE are also far apart.

\newpage

\begin{table}[hbt!]
\caption{Estimated first stage ($\hat\tau_A$) and second stage ($\hat\tau_{AC},\hat\tau_{BE}$) allocation ratios along with corresponding SSE, ASE and CP based on 5000 simulations. $\tau_A$, $\tau_{AC}$, and $\tau_{BE}$ denote true values of optimum allocation ratios. Here, $\gamma_A = 0.4$, and $\gamma_B=0.3$, and the sample size is \textbf{1000} using objective function as \textbf{Simple Difference}. It also shows the total expected number of failures at the end of SMART using optimal allocation (proposed method) and equal randomization. \\}
\resizebox{\columnwidth}{!}{%
\renewcommand{\arraystretch}{2}%
\begin{tabular}{|c|c|c|c|c|c|c|}
\hline
No. & $(p_{AA'},p_{AC},p_{AD})$ & \multirow{2}{*}{$\tau_{A}\ (\hat{\tau}_A,\ SSE,\ ASE,\ \widehat{CP})$} & \multirow{2}{*}{$\tau_{AC}\ (\hat{\tau}_{AC},\ SSE,\ ASE,\ \widehat{CP})$} & \multirow{2}{*}{$\tau_{BE}\ (\hat{\tau}_{BE},\ SSE,\ ASE,\ \widehat{CP})$} & \multicolumn{2}{c|}{Expected number of failures} \\ \cline{6-7} 
& $(p_{BB'},p_{BE},p_{BF})$ & & & & Optimal & Equal \\ \hline

1 & (0.20, 0.15, 0.15) & \multirow{2}{*}{0.521 (0.518, 0.033, 0.032, 0.946)} & \multirow{2}{*}{1.000 (1.025, 0.242, 0.218, 0.949)} & \multirow{2}{*}{0.931 (0.931, 0.029, 0.029, 0.953)} & \multirow{2}{*}{532} & \multirow{2}{*}{603} \\ & (0.45, 0.65, 0.75) &  & & & &  \\ \hline
2 & (0.30, 0.80, 0.20) & \multirow{2}{*}{1.002 (1.004, 0.030, 0.030, 0.949)} & \multirow{2}{*}{2.000 (2.073, 0.402, 0.410, 0.955)} & \multirow{2}{*}{1.044 (1.046, 0.049, 0.049, 0.952)} & \multirow{2}{*}{520} & \multirow{2}{*}{553} \\ & (0.25, 0.60, 0.55) &  & & & &  \\ \hline
3 & (0.80, 0.95, 0.85) & \multirow{2}{*}{2.025 (2.039, 0.111, 0.110, 0.954)} & \multirow{2}{*}{1.057 (1.058, 0.018, 0.018, 0.947)} & \multirow{2}{*}{1.000 (1.018, 0.207, 0.191, 0.951)} & \multirow{2}{*}{356} & \multirow{2}{*}{464} \\ & (0.35, 0.15, 0.15) & & & & & \\ \hline
4 & (0.30, 0.20, 0.80) & \multirow{2}{*}{1.109 (1.109, 0.036, 0.036, 0.954)} & \multirow{2}{*}{0.500 (0.492, 0.058, 0.050, 0.926)} & \multirow{2}{*}{0.500 (0.490, 0.069, 0.059, 0.926)} & \multirow{2}{*}{558} & \multirow{2}{*}{622} \\ & (0.25, 0.15, 0.60) & & & & & \\ \hline
5 & (0.30, 0.20, 0.20) & \multirow{2}{*}{0.686 (0.683, 0.033, 0.033, 0.947)} & \multirow{2}{*}{1.000 (1.011, 0.154, 0.146, 0.948)} & \multirow{2}{*}{0.500 (0.493, 0.059, 0.053, 0.934)} & \multirow{2}{*}{596} & \multirow{2}{*}{651} \\ & (0.65, 0.15, 0.60) & & & & & \\ \hline
6 & (0.30, 0.80, 0.20) & \multirow{2}{*}{0.985 (0.987, 0.028, 0.028, 0.949)} & \multirow{2}{*}{2.000 (2.077, 0.416, 0.428, 0.956)} & \multirow{2}{*}{1.000 (1.001, 0.045, 0.044, 0.951)} & \multirow{2}{*}{511} & \multirow{2}{*}{544}\\ & (0.25, 0.60, 0.60) & & & & & \\ \hline
7 & (0.30, 0.80, 0.80) & \multirow{2}{*}{1.085 (1.083, 0.028, 0.028, 0.954)} & \multirow{2}{*}{1.000 (1.000, 0.028, 0.029, 0.955)} & \multirow{2}{*}{0.500 (0.491, 0.067, 0.059, 0.934)} & \multirow{2}{*}{440} & \multirow{2}{*}{471} \\ & (0.65, 0.15, 0.60)  & & & & & \\ \hline
8 & (0.30, 0.80, 0.80) & \multirow{2}{*}{1.414 (1.417, 0.051, 0.051, 0.953)} & \multirow{2}{*}{1.000 (1.001, 0.027, 0.027, 0.958)} & \multirow{2}{*}{1.000 (1.017, 0.176, 0.161, 0.954)} & \multirow{2}{*}{524} & \multirow{2}{*}{549} \\ & (0.65, 0.15, 0.15) & & & & &\\ \hline
9 & (0.30, 0.20, 0.20) & \multirow{2}{*}{0.686 (0.683, 0.033, 0.033, 0.949)} & \multirow{2}{*}{1.000 (1.009, 0.147, 0.138, 0.949)} & \multirow{2}{*}{2.000 (2.061, 0.324, 0.290, 0.952)} & \multirow{2}{*}{596} & \multirow{2}{*}{651} \\ & (0.65, 0.60, 0.15)  & & & & & \\ \hline 
\multicolumn{3}{|c}{Very high/low success probability values} & \multicolumn{4}{c|}{} \\ \hline  
10 & (0.10, 0.10, 0.10) & \multirow{2}{*}{1.000 (1.005, 0.099, 0.098, 0.953)} & \multirow{2}{*}{1.000 (1.020, 0.241, 0.218, 0.952)} & \multirow{2}{*}{1.000 (1.012, 0.210, 0.187, 0.941)} & \multirow{2}{*}{900} & \multirow{2}{*}{900} \\ & (0.10, 0.10, 0.10) & & & & & \\ \hline
11 & (0.05, 0.05, 0.05) & \multirow{2}{*}{1.000 (1.011, 0.149, 0.147, 0.956)} & \multirow{2}{*}{1.000 (1.060, 0.383, 0.391, 0.942)} & \multirow{2}{*}{1.000 (1.050, 0.342, 0.340, 0.944)} & \multirow{2}{*}{950} & \multirow{2}{*}{950} \\ & (0.05, 0.05, 0.05) & & & & &\\ \hline
12 & (0.90, 0.90, 0.90) & \multirow{2}{*}{1.000 (1.000, 0.010, 0.011, 0.953)} & \multirow{2}{*}{1.000 (1.001, 0.019, 0.019, 0.952)} & \multirow{2}{*}{1.000 (1.000, 0.018, 0.012, 0.950)} & \multirow{2}{*}{100} & \multirow{2}{*}{100} \\ & (0.90, 0.90, 0.90) & &  & & &\\ \hline
13 & (0.95, 0.95, 0.95) & \multirow{2}{*}{1.000 (1.000, 0.007, 0.007, 0.948)} & \multirow{2}{*}{1.000 (1.000, 0.013, 0.013, 0.953)} & \multirow{2}{*}{1.000 (1.000, 0.013, 0.012, 0.949)} & \multirow{2}{*}{50} & \multirow{2}{*}{50} \\ & (0.95, 0.95, 0.95) & & & & &\\ \hline
14 & (0.35, 0.95, 0.05) & \multirow{2}{*}{0.943 (0.947, 0.028, 0.029, 0.962)} & \multirow{2}{*}{4.359 (5.670, 2.363, 4.608, 0.937)} & \multirow{2}{*}{3.000 (3.371, 1.174, 1.380, 0.950)} & \multirow{2}{*}{340} & \multirow{2}{*}{508} \\ & (0.65, 0.90, 0.10) & & & & &\\ \hline
15 & (0.45, 0.05, 0.95) & \multirow{2}{*}{1.072 (1.074, 0.033, 0.033, 0.957)} & \multirow{2}{*}{0.229 (0.206, 0.071, 0.073, 0.980)} & \multirow{2}{*}{3.000 (3.430, 1.277, 1.578, 0.949)} & \multirow{2}{*}{190} & \multirow{2}{*}{275} \\ & (0.25, 0.90, 0.10) & & & & &\\ \hline
16 & (0.95, 0.95, 0.05) & \multirow{2}{*}{1.057 (1.059, 0.022, 0.023, 0.960)} & \multirow{2}{*}{4.359 (5.605, 2.308, 4.358, 0.940)} & \multirow{2}{*}{0.333 (0.316, 0.070, 0.058, 0.904)} & \multirow{2}{*}{89} & \multirow{2}{*}{174} \\ & (0.90, 0.10, 0.90) & & & & &\\ \hline
\end{tabular}
\label{tab : Tau with other statistics for diff function with 1000}%
}
	\end{table}

\begin{table}[hbt!]
\caption{Estimated first stage ($\hat\tau_A$) and second stage ($\hat\tau_{AC},\hat\tau_{BE}$) allocation ratios along with corresponding SSE, ASE and CP based on 5000 simulations. $\tau_A$, $\tau_{AC}$, and $\tau_{BE}$ denote true values of optimum allocation ratios. Here, $\gamma_A = 0.4$, and $\gamma_B=0.3$, and the sample size is \textbf{2000} using objective function as \textbf{Simple Difference}. It also shows the total expected number of failures at the end of SMART using optimal allocation (proposed method) and equal randomization. \\}
\resizebox{\columnwidth}{!}{%
\renewcommand{\arraystretch}{2}%
\begin{tabular}{|c|c|c|c|c|c|c|}
\hline
No. & $(p_{AA'},p_{AC},p_{AD})$ & \multirow{2}{*}{$\tau_{A}\ (\hat{\tau}_A,\ SSE,\ ASE,\ \widehat{CP})$} & \multirow{2}{*}{$\tau_{AC}\ (\hat{\tau}_{AC},\ SSE,\ ASE,\ \widehat{CP})$} & \multirow{2}{*}{$\tau_{BE}\ (\hat{\tau}_{BE},\ SSE,\ ASE,\ \widehat{CP})$} & \multicolumn{2}{c|}{Expected number of failures} \\ \cline{6-7} 
& $(p_{BB'},p_{BE},p_{BF})$ & & & & Optimal & Equal \\ \hline


1 & (0.20, 0.15, 0.15) & \multirow{2}{*}{0.521 (0.519, 0.023, 0.023, 0.950)} & \multirow{2}{*}{1.000 (1.009, 0.132, 0.125, 0.948)} & \multirow{2}{*}{0.931 (0.931, 0.020, 0.020, 0.952)} & \multirow{2}{*}{1061} & \multirow{2}{*}{1205} \\ & (0.45, 0.65, 0.75) &  & & & &  \\ \hline
2 & (0.30, 0.80, 0.20) & \multirow{2}{*}{1.002 (1.003, 0.021, 0.021, 0.945)} & \multirow{2}{*}{2.000 (2.022, 0.158, 0.152, 0.952)} & \multirow{2}{*}{1.044 (1.045, 0.034, 0.034, 0.950)} & \multirow{2}{*}{1040} & \multirow{2}{*}{1103} \\ & (0.25, 0.60, 0.55) &  & & & &  \\ \hline
3 & (0.80, 0.95, 0.85) & \multirow{2}{*}{2.025 (2.032, 0.077, 0.076, 0.951)} & \multirow{2}{*}{1.057 (1.058, 0.013, 0.013, 0.947)} & \multirow{2}{*}{1.000 (1.003, 0.119, 0.116, 0.949)} & \multirow{2}{*}{710} & \multirow{2}{*}{929} \\ & (0.35, 0.15, 0.15) & & & & & \\ \hline
4 & (0.30, 0.20, 0.80) & \multirow{2}{*}{1.109 (1.110, 0.025, 0.025, 0.952)} & \multirow{2}{*}{0.500 (0.497, 0.037, 0.035, 0.942)} & \multirow{2}{*}{0.500 (0.497, 0.044, 0.041, 0.940)} & \multirow{2}{*}{1120} & \multirow{2}{*}{1243} \\ & (0.25, 0.15, 0.60) & & & & & \\ \hline
5 & (0.30, 0.20, 0.20) & \multirow{2}{*}{0.686 (0.684, 0.023, 0.023, 0.952)} & \multirow{2}{*}{1.000 (1.003, 0.096, 0.093, 0.947)} & \multirow{2}{*}{0.500 (0.497, 0.039, 0.037, 0.943)} & \multirow{2}{*}{1196} & \multirow{2}{*}{1302} \\ & (0.65, 0.15, 0.60) & & & & & \\ \hline
6 & (0.30, 0.80, 0.20) & \multirow{2}{*}{0.985 (0.985, 0.020, 0.020, 0.948)} & \multirow{2}{*}{2.000 (2.021, 0.161, 0.153, 0.951)} & \multirow{2}{*}{1.000 (1.001, 0.031, 0.031, 0.954)} & \multirow{2}{*}{1023} & \multirow{2}{*}{1086}\\ & (0.25, 0.60, 0.60) & & & & & \\ \hline
7 & (0.30, 0.80, 0.80) & \multirow{2}{*}{1.085 (1.084, 0.020, 0.020, 0.950)} & \multirow{2}{*}{1.000 (1.001, 0.020, 0.020, 0.954)} & \multirow{2}{*}{0.500 (0.497, 0.044, 0.041, 0.937)} & \multirow{2}{*}{883} & \multirow{2}{*}{942} \\ & (0.65, 0.15, 0.60)  & & & & & \\ \hline
8 & (0.30, 0.80, 0.80) & \multirow{2}{*}{1.414 (1.416, 0.036, 0.036, 0.947)} & \multirow{2}{*}{1.000 (1.000, 0.019, 0.019, 0.952)} & \multirow{2}{*}{1.000 (1.005, 0.102, 0.102, 0.956)} & \multirow{2}{*}{1049} & \multirow{2}{*}{1099} \\ & (0.65, 0.15, 0.15) & & & & &\\ \hline
9 & (0.30, 0.20, 0.20) & \multirow{2}{*}{0.686 (0.684, 0.023, 0.023, 0.953)} & \multirow{2}{*}{1.000 (1.003, 0.095, 0.093, 0.949)} & \multirow{2}{*}{2.000 (2.025, 0.163, 0.156, 0.951)} & \multirow{2}{*}{1195} & \multirow{2}{*}{1302} \\ & (0.65, 0.60, 0.15)  & & & & & \\ \hline 
\multicolumn{3}{|c}{Very high/low success probability values} & \multicolumn{4}{c|}{} \\ \hline  
10 & (0.10, 0.10, 0.10) & \multirow{2}{*}{1.000 (1.001, 0.068, 0.068, 0.954)} & \multirow{2}{*}{1.000 (1.009, 0.135, 0.130, 0.947)} & \multirow{2}{*}{1.000 (1.009, 0.126, 0.120, 0.950)} & \multirow{2}{*}{1800} & \multirow{2}{*}{1800} \\ & (0.10, 0.10, 0.10) & & & & & \\ \hline
11 & (0.05, 0.05, 0.05) & \multirow{2}{*}{1.000 (1.005, 0.103, 0.100, 0.950)} & \multirow{2}{*}{1.000 (1.031, 0.225, 0.212, 0.955)} & \multirow{2}{*}{1.000 (1.016, 0.201, 0.189, 0.949)} & \multirow{2}{*}{1900} & \multirow{2}{*}{1900} \\ & (0.05, 0.05, 0.05) & & & & &\\ \hline
12 & (0.90, 0.90, 0.90) & \multirow{2}{*}{1.000 (1.000, 0.007, 0.007, 0.952)} & \multirow{2}{*}{1.000 (1.000, 0.014, 0.014, 0.955)} & \multirow{2}{*}{1.000 (1.000, 0.013, 0.013, 0.951)} & \multirow{2}{*}{200} & \multirow{2}{*}{200} \\ & (0.90, 0.90, 0.90) & &  & & &\\ \hline
13 & (0.95, 0.95, 0.95) & \multirow{2}{*}{1.000 (1.000, 0.005, 0.005, 0.953)} & \multirow{2}{*}{1.000 (1.000, 0.010, 0.009, 0.944)} & \multirow{2}{*}{1.000 (1.000, 0.009, 0.009, 0.953)} & \multirow{2}{*}{100} & \multirow{2}{*}{100} \\ & (0.95, 0.95, 0.95) & & & & &\\ \hline
14 & (0.35, 0.95, 0.05) & \multirow{2}{*}{0.943 (0.947, 0.020, 0.019, 0.952)} & \multirow{2}{*}{4.359 (5.123, 1.716, 2.262, 0.949)} & \multirow{2}{*}{3.000 (3.133, 0.531, 0.464, 0.954)} & \multirow{2}{*}{685} & \multirow{2}{*}{1015} \\ & (0.65, 0.90, 0.10) & & & & &\\ \hline
15 & (0.45, 0.05, 0.95) & \multirow{2}{*}{1.072 (1.074, 0.022, 0.021, 0.951)} & \multirow{2}{*}{0.229 (0.216, 0.053, 0.049, 0.980)} & \multirow{2}{*}{3.000 (3.140, 0.559, 0.504, 0.956)} & \multirow{2}{*}{764} & \multirow{2}{*}{1096} \\ & (0.25, 0.90, 0.10) & & & & &\\ \hline
16 & (0.95, 0.95, 0.05) & \multirow{2}{*}{1.057 (1.059, 0.016, 0.015, 0.949)} & \multirow{2}{*}{4.359 (5.041, 1.640, 2.089, 0.939)} & \multirow{2}{*}{0.333 (0.326, 0.045, 0.039, 0.927)} & \multirow{2}{*}{368} & \multirow{2}{*}{699} \\ & (0.90, 0.10, 0.90) & & & & &\\ \hline
\end{tabular}
\label{tab : Tau with other statistics for diff function with 2000}%
}
	\end{table}

\begin{table}[hbt!]
\caption{Estimated first stage ($\hat\tau_A$) and second stage ($\hat\tau_{AC},\hat\tau_{BE}$) allocation ratios along with corresponding SSE, ASE and CP based on 5000 simulations. $\tau_A$, $\tau_{AC}$, and $\tau_{BE}$ denote true values of optimum allocation ratios. Here, $\gamma_A = 0.4$, and $\gamma_B=0.3$, and the sample size is \textbf{500} using objective function as \textbf{Odds Ratio}. It also shows the total expected number of failures at the end of SMART using optimal allocation (proposed method) and equal randomization. \\}
\resizebox{\columnwidth}{!}{%
\renewcommand{\arraystretch}{2}%
\begin{tabular}{|c|c|c|c|c|c|c|}
\hline
No. & $(p_{AA'},p_{AC},p_{AD})$ & \multirow{2}{*}{$\tau_{A}\ (\hat{\tau}_A,\ SSE,\ ASE,\ \widehat{CP})$} & \multirow{2}{*}{$\tau_{AC}\ (\hat{\tau}_{AC},\ SSE,\ ASE,\ \widehat{CP})$} & \multirow{2}{*}{$\tau_{BE}\ (\hat{\tau}_{BE},\ SSE,\ ASE,\ \widehat{CP})$} & \multicolumn{2}{c|}{Expected number of failures} \\ \cline{6-7} 
& $(p_{BB'},p_{BE},p_{BF})$ & & & & Optimal & Equal \\ \hline


1 & (0.20, 0.15, 0.15) & \multirow{2}{*}{0.864 (0.866, 0.060, 0.060, 0.950)} & \multirow{2}{*}{1.000 (1.013, 0.143, 0.146, 0.971)} & \multirow{2}{*}{0.767 (0.780, 0.140, 0.141, 0.944)} & \multirow{2}{*}{293} & \multirow{2}{*}{302} \\ & (0.45, 0.65, 0.75) &  & & & &  \\ \hline
2 & (0.30, 0.80, 0.20) & \multirow{2}{*}{1.002 (1.003, 0.035, 0.035, 0.954)} & \multirow{2}{*}{2.000 (2.025, 0.409, 0.419, 0.938)} & \multirow{2}{*}{1.082 (1.077, 0.124, 0.125, 0.945} & \multirow{2}{*}{263} & \multirow{2}{*}{277} \\ & (0.25, 0.60, 0.55) &  & & & &  \\ \hline
3 & (0.80, 0.95, 0.85) & \multirow{2}{*}{2.804 (2.785, 0.389, 0.393, 0.932)} & \multirow{2}{*}{2.838 (3.385, 2.062, 2.146, 0.923)} & \multirow{2}{*}{1.000 (1.014, 0.182, 0.191, 0.974)} & \multirow{2}{*}{158} & \multirow{2}{*}{232} \\ & (0.35, 0.15, 0.15) & & & & & \\ \hline
4 & (0.30, 0.20, 0.80) & \multirow{2}{*}{1.057 (1.058, 0.028, 0.028, 0.954)} & \multirow{2}{*}{0.500 (0.512, 0.099, 0.098, 0.942)} & \multirow{2}{*}{0.941 (0.958, 0.123, 0.125, 0.966)} & \multirow{2}{*}{296} & \multirow{2}{*}{311} \\ & (0.25, 0.15, 0.60) & & & & & \\ \hline
5 & (0.30, 0.20, 0.20) & \multirow{2}{*}{0.940 (0.939, 0.037, 0.038, 0.952)} & \multirow{2}{*}{1.000 (1.004, 0.094, 0.095, 0.977)} & \multirow{2}{*}{0.941 (0.954, 0.116, 0.120, 0.963)} & \multirow{2}{*}{322} & \multirow{2}{*}{325} \\ & (0.65, 0.15, 0.60) & & & & & \\ \hline
6 & (0.30, 0.80, 0.20) & \multirow{2}{*}{0.986 (0.988, 0.037, 0.037, 0.955)} & \multirow{2}{*}{2.000 (2.024, 0.411, 0.420, 0.934)} & \multirow{2}{*}{1.000 (1.008, 0.127, 0.127, 0.948)} & \multirow{2}{*}{259} & \multirow{2}{*}{272}\\ & (0.25, 0.60, 0.60) & & & & & \\ \hline
7 & (0.30, 0.80, 0.80) & \multirow{2}{*}{1.129 (1.124, 0.064, 0.063, 0.944)} & \multirow{2}{*}{1.000 (1.048, 0.303, 0.308, 0.941)} & \multirow{2}{*}{0.941 (0.958, 0.127, 0.127, 0.962)} & \multirow{2}{*}{233} & \multirow{2}{*}{235} \\ & (0.65, 0.15, 0.60)  & & & & & \\ \hline
8 & (0.30, 0.80, 0.80) & \multirow{2}{*}{1.237 (1.234, 0.053, 0.054, 0.934)} & \multirow{2}{*}{1.000 (1.035, 0.295, 0.298, 0.936)} & \multirow{2}{*}{1.000 (1.007, 0.133, 0.137, 0.964)} & \multirow{2}{*}{267} & \multirow{2}{*}{274} \\ & (0.65, 0.15, 0.15) & & & & &\\ \hline
9 & (0.30, 0.20, 0.20) & \multirow{2}{*}{0.940 (0.938, 0.037, 0.038, 0.954)} & \multirow{2}{*}{1.000 (1.006, 0.093, 0.095, 0.982)} & \multirow{2}{*}{1.063 (1.064, 0.129, 0.131, 0.955)} & \multirow{2}{*}{322} & \multirow{2}{*}{325} \\ & (0.65, 0.60, 0.15)  & & & & & \\ \hline 
\multicolumn{3}{|c}{Very high/low success probability values} & \multicolumn{4}{c|}{} \\ \hline  
10 & (0.10, 0.10, 0.10) & \multirow{2}{*}{1.000 (1.007, 0.107, 0.109, 0.953)} & \multirow{2}{*}{1.000 (1.022, 0.218, 0.219, 0.956)} & \multirow{2}{*}{1.000 (1.021, 0.199, 0.197, 0.954)} & \multirow{2}{*}{450} & \multirow{2}{*}{450} \\ & (0.10, 0.10, 0.10) & & & & & \\ \hline
11 & (0.05, 0.05, 0.05) & \multirow{2}{*}{1.000 (1.021, 0.190, 0.198, 0.956)} & \multirow{2}{*}{1.000 (1.064, 0.407, 0.433, 0.936)} & \multirow{2}{*}{1.000 (1.058, 0.367, 0.383, 0.933)} & \multirow{2}{*}{475} & \multirow{2}{*}{475} \\ & (0.05, 0.05, 0.05) & & & & &\\ \hline
12 & (0.90, 0.90, 0.90) & \multirow{2}{*}{1.000 (1.041, 0.285, 0.278, 0.931)} & \multirow{2}{*}{1.000 (1.152, 0.718, 0.694, 0.904)} & \multirow{2}{*}{1.000 (1.118, 0.568, 0.558, 0.908)} & \multirow{2}{*}{50} & \multirow{2}{*}{50} \\ & (0.90, 0.90, 0.90) & &  & & &\\ \hline
13 & (0.95, 0.95, 0.95) & \multirow{2}{*}{1.000 (1.107, 0.513, 0.506, 0.913)} & \multirow{2}{*}{1.000 (1.509, 1.892, 2.408, 0.870)} & \multirow{2}{*}{1.000 (1.377, 1.562, 1.876, 0.866)} & \multirow{2}{*}{25} & \multirow{2}{*}{25} \\ & (0.95, 0.95, 0.95) & & & & &\\ \hline
14 & (0.35, 0.95, 0.05) & \multirow{2}{*}{0.855 (0.858, 0.079, 0.086, 0.967)} & \multirow{2}{*}{4.359 (5.216, 3.940, 5.206, 0.914)} & \multirow{2}{*}{3.000 (3.122, 1.041, 1.048, 0.930)} & \multirow{2}{*}{186} & \multirow{2}{*}{254} \\ & (0.65, 0.90, 0.10) & & & & &\\ \hline
15 & (0.45, 0.05, 0.95) & \multirow{2}{*}{1.157 (1.162, 0.096, 0.107, 0.956)} & \multirow{2}{*}{0.229 (0.262, 0.141, 0.173, 0.914)} & \multirow{2}{*}{3.000 (3.143, 1.114, 1.176, 0.927)} & \multirow{2}{*}{207} & \multirow{2}{*}{274} \\ & (0.25, 0.90, 0.10) & & & & &\\ \hline
16 & (0.95, 0.95, 0.05) & \multirow{2}{*}{1.506 (1.549, 0.326, 0.397, 0.938)} & \multirow{2}{*}{4.359 (4.979, 3.182, 3.761, 0.921)} & \multirow{2}{*}{0.333 (0.359, 0.134, 0.139, 0.943)} & \multirow{2}{*}{103} & \multirow{2}{*}{173} \\ & (0.90, 0.10, 0.90) & & & & &\\ \hline
\end{tabular}
\label{tab : Tau with other statistics for odds ratio function with 500}%
}
	\end{table}

\begin{table}[hbt!]
\caption{Estimated first stage ($\hat\tau_A$) and second stage ($\hat\tau_{AC},\hat\tau_{BE}$) allocation ratios along with corresponding SSE, ASE and CP based on 5000 simulations. $\tau_A$, $\tau_{AC}$, and $\tau_{BE}$ denote true values of optimum allocation ratios. Here, $\gamma_A = 0.4$, and $\gamma_B=0.3$, and the sample size is \textbf{1000} using objective function as \textbf{Odds Ratio}. It also shows the total expected number of failures at the end of SMART using optimal allocation (proposed method) and equal randomization. \\}
\resizebox{\columnwidth}{!}{%
\renewcommand{\arraystretch}{2}%
\begin{tabular}{|c|c|c|c|c|c|c|}
\hline
No. & $(p_{AA'},p_{AC},p_{AD})$ & \multirow{2}{*}{$\tau_{A}\ (\hat{\tau}_A,\ SSE,\ ASE,\ \widehat{CP})$} & \multirow{2}{*}{$\tau_{AC}\ (\hat{\tau}_{AC},\ SSE,\ ASE,\ \widehat{CP})$} & \multirow{2}{*}{$\tau_{BE}\ (\hat{\tau}_{BE},\ SSE,\ ASE,\ \widehat{CP})$} & \multicolumn{2}{c|}{Expected number of failures} \\ \cline{6-7} 
& $(p_{BB'},p_{BE},p_{BF})$ & & & & Optimal & Equal \\ \hline


1 & (0.20, 0.15, 0.15) & \multirow{2}{*}{0.864 (0.865, 0.042, 0.042, 0.946)} & \multirow{2}{*}{1.000 (1.004, 0.098, 0.097, 0.957)} & \multirow{2}{*}{0.767 (0.774, 0.099, 0.098, 0.947)} & \multirow{2}{*}{585} & \multirow{2}{*}{603} \\ & (0.45, 0.65, 0.75) &  & & & &  \\ \hline
2 & (0.30, 0.80, 0.20) & \multirow{2}{*}{1.003 (1.002, 0.025, 0.025, 0.954)} & \multirow{2}{*}{2.000 (2.010, 0.274, 0.280, 0.947)} & \multirow{2}{*}{1.082 (1.080, 0.087, 0.086, 0.946} & \multirow{2}{*}{524} & \multirow{2}{*}{553} \\ & (0.25, 0.60, 0.55) &  & & & &  \\ \hline
3 & (0.80, 0.95, 0.85) & \multirow{2}{*}{2.804 (2.795, 0.275, 0.272, 0.937)} & \multirow{2}{*}{2.838 (3.071, 1.095, 1.071, 0.938)} & \multirow{2}{*}{1.000 (1.006, 0.121, 0.123, 0.971)} & \multirow{2}{*}{308} & \multirow{2}{*}{464} \\ & (0.35, 0.15, 0.15) & & & & & \\ \hline
4 & (0.30, 0.20, 0.80) & \multirow{2}{*}{1.057 (1.058, 0.019, 0.019, 0.957)} & \multirow{2}{*}{0.500 (0.505, 0.067, 0.067, 0.952)} & \multirow{2}{*}{0.941 (0.949, 0.083, 0.084, 0.960)} & \multirow{2}{*}{590} & \multirow{2}{*}{622} \\ & (0.25, 0.15, 0.60) & & & & & \\ \hline
5 & (0.30, 0.20, 0.20) & \multirow{2}{*}{0.940 (0.940, 0.027, 0.026, 0.950)} & \multirow{2}{*}{1.000 (1.002, 0.063, 0.062, 0.966)} & \multirow{2}{*}{0.941 (0.948, 0.081, 0.081, 0.957)} & \multirow{2}{*}{645} & \multirow{2}{*}{650} \\ & (0.65, 0.15, 0.60) & & & & & \\ \hline
6 & (0.30, 0.80, 0.20) & \multirow{2}{*}{0.986 (0.987, 0.026, 0.026, 0.952)} & \multirow{2}{*}{2.000 (2.007, 0.278, 0.280, 0.938)} & \multirow{2}{*}{1.000 (1.002, 0.087, 0.088, 0.956)} & \multirow{2}{*}{515} & \multirow{2}{*}{544}\\ & (0.25, 0.60, 0.60) & & & & & \\ \hline
7 & (0.30, 0.80, 0.80) & \multirow{2}{*}{1.129 (1.128, 0.045, 0.045, 0.947)} & \multirow{2}{*}{1.000 (1.019, 0.203, 0.206, 0.943)} & \multirow{2}{*}{0.941 (0.951, 0.087, 0.086, 0.954)} & \multirow{2}{*}{466} & \multirow{2}{*}{471} \\ & (0.65, 0.15, 0.60)  & & & & & \\ \hline
8 & (0.30, 0.80, 0.80) & \multirow{2}{*}{1.237 (1.236, 0.037, 0.038, 0.941)} & \multirow{2}{*}{1.000 (1.017, 0.202, 0.201, 0.942)} & \multirow{2}{*}{1.000 (1.004, 0.091, 0.091, 0.957)} & \multirow{2}{*}{534} & \multirow{2}{*}{549} \\ & (0.65, 0.15, 0.15) & & & & &\\ \hline
9 & (0.30, 0.20, 0.20) & \multirow{2}{*}{0.940 (0.940, 0.026, 0.026, 0.951)} & \multirow{2}{*}{1.000 (1.001, 0.063, 0.062, 0.964)} & \multirow{2}{*}{1.063 (1.062, 0.091, 0.090, 0.948)} & \multirow{2}{*}{645} & \multirow{2}{*}{651} \\ & (0.65, 0.60, 0.15)  & & & & & \\ \hline 
\multicolumn{3}{|c}{Very high/low success probability values} & \multicolumn{4}{c|}{} \\ \hline  
10 & (0.10, 0.10, 0.10) & \multirow{2}{*}{1.000 (1.003, 0.075, 0.076, 0.953)} & \multirow{2}{*}{1.000 (1.008, 0.143, 0.142, 0.955)} & \multirow{2}{*}{1.000 (1.010, 0.132, 0.131, 0.947)} & \multirow{2}{*}{900} & \multirow{2}{*}{900} \\ & (0.10, 0.10, 0.10) & & & & & \\ \hline
11 & (0.05, 0.05, 0.05) & \multirow{2}{*}{1.000 (1.009, 0.128, 0.131, 0.956)} & \multirow{2}{*}{1.000 (1.027, 0.251, 0.253, 0.945)} & \multirow{2}{*}{1.000 (1.025, 0.233, 0.231, 0.945)} & \multirow{2}{*}{950} & \multirow{2}{*}{950} \\ & (0.05, 0.05, 0.05) & & & & &\\ \hline
12 & (0.90, 0.90, 0.90) & \multirow{2}{*}{1.000 (1.020, 0.190, 0.187, 0.943)} & \multirow{2}{*}{1.000 (1.060, 0.384, 0.376, 0.926)} & \multirow{2}{*}{1.000 (1.048, 0.346, 0.339, 0.922)} & \multirow{2}{*}{100} & \multirow{2}{*}{100} \\ & (0.90, 0.90, 0.90) & &  & & &\\ \hline
13 & (0.95, 0.95, 0.95) & \multirow{2}{*}{1.000 (1.050, 0.321, 0.304, 0.927)} & \multirow{2}{*}{1.000 (1.173, 0.904, 0.901, 0.893)} & \multirow{2}{*}{1.000 (1.143, 0.784, 0.731, 0.906)} & \multirow{2}{*}{50} & \multirow{2}{*}{50} \\ & (0.95, 0.95, 0.95) & & & & &\\ \hline
14 & (0.35, 0.95, 0.05) & \multirow{2}{*}{0.855 (0.857, 0.056, 0.058, 0.960)} & \multirow{2}{*}{4.359 (4.724, 2.135, 2.285, 0.927)} & \multirow{2}{*}{3.000 (3.058, 0.669, 0.669, 0.938)} & \multirow{2}{*}{362} & \multirow{2}{*}{508} \\ & (0.65, 0.90, 0.10) & & & & &\\ \hline
15 & (0.45, 0.05, 0.95) & \multirow{2}{*}{1.157 (1.161, 0.071, 0.072, 0.952)} & \multirow{2}{*}{0.229 (0.248, 0.097, 0.106, 0.927)} & \multirow{2}{*}{3.000 (3.074, 0.728, 0.734, 0.941)} & \multirow{2}{*}{403} & \multirow{2}{*}{548} \\ & (0.25, 0.90, 0.10) & & & & &\\ \hline
16 & (0.95, 0.95, 0.05) & \multirow{2}{*}{1.506 (1.527, 0.233, 0.246, 0.938)} & \multirow{2}{*}{4.359 (4.639, 1.788, 1.817, 0.936)} & \multirow{2}{*}{0.333 (0.345, 0.083, 0.088, 0.944)} & \multirow{2}{*}{198} & \multirow{2}{*}{349} \\ & (0.90, 0.10, 0.90) & & & & &\\ \hline
\end{tabular}
\label{tab : Tau with other statistics for odds ratio function with 1000}%
}
	\end{table}

\begin{table}[hbt!]
\caption{Estimated first stage ($\hat\tau_A$) and second stage ($\hat\tau_{AC},\hat\tau_{BE}$) allocation ratios along with corresponding SSE, ASE and CP based on 5000 simulations. $\tau_A$, $\tau_{AC}$, and $\tau_{BE}$ denote true values of optimum allocation ratios. Here, $\gamma_A = 0.4$, and $\gamma_B=0.3$, and the sample size is \textbf{2000} using objective function as \textbf{Odds Ratio}. It also shows the total expected number of failures at the end of SMART using optimal allocation (proposed method) and equal randomization. \\}
\resizebox{\columnwidth}{!}{%
\renewcommand{\arraystretch}{2}%
\begin{tabular}{|c|c|c|c|c|c|c|}
\hline
No. & $(p_{AA'},p_{AC},p_{AD})$ & \multirow{2}{*}{$\tau_{A}\ (\hat{\tau}_A,\ SSE,\ ASE,\ \widehat{CP})$} & \multirow{2}{*}{$\tau_{AC}\ (\hat{\tau}_{AC},\ SSE,\ ASE,\ \widehat{CP})$} & \multirow{2}{*}{$\tau_{BE}\ (\hat{\tau}_{BE},\ SSE,\ ASE,\ \widehat{CP})$} & \multicolumn{2}{c|}{Expected number of failures} \\ \cline{6-7} 
& $(p_{BB'},p_{BE},p_{BF})$ & & & & Optimal & Equal \\ \hline


1 & (0.20, 0.15, 0.15) & \multirow{2}{*}{0.864 (0.865, 0.030, 0.030, 0.953)} & \multirow{2}{*}{1.000 (1.003, 0.067, 0.067, 0.955)} & \multirow{2}{*}{0.767 (0.770, 0.069, 0.069, 0.947)} & \multirow{2}{*}{1169} & \multirow{2}{*}{1205} \\ & (0.45, 0.65, 0.75) &  & & & &  \\ \hline
2 & (0.30, 0.80, 0.20) & \multirow{2}{*}{1.002 (1.002, 0.018, 0.017, 0.949)} & \multirow{2}{*}{2.000 (2.002, 0.196, 0.193, 0.939)} & \multirow{2}{*}{1.082 (1.080, 0.061, 0.061, 0.948} & \multirow{2}{*}{1045} & \multirow{2}{*}{1103} \\ & (0.25, 0.60, 0.55) &  & & & &  \\ \hline
3 & (0.80, 0.95, 0.85) & \multirow{2}{*}{2.804 (2.801, 0.192, 0.190, 0.942)} & \multirow{2}{*}{2.838 (2.928, 0.693, 0.677, 0.938)} & \multirow{2}{*}{1.000 (1.003, 0.083, 0.084, 0.960)} & \multirow{2}{*}{610} & \multirow{2}{*}{929} \\ & (0.35, 0.15, 0.15) & & & & & \\ \hline
4 & (0.30, 0.20, 0.80) & \multirow{2}{*}{1.057 (1.057, 0.013, 0.013, 0.949)} & \multirow{2}{*}{0.500 (0.504, 0.047, 0.047, 0.949)} & \multirow{2}{*}{0.941 (0.947, 0.059, 0.059, 0.951)} & \multirow{2}{*}{1178} & \multirow{2}{*}{1243} \\ & (0.25, 0.15, 0.60) & & & & & \\ \hline
5 & (0.30, 0.20, 0.20) & \multirow{2}{*}{0.940 (0.940, 0.019, 0.019, 0.951)} & \multirow{2}{*}{1.000 (1.001, 0.043, 0.043, 0.958)} & \multirow{2}{*}{0.941 (0.945, 0.056, 0.057, 0.951)} & \multirow{2}{*}{1291} & \multirow{2}{*}{1302} \\ & (0.65, 0.15, 0.60) & & & & & \\ \hline
6 & (0.30, 0.80, 0.20) & \multirow{2}{*}{0.986 (0.986, 0.018, 0.018, 0.947)} & \multirow{2}{*}{2.000 (2.006, 0.192, 0.194, 0.946)} & \multirow{2}{*}{1.000 (1.004, 0.063, 0.062, 0.945)} & \multirow{2}{*}{1028} & \multirow{2}{*}{1086}\\ & (0.25, 0.60, 0.60) & & & & & \\ \hline
7 & (0.30, 0.80, 0.80) & \multirow{2}{*}{1.129 (1.128, 0.031, 0.031, 0.950)} & \multirow{2}{*}{1.000 (1.014, 0.140, 0.143, 0.951)} & \multirow{2}{*}{0.941 (0.946, 0.060, 0.060, 0.949)} & \multirow{2}{*}{931} & \multirow{2}{*}{942} \\ & (0.65, 0.15, 0.60)  & & & & & \\ \hline
8 & (0.30, 0.80, 0.80) & \multirow{2}{*}{1.237 (1.237, 0.026, 0.027, 0.953)} & \multirow{2}{*}{1.000 (1.009, 0.135, 0.139, 0.956)} & \multirow{2}{*}{1.000 (1.001, 0.063, 0.063, 0.955)} & \multirow{2}{*}{1068} & \multirow{2}{*}{1099} \\ & (0.65, 0.15, 0.15) & & & & &\\ \hline
9 & (0.30, 0.20, 0.20) & \multirow{2}{*}{0.940 (0.940, 0.019, 0.019, 0.951)} & \multirow{2}{*}{1.000 (1.001, 0.043, 0.043, 0.955)} & \multirow{2}{*}{1.063 (1.062, 0.063, 0.063, 0.953)} & \multirow{2}{*}{1291} & \multirow{2}{*}{1302} \\ & (0.65, 0.60, 0.15)  & & & & & \\ \hline 
\multicolumn{3}{|c}{Very high/low success probability values} & \multicolumn{4}{c|}{} \\ \hline  
10 & (0.10, 0.10, 0.10) & \multirow{2}{*}{1.000 (1.002, 0.052, 0.053, 0.951)} & \multirow{2}{*}{1.000 (1.006, 0.098, 0.098, 0.953)} & \multirow{2}{*}{1.000 (1.004, 0.091, 0.090, 0.948)} & \multirow{2}{*}{1800} & \multirow{2}{*}{1800} \\ & (0.10, 0.10, 0.10) & & & & & \\ \hline
11 & (0.05, 0.05, 0.05) & \multirow{2}{*}{1.000 (1.005, 0.088, 0.090, 0.954)} & \multirow{2}{*}{1.000 (1.012, 0.168, 0.168, 0.949)} & \multirow{2}{*}{1.000 (1.012, 0.155, 0.154, 0.948)} & \multirow{2}{*}{1900} & \multirow{2}{*}{1900} \\ & (0.05, 0.05, 0.05) & & & & &\\ \hline
12 & (0.90, 0.90, 0.90) & \multirow{2}{*}{1.000 (1.008, 0.127, 0.129, 0.952)} & \multirow{2}{*}{1.000 (1.026, 0.248, 0.246, 0.942)} & \multirow{2}{*}{1.000 (1.028, 0.229, 0.227, 0.937)} & \multirow{2}{*}{200} & \multirow{2}{*}{200} \\ & (0.90, 0.90, 0.90) & &  & & &\\ \hline
13 & (0.95, 0.95, 0.95) & \multirow{2}{*}{1.000 (1.021, 0.321, 0.207, 0.936)} & \multirow{2}{*}{1.000 (1.077, 0.427, 0.409, 0.923)} & \multirow{2}{*}{1.000 (1.068, 0.392, 0.370, 0.924)} & \multirow{2}{*}{100} & \multirow{2}{*}{100} \\ & (0.95, 0.95, 0.95) & & & & &\\ \hline
14 & (0.35, 0.95, 0.05) & \multirow{2}{*}{0.855 (0.856, 0.040, 0.040, 0.955)} & \multirow{2}{*}{4.359 (4.499, 1.303, 1.346, 0.936)} & \multirow{2}{*}{3.000 (3.027, 0.458, 0.455, 0.942)} & \multirow{2}{*}{715} & \multirow{2}{*}{1015} \\ & (0.65, 0.90, 0.10) & & & & &\\ \hline
15 & (0.45, 0.05, 0.95) & \multirow{2}{*}{1.157 (1.158, 0.050, 0.050, 0.947)} & \multirow{2}{*}{0.229 (0.237, 0.064, 0.066, 0.938)} & \multirow{2}{*}{3.000 (3.036, 0.491, 0.495, 0.950)} & \multirow{2}{*}{795} & \multirow{2}{*}{1096} \\ & (0.25, 0.90, 0.10) & & & & &\\ \hline
16 & (0.95, 0.95, 0.05) & \multirow{2}{*}{1.506 (1.514, 0.159, 0.162, 0.945)} & \multirow{2}{*}{4.359 (4.484, 1.115, 1.149, 0.949)} & \multirow{2}{*}{0.333 (0.339, 0.058, 0.059, 0.949)} & \multirow{2}{*}{387} & \multirow{2}{*}{699} \\ & (0.90, 0.10, 0.90) & & & & &\\ \hline
\end{tabular}
\label{tab : Tau with other statistics for odds ratio function with 2000}%
}
	\end{table}

\begin{table}[hbt!]
\caption{Estimated first stage ($\hat\tau_A$) and second stage ($\hat\tau_{AC},\hat\tau_{BE}$) allocation ratios along with corresponding SSE, ASE and CP based on 5000 simulations. $\tau_A$, $\tau_{AC}$, and $\tau_{BE}$ denote true values of optimum allocation ratios. Here, $\gamma_A = 0.4$, and $\gamma_B=0.3$, and the sample size is \textbf{500} using objective function as \textbf{Relative Risk}. It also shows the total expected number of failures at the end of SMART using optimal allocation (proposed method) and equal randomization. \\}
\resizebox{\columnwidth}{!}{%
\renewcommand{\arraystretch}{2}%
\begin{tabular}{|c|c|c|c|c|c|c|}
\hline
No. & $(p_{AA'},p_{AC},p_{AD})$ & \multirow{2}{*}{$\tau_{A}\ (\hat{\tau}_A,\ SSE,\ ASE,\ \widehat{CP})$} & \multirow{2}{*}{$\tau_{AC}\ (\hat{\tau}_{AC},\ SSE,\ ASE,\ \widehat{CP})$} & \multirow{2}{*}{$\tau_{BE}\ (\hat{\tau}_{BE},\ SSE,\ ASE,\ \widehat{CP})$} & \multicolumn{2}{c|}{Expected number of failures} \\ \cline{6-7} 
& $(p_{BB'},p_{BE},p_{BF})$ & & & & Optimal & Equal \\ \hline


1 & (0.20, 0.15, 0.15) & \multirow{2}{*}{0.235 (0.233, 0.042, 0.043, 0.954)} & \multirow{2}{*}{1.000 (1.366, 1.264, 2.672, 0.897)} & \multirow{2}{*}{0.665 (0.675, 0.160, 0.155, 0.931)} & \multirow{2}{*}{233} & \multirow{2}{*}{302} \\ & (0.45, 0.65, 0.75) &  & & & &  \\ \hline
2 & (0.30, 0.80, 0.20) & \multirow{2}{*}{1.006 (1.089, 0.170, 0.248, 0.961)} & \multirow{2}{*}{8.000 (16.450, 15.869, 58.974, 0.953)} & \multirow{2}{*}{1.175 (1.234, 0.518, 0.774, 0.926} & \multirow{2}{*}{241} & \multirow{2}{*}{277} \\ & (0.25, 0.60, 0.55) &  & & & &  \\ \hline
3 & (0.80, 0.40, 0.60) & \multirow{2}{*}{1.512 (1.539, 0.214, 0.214, 0.948)} & \multirow{2}{*}{0.544 (0.545, 0.144, 0.132, 0.918)} & \multirow{2}{*}{1.889 (2.191, 1.926, 4.281, 0.936)} & \multirow{2}{*}{207} & \multirow{2}{*}{223} \\ & (0.35, 0.65, 0.45) & & & & & \\ \hline
4 & (0.30, 0.20, 0.80) & \multirow{2}{*}{1.301 (1.340, 0.209, 0.301, 0.972)} & \multirow{2}{*}{0.125 (0.108, 0.058, 0.053, 0.934)} & \multirow{2}{*}{0.235 (0.209, 0.118, 0.124, 0.902)} & \multirow{2}{*}{296} & \multirow{2}{*}{311} \\ & (0.25, 0.15, 0.60) & & & & & \\ \hline
5 & (0.30, 0.20, 0.20) & \multirow{2}{*}{0.442 (0.427, 0.063, 0.064, 0.932)} & \multirow{2}{*}{1.000 (1.195, 0.962, 1.598, 0.898)} & \multirow{2}{*}{0.235 (0.217, 0.080, 0.065, 0.862)} & \multirow{2}{*}{322} & \multirow{2}{*}{325} \\ & (0.65, 0.15, 0.60) & & & & & \\ \hline
6 & (0.30, 0.60, 0.45) & \multirow{2}{*}{0.846 (0.850, 0.112, 0.108, 0.937)} & \multirow{2}{*}{1.588 (1.769, 1.191, 2.355, 0.939)} & \multirow{2}{*}{1.000 (1.034, 0.279, 0.256, 0.923)} & \multirow{2}{*}{265} & \multirow{2}{*}{269}\\ & (0.25, 0.60, 0.60) & & & & & \\ \hline
7 & (0.30, 0.80, 0.80) & \multirow{2}{*}{1.329 (1.280, 0.178, 0.193, 0.939)} & \multirow{2}{*}{1.000 (1.072, 0.791, 1.999, 0.916)} & \multirow{2}{*}{0.235 (0.208, 0.097, 0.084, 0.895)} & \multirow{2}{*}{209} & \multirow{2}{*}{235} \\ & (0.65, 0.15, 0.60)  & & & & & \\ \hline
8 & (0.30, 0.80, 0.80) & \multirow{2}{*}{2.475 (2.499, 0.325, 0.326, 0.949)} & \multirow{2}{*}{1.000 (1.055, 0.358, 0.347, 0.927)} & \multirow{2}{*}{1.000 (1.176, 0.825, 1.223, 0.907)} & \multirow{2}{*}{245} & \multirow{2}{*}{274} \\ & (0.65, 0.15, 0.15) & & & & &\\ \hline
9 & (0.30, 0.20, 0.20) & \multirow{2}{*}{0.414 (0.413, 0.061, 0.060, 0.944)} & \multirow{2}{*}{1.000 (1.206, 0.986, 1.649, 0.905)} & \multirow{2}{*}{1.563 (1.636, 0.542, 0.702, 0.937)} & \multirow{2}{*}{268} & \multirow{2}{*}{325} \\ & (0.65, 0.55, 0.40)  & & & & & \\ \hline 
\multicolumn{3}{|c}{Very high/low success probability values} & \multicolumn{4}{c|}{} \\ \hline  
10 & (0.10, 0.10, 0.10) & \multirow{2}{*}{1.000 (1.017, 0.185, 0.178, 0.948)} & \multirow{2}{*}{1.000 (1.114, 0.594, 0.707, 0.920)} & \multirow{2}{*}{1.000 (1.089, 0.533, 0.579, 0.913)} & \multirow{2}{*}{450} & \multirow{2}{*}{450} \\ & (0.10, 0.10, 0.10) & & & & & \\ \hline
11 & (0.05, 0.05, 0.05) & \multirow{2}{*}{1.000 (1.026, 0.249, 0.250, 0.952)} & \multirow{2}{*}{1.000 (1.146, 0.601, 0.816, 0.951)} & \multirow{2}{*}{1.000 (1.110, 0.531, 0.674, 0.944)} & \multirow{2}{*}{475} & \multirow{2}{*}{475} \\ & (0.05, 0.05, 0.05) & & & & &\\ \hline
12 & (0.90, 0.90, 0.90) & \multirow{2}{*}{1.000 (1.065, 0.355, 0.328, 0.930)} & \multirow{2}{*}{1.000 (1.196, 0.917, 0.980, 0.892)} & \multirow{2}{*}{1.000 (1.200, 1.418, 4.662, 0.894)} & \multirow{2}{*}{50} & \multirow{2}{*}{50} \\ & (0.90, 0.90, 0.90) & &  & & &\\ \hline
13 & (0.95, 0.95, 0.95) & \multirow{2}{*}{1.000 (1.135, 0.602, 0.590, 0.905)} & \multirow{2}{*}{1.000 (1.642, 2.898, 4.726, 0.855)} & \multirow{2}{*}{1.000 (1.585, 2.842, 4.491, 0.858)} & \multirow{2}{*}{25} & \multirow{2}{*}{25} \\ & (0.95, 0.95, 0.95) & & & & &\\ \hline
14 & (0.35, 0.95, 0.05) & \multirow{2}{*}{0.760 (0.735, 0.186, 0.931, 0.997)} & \multirow{2}{*}{82.819 (200.255, 157.539, 2617.842, 0.949)} & \multirow{2}{*}{27.000 (63.353, 42.085, 269.637, 0.953)} & \multirow{2}{*}{125} & \multirow{2}{*}{254} \\ & (0.65, 0.90, 0.10) & & & & &\\ \hline
15 & (0.45, 0.05, 0.95) & \multirow{2}{*}{1.329 (1.334, 0.300, 1.539, 0.996)} & \multirow{2}{*}{0.010 (0.012, 0.010, 0.027, 0.999)} & \multirow{2}{*}{27.000 (65.787, 45.008, 339.797, 0.956)} & \multirow{2}{*}{148} & \multirow{2}{*}{274} \\ & (0.25, 0.90, 0.10) & & & & &\\ \hline
16 & (0.95, 0.95, 0.05) & \multirow{2}{*}{1.683 (1.875, 0.724, 5.531, 0.993)} & \multirow{2}{*}{82.819 (179.597, 113.899, 1462.250, 0.951)} & \multirow{2}{*}{0.037 (0.029, 0.026, 0.044, 0.999)} & \multirow{2}{*}{45} & \multirow{2}{*}{173} \\ & (0.90, 0.10, 0.90) & & & & &\\ \hline
\end{tabular}
\label{tab : Tau with other statistics for relative risk function with 500}%
}
	\end{table}

\begin{table}[hbt!]
\caption{Estimated first stage ($\hat\tau_A$) and second stage ($\hat\tau_{AC},\hat\tau_{BE}$) allocation ratios along with corresponding SSE, ASE and CP based on 5000 simulations. $\tau_A$, $\tau_{AC}$, and $\tau_{BE}$ denote true values of optimum allocation ratios. Here, $\gamma_A = 0.4$, and $\gamma_B=0.3$, and the sample size is \textbf{1000} using objective function as \textbf{Relative Risk}. It also shows the total expected number of failures at the end of SMART using optimal allocation (proposed method) and equal randomization. \\}
\resizebox{\columnwidth}{!}{%
\renewcommand{\arraystretch}{2}%
\begin{tabular}{|c|c|c|c|c|c|c|}
\hline
No. & $(p_{AA'},p_{AC},p_{AD})$ & \multirow{2}{*}{$\tau_{A}\ (\hat{\tau}_A,\ SSE,\ ASE,\ \widehat{CP})$} & \multirow{2}{*}{$\tau_{AC}\ (\hat{\tau}_{AC},\ SSE,\ ASE,\ \widehat{CP})$} & \multirow{2}{*}{$\tau_{BE}\ (\hat{\tau}_{BE},\ SSE,\ ASE,\ \widehat{CP})$} & \multicolumn{2}{c|}{Expected number of failures} \\ \cline{6-7} 
& $(p_{BB'},p_{BE},p_{BF})$ & & & & Optimal & Equal \\ \hline


1 & (0.20, 0.15, 0.15) & \multirow{2}{*}{0.235 (0.232, 0.030, 0.030, 0.943)} & \multirow{2}{*}{1.000 (1.139, 0.725, 1.007, 0.912)} & \multirow{2}{*}{0.665 (0.668, 0.109, 0.107, 0.940)} & \multirow{2}{*}{460} & \multirow{2}{*}{603} \\ & (0.45, 0.65, 0.75) &  & & & &  \\ \hline
2 & (0.30, 0.80, 0.20) & \multirow{2}{*}{1.006 (1.050, 0.116, 0.135, 0.960)} & \multirow{2}{*}{8.000 (12.591, 11.660, 28.489, 0.951)} & \multirow{2}{*}{1.175 (1.199, 0.218, 0.212, 0.943} & \multirow{2}{*}{479} & \multirow{2}{*}{553} \\ & (0.25, 0.60, 0.55) &  & & & &  \\ \hline
3 & (0.80, 0.40, 0.60) & \multirow{2}{*}{1.512 (1.527, 0.151, 0.147, 0.943)} & \multirow{2}{*}{0.544 (0.546, 0.095, 0.092, 0.933)} & \multirow{2}{*}{1.889 (1.969, 0.557, 0.660, 0.945)} & \multirow{2}{*}{413} & \multirow{2}{*}{445} \\ & (0.35, 0.15, 0.15) & & & & & \\ \hline
4 & (0.30, 0.20, 0.80) & \multirow{2}{*}{1.301 (1.326, 0.148, 0.163, 0.961)} & \multirow{2}{*}{0.125 (0.114, 0.044, 0.036, 0.860)} & \multirow{2}{*}{0.235 (0.220, 0.070, 0.057, 0.882)} & \multirow{2}{*}{492} & \multirow{2}{*}{622} \\ & (0.25, 0.15, 0.60) & & & & & \\ \hline
5 & (0.30, 0.20, 0.20) & \multirow{2}{*}{0.442 (0.435, 0.044, 0.043, 0.937)} & \multirow{2}{*}{1.000 (1.051, 0.391, 0.407, 0.931)} & \multirow{2}{*}{0.235 (0.228, 0.052, 0.044, 0.909)} & \multirow{2}{*}{537} & \multirow{2}{*}{651} \\ & (0.65, 0.15, 0.60) & & & & & \\ \hline
6 & (0.30, 0.60, 0.45) & \multirow{2}{*}{0.846 (0.847, 0.077, 0.076, 0.944)} & \multirow{2}{*}{1.588 (1.635, 0.339, 0.323, 0.946)} & \multirow{2}{*}{1.000 (1.016, 0.178, 0.174, 0.941)} & \multirow{2}{*}{529} & \multirow{2}{*}{536}\\ & (0.25, 0.60, 0.60) & & & & & \\ \hline
7 & (0.30, 0.80, 0.80) & \multirow{2}{*}{1.329 (1.305, 0.126, 0.123, 0.936)} & \multirow{2}{*}{1.000 (1.035, 0.272, 0.263, 0.934)} & \multirow{2}{*}{0.235 (0.220, 0.068, 0.057, 0.887)} & \multirow{2}{*}{418} & \multirow{2}{*}{471} \\ & (0.65, 0.15, 0.60)  & & & & & \\ \hline
8 & (0.30, 0.80, 0.80) & \multirow{2}{*}{2.475 (2.494, 0.228, 0.226, 0.950)} & \multirow{2}{*}{1.000 (1.028, 0.234, 0.231, 0.939)} & \multirow{2}{*}{1.000 (1.146, 0.343, 0.319, 0.932)} & \multirow{2}{*}{489} & \multirow{2}{*}{549} \\ & (0.65, 0.15, 0.15) & & & & &\\ \hline
9 & (0.30, 0.20, 0.20) & \multirow{2}{*}{0.414 (0.411, 0.042, 0.042, 0.945)} & \multirow{2}{*}{1.000 (1.044, 0.353, 0.355, 0.935)} & \multirow{2}{*}{1.563 (1.586, 0.230, 0.220, 0.941)} & \multirow{2}{*}{549} & \multirow{2}{*}{616} \\ & (0.65, 0.55, 0.40)  & & & & & \\ \hline 
\multicolumn{3}{|c}{Very high/low success probability values} & \multicolumn{4}{c|}{} \\ \hline  
10 & (0.10, 0.10, 0.10) & \multirow{2}{*}{1.000 (1.007, 0.122, 0.119, 0.944)} & \multirow{2}{*}{1.000 (1.037, 0.295, 0.267, 0.938)} & \multirow{2}{*}{1.000 (1.030, 0.266, 0.240, 0.941)} & \multirow{2}{*}{900} & \multirow{2}{*}{900} \\ & (0.10, 0.10, 0.10) & & & & & \\ \hline
11 & (0.05, 0.05, 0.05) & \multirow{2}{*}{1.000 (1.017, 0.167, 0.163, 0.956)} & \multirow{2}{*}{1.000 (1.062, 0.398, 0.413, 0.938)} & \multirow{2}{*}{1.000 (1.062, 0.383, 0.382, 0.936)} & \multirow{2}{*}{950} & \multirow{2}{*}{950} \\ & (0.05, 0.05, 0.05) & & & & &\\ \hline
12 & (0.90, 0.90, 0.90) & \multirow{2}{*}{1.000 (1.023, 0.221, 0.212, 0.935)} & \multirow{2}{*}{1.000 (1.078, 0.480, 0.446, 0.914)} & \multirow{2}{*}{1.000 (1.064, 0.422, 0.395, 0.921)} & \multirow{2}{*}{100} & \multirow{2}{*}{100} \\ & (0.90, 0.90, 0.90) & &  & & &\\ \hline
13 & (0.95, 0.95, 0.95) & \multirow{2}{*}{1.000 (1.061, 0.379, 0.334, 0.920)} & \multirow{2}{*}{1.000 (1.256, 2.392, 3.991, 0.887)} & \multirow{2}{*}{1.000 (1.188, 1.180, 1.168, 0.899)} & \multirow{2}{*}{50} & \multirow{2}{*}{50} \\ & (0.95, 0.95, 0.95) & & & & &\\ \hline
14 & (0.35, 0.95, 0.05) & \multirow{2}{*}{0.760 (0.755, 0.166, 0.599, 0.995)} & \multirow{2}{*}{82.819 (172.191, 89.869, 976.680, 0.956)} & \multirow{2}{*}{27.000 (54.425, 36.587, 153.248, 0.955)} & \multirow{2}{*}{242} & \multirow{2}{*}{508} \\ & (0.65, 0.90, 0.10) & & & & &\\ \hline
15 & (0.45, 0.05, 0.95) & \multirow{2}{*}{1.329 (1.343, 0.274, 1.005, 0.996)} & \multirow{2}{*}{0.010 (0.009, 0.008, 0.019, 0.999)} & \multirow{2}{*}{27.000 (59.231, 39.071, 197.466, 0.949)} & \multirow{2}{*}{288} & \multirow{2}{*}{548} \\ & (0.25, 0.90, 0.10) & & & & &\\ \hline
16 & (0.95, 0.95, 0.05) & \multirow{2}{*}{1.683 (1.900, 0.675, 3.536, 0.990)} & \multirow{2}{*}{82.819 (163.388, 80.285, 695.320, 0.958)} & \multirow{2}{*}{0.037 (0.030, 0.021, 0.030, 0.999)} & \multirow{2}{*}{83} & \multirow{2}{*}{348} \\ & (0.90, 0.10, 0.90) & & & & &\\ \hline
\end{tabular}
\label{tab : Tau with other statistics for relative risk function with 1000}%
}
	\end{table}

\begin{table}[ht!]
\caption{Estimated first stage ($\hat\tau_A$) and second stage ($\hat\tau_{AC},\hat\tau_{BE}$) allocation ratios along with corresponding SSE, ASE and CP based on 5000 simulations. $\tau_A$, $\tau_{AC}$, and $\tau_{BE}$ denote true values of optimum allocation ratios. Here, $\gamma_A = 0.4$, and $\gamma_B=0.3$, and the sample size is \textbf{2000} using objective function as \textbf{Relative Risk}. It also shows the total expected number of failures at the end of SMART using optimal allocation (proposed method) and equal randomization. \\}
\resizebox{\columnwidth}{!}{%
\renewcommand{\arraystretch}{2}%
\begin{tabular}{|c|c|c|c|c|c|c|}
\hline
No. & $(p_{AA'},p_{AC},p_{AD})$ & \multirow{2}{*}{$\tau_{A}\ (\hat{\tau}_A,\ SSE,\ ASE,\ \widehat{CP})$} & \multirow{2}{*}{$\tau_{AC}\ (\hat{\tau}_{AC},\ SSE,\ ASE,\ \widehat{CP})$} & \multirow{2}{*}{$\tau_{BE}\ (\hat{\tau}_{BE},\ SSE,\ ASE,\ \widehat{CP})$} & \multicolumn{2}{c|}{Expected number of failures} \\ \cline{6-7} 
& $(p_{BB'},p_{BE},p_{BF})$ & & & & Optimal & Equal \\ \hline


1 & (0.20, 0.15, 0.15) & \multirow{2}{*}{0.235 (0.233, 0.021, 0.021, 0.945)} & \multirow{2}{*}{1.000 (1.039, 0.318, 0.298, 0.931)} & \multirow{2}{*}{0.665 (0.668, 0.076, 0.075, 0.947)} & \multirow{2}{*}{916} & \multirow{2}{*}{1205} \\ & (0.45, 0.65, 0.75) &  & & & &  \\ \hline
2 & (0.30, 0.80, 0.20) & \multirow{2}{*}{1.006 (1.023, 0.070, 0.069, 0.955)} & \multirow{2}{*}{8.000 (9.580, 5.909, 9.278, 0.958)} & \multirow{2}{*}{1.175 (1.185, 0.146, 0.145, 0.949} & \multirow{2}{*}{958} & \multirow{2}{*}{1104} \\ & (0.25, 0.60, 0.55) &  & & & &  \\ \hline
3 & (0.80, 0.40, 0.60) & \multirow{2}{*}{1.512 (1.518, 0.105, 0.103, 0.945)} & \multirow{2}{*}{0.544 (0.544, 0.785, 0.769, 0.945)} & \multirow{2}{*}{1.889 (1.927, 0.686, 0.965, 0.910)} & \multirow{2}{*}{365} & \multirow{2}{*}{930} \\ & (0.35, 0.65, 0.45) & & & & & \\ \hline
4 & (0.30, 0.20, 0.80) & \multirow{2}{*}{1.301 (1.309, 0.083, 0.081, 0.956)} & \multirow{2}{*}{0.125 (0.120, 0.029, 0.025, 0.900)} & \multirow{2}{*}{0.235 (0.228, 0.046, 0.040, 0.908)} & \multirow{2}{*}{983} & \multirow{2}{*}{1243} \\ & (0.25, 0.15, 0.60) & & & & & \\ \hline
5 & (0.30, 0.20, 0.20) & \multirow{2}{*}{0.442 (0.439, 0.030, 0.030, 0.943)} & \multirow{2}{*}{1.000 (1.016, 0.178, 0.166, 0.941)} & \multirow{2}{*}{0.235 (0.233, 0.034, 0.031, 0.927)} & \multirow{2}{*}{1077} & \multirow{2}{*}{1302} \\ & (0.65, 0.15, 0.60) & & & & & \\ \hline
6 & (0.30, 0.60, 0.45) & \multirow{2}{*}{0.846 (0.846, 0.053, 0.053, 0.951)} & \multirow{2}{*}{1.588 (1.609, 0.223, 0.219, 0.947)} & \multirow{2}{*}{1.000 (1.004, 0.120, 0.120, 0.945)} & \multirow{2}{*}{1056} & \multirow{2}{*}{1071}\\ & (0.25, 0.60, 0.60) & & & & & \\ \hline
7 & (0.30, 0.80, 0.80) & \multirow{2}{*}{1.329 (1.321, 0.083, 0.083, 0.946)} & \multirow{2}{*}{1.000 (1.012, 0.178, 0.178, 0.944)} & \multirow{2}{*}{0.235 (0.230, 0.044, 0.040, 0.922)} & \multirow{2}{*}{836} & \multirow{2}{*}{941} \\ & (0.65, 0.15, 0.60)  & & & & & \\ \hline
8 & (0.30, 0.80, 0.80) & \multirow{2}{*}{2.475 (2.489, 0.161, 0.159, 0.946)} & \multirow{2}{*}{1.000 (1.011, 0.157, 0.158, 0.947)} & \multirow{2}{*}{1.000 (1.018, 0.182, 0.171, 0.941)} & \multirow{2}{*}{975} & \multirow{2}{*}{1099} \\ & (0.65, 0.15, 0.15) & & & & &\\ \hline
9 & (0.30, 0.20, 0.20) & \multirow{2}{*}{0.414 (0.413, 0.030, 0.030, 0.942)} & \multirow{2}{*}{1.000 (1.017, 0.179, 0.169, 0.942)} & \multirow{2}{*}{1.563 (1.576, 0.157, 0.153, 0.946)} & \multirow{2}{*}{1098} & \multirow{2}{*}{1232} \\ & (0.65, 0.55, 0.40)  & & & & & \\ \hline 
\multicolumn{3}{|c}{Very high/low success probability values} & \multicolumn{4}{c|}{} \\ \hline  
10 & (0.10, 0.10, 0.10) & \multirow{2}{*}{1.000 (1.004, 0.085, 0.083, 0.943)} & \multirow{2}{*}{1.000 (1.013, 0.165, 0.159, 0.945)} & \multirow{2}{*}{1.000 (1.009, 0.152, 0.145, 0.947)} & \multirow{2}{*}{1800} & \multirow{2}{*}{1800} \\ & (0.10, 0.10, 0.10) & & & & & \\ \hline
11 & (0.05, 0.05, 0.05) & \multirow{2}{*}{1.000 (1.006, 0.113, 0.111, 0.948)} & \multirow{2}{*}{1.000 (1.031, 0.250, 0.234, 0.949)} & \multirow{2}{*}{1.000 (1.020, 0.223, 0.209, 0.944)} & \multirow{2}{*}{1900} & \multirow{2}{*}{1900} \\ & (0.05, 0.05, 0.05) & & & & &\\ \hline
12 & (0.90, 0.90, 0.90) & \multirow{2}{*}{1.000 (1.008, 0.146, 0.145, 0.940)} & \multirow{2}{*}{1.000 (1.036, 0.294, 0.281, 0.928)} & \multirow{2}{*}{1.000 (1.033, 0.266, 257, 0.940)} & \multirow{2}{*}{200} & \multirow{2}{*}{200} \\ & (0.90, 0.90, 0.90) & &  & & &\\ \hline
13 & (0.95, 0.95, 0.95) & \multirow{2}{*}{1.000 (1.019, 0.219, 0.211, 0.936)} & \multirow{2}{*}{1.000 (1.087, 0.489, 0.448, 0.924)} & \multirow{2}{*}{1.000 (1.073, 0.420, 0.396, 0.922)} & \multirow{2}{*}{100} & \multirow{2}{*}{100} \\ & (0.95, 0.95, 0.95) & & & & &\\ \hline
14 & (0.35, 0.95, 0.05) & \multirow{2}{*}{0.760 (0.774, 0.149, 0.399, 0.992)} & \multirow{2}{*}{82.819 (158.768, 76.379, 568.527, 0.950)} & \multirow{2}{*}{27.000 (45.587, 30.845, 84.325, 0.954)} & \multirow{2}{*}{477} & \multirow{2}{*}{1015} \\ & (0.65, 0.90, 0.10) & & & & &\\ \hline
15 & (0.45, 0.05, 0.95) & \multirow{2}{*}{1.329 (1.362, 0.246, 0.667, 0.989)} & \multirow{2}{*}{0.010 (0.009, 0.007, 0.013, 0.999)} & \multirow{2}{*}{27.000 (50.493, 33.603, 111.278, 0.958)} & \multirow{2}{*}{566} & \multirow{2}{*}{1096} \\ & (0.25, 0.90, 0.10) & & & & &\\ \hline
16 & (0.95, 0.95, 0.05) & \multirow{2}{*}{1.683 (1.910, 0.622, 2.322, 0.980)} & \multirow{2}{*}{82.819 (152.549, 70.758, 425.214, 0.949)} & \multirow{2}{*}{0.037 (0.030, 0.018, 0.020, 0.998)} & \multirow{2}{*}{160} & \multirow{2}{*}{699} \\ & (0.90, 0.10, 0.90) & & & & &\\ \hline
\end{tabular}
\label{tab : Tau with other statistics for relative risk function with 2000}%
}
	\end{table}
 
\clearpage

\subsection{\centering \textbf{Application to M-Bridge Data: Odds Ratio and Relative Risk}} 
In the main paper, we have demonstrated an application of the developed adaptive allocation (randomization) procedure using M-Bridge data \citep{patrick2021main} with the simple difference of the success probabilities (separately for the first and second stages) as the objective function. Here, we show the application of the developed procedure using the M-Bridge data for the two other objective functions, namely odds ratio and relative risk.

\begin{figure}[!hbt]
     \centering
     \begin{subfigure}[b]{0.3\textwidth}
         \centering
         \includegraphics[width=\textwidth]{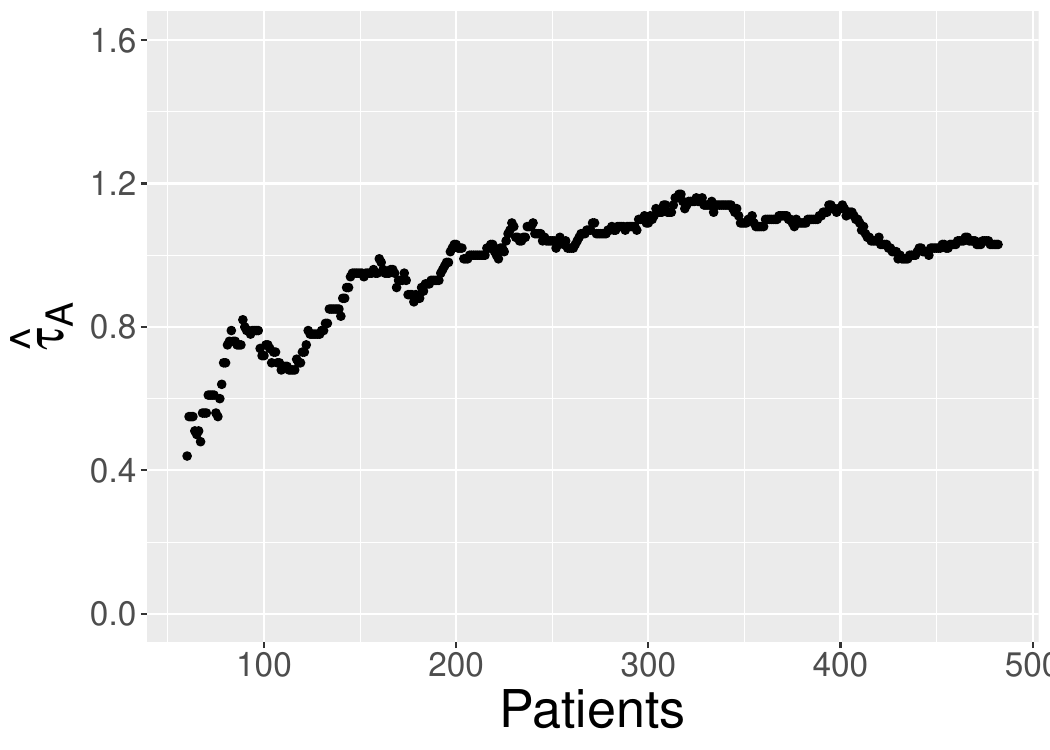}
         \caption{$\hat \tau_{A}$: The estimated first stage optimal allocation ratio.\\}
         \label{fig: mbridge tau a for odds ratio}
     \end{subfigure}
     \hfill
     \begin{subfigure}[b]{0.3\textwidth}
         \centering
         \includegraphics[width=\textwidth]{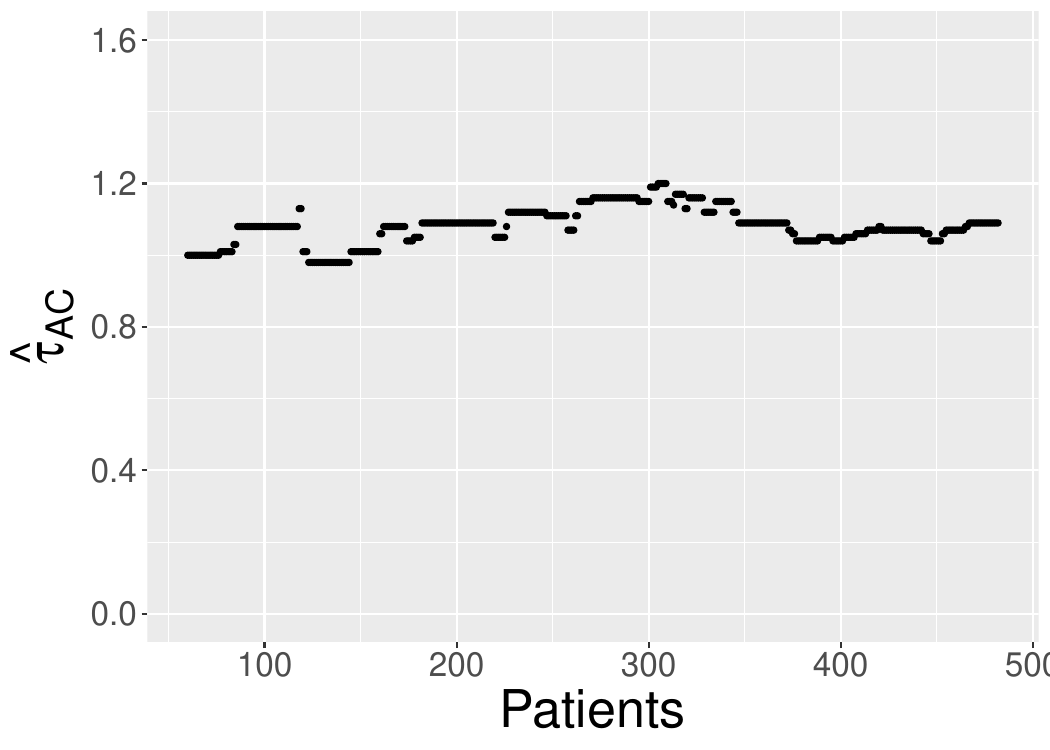}
         \caption{$\hat \tau_{AC}$: The estimated second stage optimal allocation ratio for non-responders who obtained $A$ at the first stage.}
         \label{fig:mbridge tau ac for odds ratio}
     \end{subfigure}     
     \hfill
     \begin{subfigure}[b]{0.3\textwidth}
         \centering
         \includegraphics[width=\textwidth]{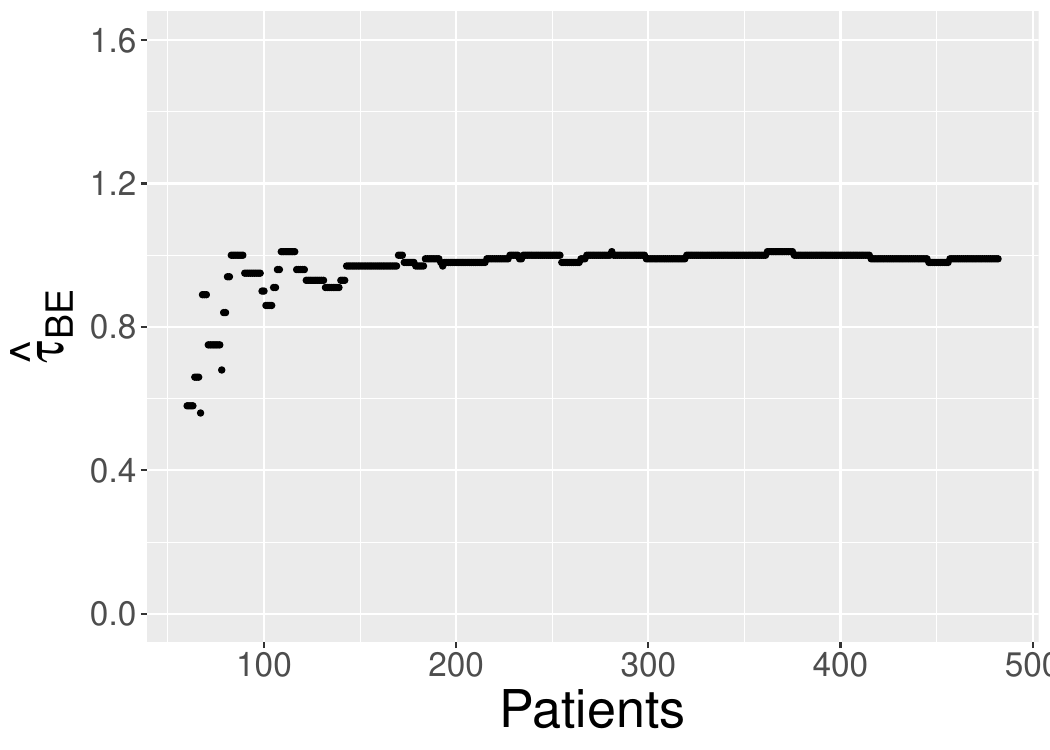}
         \caption{$\hat \tau_{BE}$: The estimated second stage optimal allocation ratio for non-responders who obtained $B$ at the first stage.}
         \label{fig:mbridge tau be for odds ratio}
     \end{subfigure}
        \caption{Convergence patterns of estimated (black dots) optimal allocation ratios $\hat \tau_{A}$, $\hat \tau_{AC}$, and $\hat \tau_{BE}$ in the M-Bridge study with the objective function as odds ratio.}
        \label{fig:convergence of all tau for mbridge for odds ratio}
\end{figure}

\begin{figure}[!hbt]
     \centering
     \begin{subfigure}[b]{0.3\textwidth}
         \centering
         \includegraphics[width=\textwidth]{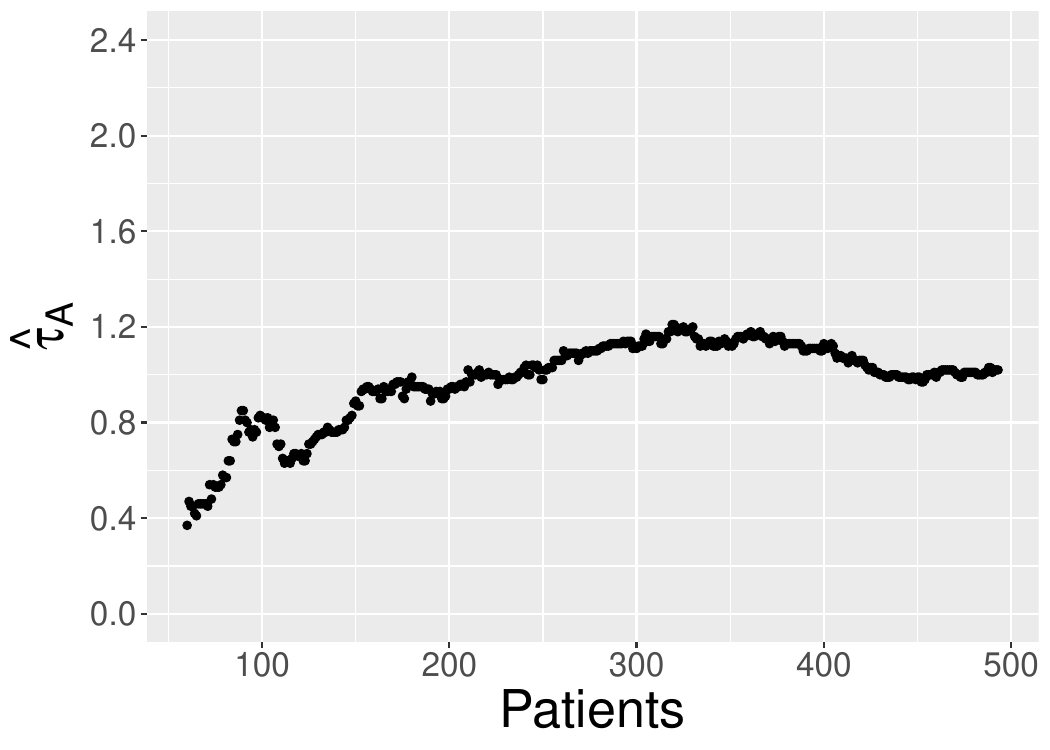}
         \caption{$\hat \tau_{A}$: The estimated first stage optimal allocation ratio.\\}
         \label{fig: mbridge tau a for relative risk}
     \end{subfigure}
     \hfill
     \begin{subfigure}[b]{0.3\textwidth}
         \centering
         \includegraphics[width=\textwidth]{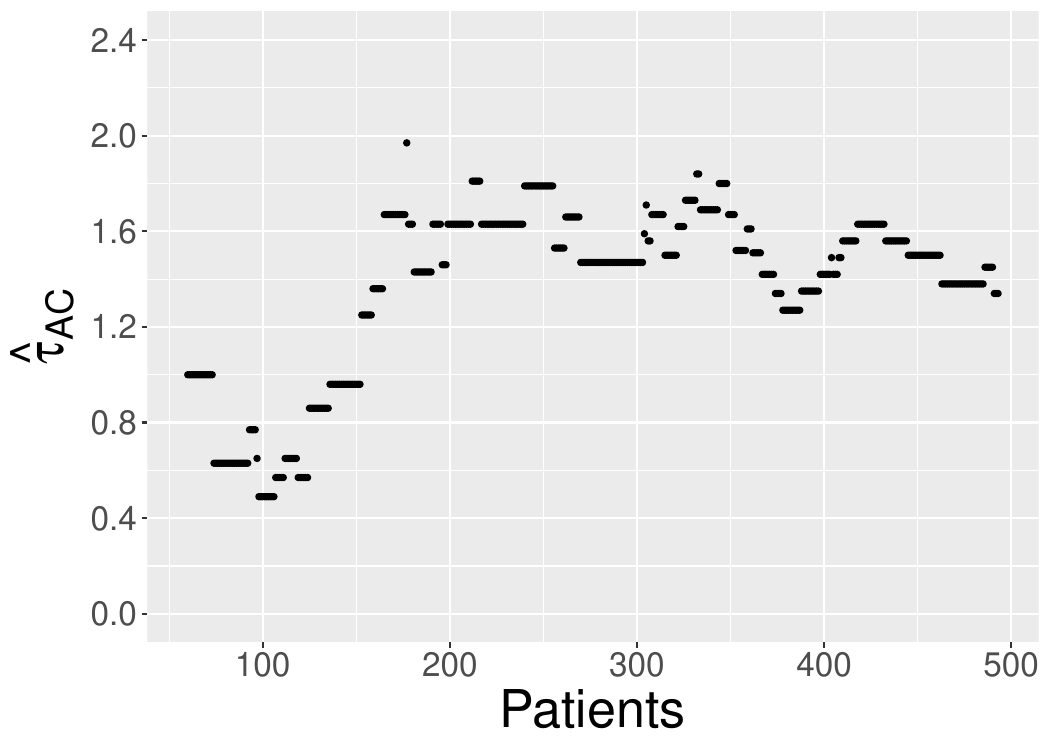}
         \caption{$\hat \tau_{AC}$: The estimated second stage optimal allocation ratio for non-responders who obtained $A$ at the first stage.}
         \label{fig:mbridge tau ac for relative risk}
     \end{subfigure}     
     \hfill
     \begin{subfigure}[b]{0.3\textwidth}
         \centering
         \includegraphics[width=\textwidth]{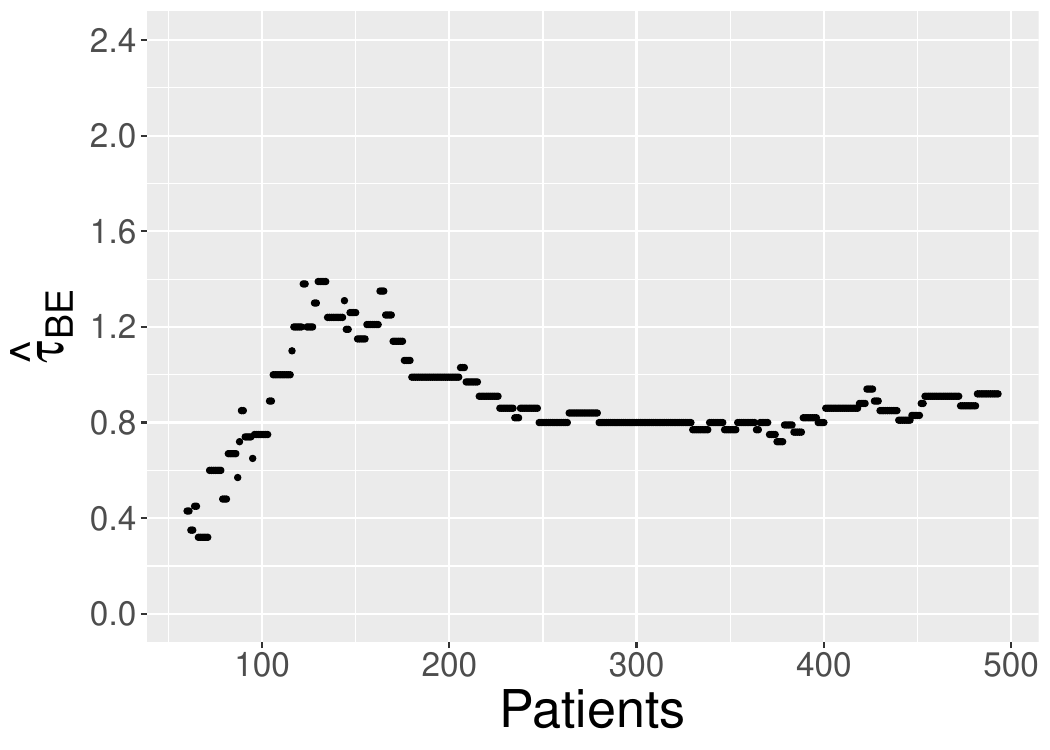}
         \caption{$\hat \tau_{BE}$: The estimated second stage optimal allocation ratio for non-responders who obtained $B$ at the first stage.}
         \label{fig:mbridge tau be for relative risk}
     \end{subfigure}
        \caption{Convergence patterns of estimated (black dots) optimal allocation ratios $\hat \tau_{A}$, $\hat \tau_{AC}$, and $\hat \tau_{BE}$ in the M-Bridge study with the objective function as relative risk.}
        \label{fig:convergence of all tau for mbridge for relative risk}
\end{figure}

Figures \ref{fig:convergence of all tau for mbridge for odds ratio} and \ref{fig:convergence of all tau for mbridge for relative risk} report the convergence patterns of each of the three allocation ratios for the objective function of odds ratio, and relative risk, respectively. On the other hand, Tables \ref{tab : dtr alloc for mbridge study for odds} and \ref{tab : dtr alloc for mbridge study for relative risk} report the allocated patients to the different dynamic treatment regimes and compare the result with equal (M-Bridge) allocation procedure. Following the optimal adaptive allocation process using both objective functions, Tables \ref{tab : dtr alloc for mbridge study for odds} and \ref{tab : dtr alloc for mbridge study for relative risk} show there is an improvement over the M-Bridge (equal) allocation.

	\begin{table}[!ht]
\caption{Allocated participants and proportion of failures (in parentheses) following optimal adaptive allocation (OAA) and 1:1 allocation in M-Bridge study. The \textbf{odds ratio} of the success probabilities is used as the objective function $g(\cdot, \cdot)$. The OAA has to stop after 482 participants as treatment sequence $\{B, B'\}$ of the M-Bridge SMART has no available participants. The proportion of failures for $d_i$ is $q_{d_{i}}$, whereas the proportion of failure (in the last row) is the ratio of the total number of failures to total participants.   \\}
\resizebox{\columnwidth}{!}{%
\renewcommand{\arraystretch}{2}%
\begin{tabular}{|c|c|c|c|c|c|}
\hline
\multirow{2}{*}{$DTR$} & \multicolumn{5}{c|}{Responder (R) + Non-Responder (NR) = Total (Proportion of failures) }\\
\cline{2-6} 
 & \multicolumn{2}{c|}{Optimal Adaptive Allocation (OAA)} & \multicolumn{2}{c|}{M-Bridge Allocation} & M-Bridge Study allocation (end of study) \\ \cline{2-6} 
& Participants with OAA  & Remaining participants & Till 482 participants & Remaining 39 participants & All participants \\ \cline{2-6} 
\hline
$d_1$ & 178 + 36 = 214 (0.231) & 9 + 1 = 10 (0.635) & 171 + 36 = 207 (0.227) & 16 + 1 = 17 (0.580) & 187 + 37 = 224 (0.235) \\ \hline
$d_2$ & 178 + 22 = 200 (0.270) & 9 + 16 = 25 (0.706) & 171 + 35 = 206 (0.271) & 16 + 3 = 19 (0.580) & 187 + 38 = 225 (0.284) \\ \hline
$d_3$ & 176 + 33 = 209 (0.266) & 0 + 3 = 3 (0.107) & 165 + 30 = 195 (0.282) & 11 + 6 = 17 (0.770) & 176 + 36 = 212 (0.266) \\ \hline
$d_4$ & 176 + 37 = 213 (0.252) & 0 + 10 = 10 (0.128) & 165 + 45 = 210 (0.250) & 11 + 2 = 13 (0.716) & 176 + 47 = 223 (0.252) \\ \hline
$Total$ & 482 (0.234) & 39 (0.564) & 482 (0.256) & 39 (0.308) & 521 (0.259) \\ \hline
\end{tabular}
\label{tab : dtr alloc for mbridge study for odds}%
}
\end{table}

	\begin{table}[!ht]
\caption{Allocated participants and proportion of failures (in parentheses) following optimal adaptive allocation (OAA) and 1:1 allocation in M-Bridge study. The \textbf{relative risk} of the success probabilities is used as the objective function $g(\cdot, \cdot)$. The OAA has to stop after 493 participants as treatment sequence $\{B, B'\}$ of the M-Bridge SMART has no available participants. The proportion of failures for $d_i$ is $q_{d_{i}}$, whereas the proportion of failure (in the last row) is the ratio of the total number of failures to total participants.   \\}
\resizebox{\columnwidth}{!}{%
\renewcommand{\arraystretch}{2}%
\begin{tabular}{|c|c|c|c|c|c|}
\hline
\multirow{2}{*}{$DTR$} & \multicolumn{5}{c|}{Responder (R) + Non-Responder (NR) = Total (Proportion of failures) }\\
\cline{2-6} 
 & \multicolumn{2}{c|}{Optimal Adaptive Allocation (OAA)} & \multicolumn{2}{c|}{M-Bridge Allocation} & M-Bridge Study allocation (end of study) \\ \cline{2-6} 
& Participants with OAA  & Remaining participants & Till 493 participants & Remaining 28 participants & All participants \\ \cline{2-6} 
\hline
$d_1$ & 183 + 35 = 218 (0.237) & 4 + 2 = 10 (0.857) & 177 + 36 = 213 (0.232) & 10 + 1 = 11 (0.643) & 187 + 37 = 224 (0.235) \\ \hline
$d_2$ & 183 + 26 = 209 (0.267) & 4 + 12 = 25 (0.762) & 177 + 36 = 213 (0.279) & 10 + 2 = 12 (0.643) & 187 + 38 = 225 (0.284) \\ \hline
$d_3$ & 176 + 34 = 210 (0.260) & 0 + 2 = 2 (0.000) & 168 + 31 = 199 (0.278) & 8 + 5 = 13 (0.787) & 176 + 36 = 212 (0.266) \\ \hline
$d_4$ & 176 + 39 = 215 (0.251) & 0 + 8 = 8 (0.120) & 168 + 45 = 213 (0.253) & 8 + 2 = 10 (0.755) & 176 + 47 = 223 (0.252) \\ \hline
$Total$ & 493 (0.237) & 28 (0.643) & 493 (0.258) & 28 (0.286) & 521 (0.259) \\ \hline
\end{tabular}
\label{tab : dtr alloc for mbridge study for relative risk}%
}
\end{table}

\end{document}